\documentclass[twocolumn]{aastex63}
\usepackage{amsmath}
\usepackage{rotating}
\usepackage{epsf}
\usepackage{psfig}
\usepackage{color}
\usepackage [english]{babel}
\usepackage [autostyle, english = american]{csquotes}
\MakeOuterQuote{"}
\pdfoutput=1 
\usepackage{graphicx}
\usepackage[inline]{enumitem}
\usepackage{verbatim}
\usepackage{color}
\usepackage{longtable}

\shorttitle{Taffy in CO with ALMA}
\shortauthors{Appleton et al.}

\newcommand{\kms}{km~s$^{-1}$}

\begin{document}

\title{The CO emission in the Taffy Galaxies (UGC 12914/5) at 60pc resolution-I: The battle for star formation in the turbulent Taffy Bridge}

\author{P. N. Appleton}
\affiliation{Caltech/IPAC, MC 314-6, 1200 E. California Blvd., Pasadena, CA 91125, USA. apple@ipac.caltech.edu}
\author{B. Emonts}
\affiliation{National Radio Astronomy Observatory, 520 Edgemont Road, Charlottesville, VA 22903, USA}
\author{U. Lisenfeld}
\affiliation{Departamento de Fisica Teorica y del Cosmos, Universidad de Granada, Spain and Instituto Carlos I de Fisica Teorica y Computacional, Universidad de Granada, Spain}
\author{E. Falgarone}
\affiliation{LERMA/LRA, Observatoire de Paris, PSL Research University, CNRS, Sorbonne Universites, UPMC Universite Paris 06, Ecole normale superieure, 75005 Paris, France}
\author{P. Guillard}
\affiliation{Sorbonne Universit\'es, UPMC Paris 6 et CNRS, UMR 7095, Institut d'Astrophysique de Paris, 98 bis Bd Arago, 75014 Paris, France}
\author{F. Boulanger}
\affiliation{UPMC Universite Paris 06, Ecole normale superieure, 75005 Paris, France}
\author{J. Braine}
\affiliation{Laboratoire d'Astrophysique de Bordeaux, Univ. Bordeaux, CNRS, B18N, all\'ee Geoffroy Saint-Hilaire, 33615 Pessac, France}
\author{P. Ogle}
\affiliation{Space Telescope Science Institute, 3700, San Martin Drive, Baltimore, MD21218,  USA}
\author{C. Struck}
\affiliation{Dept. of Physics and Astronomy, Iowa State University, Ames, Iowa, 50011}
\author{B. Vollmer}
\affiliation{Universit\'e de Strasbourg, F-67000 Strasbourg, France}
\author{T. Yeager}
\affiliation{Lawrence Livermore Laboratories, 7000 East Avenue, Livermore, CA 94550}
\submitjournal{ApJ}
\accepted{March 28 2022}

\begin{abstract}
We present ALMA observations at a spatial resolution of 0.2 arcsec (60 pc) of CO emission from the Taffy galaxies (UGC 12914/5).  The observations are compared with narrow-band Pa$\alpha$, mid-IR, radio continuum and X-ray imaging, plus optical spectroscopy. The galaxies have undergone a recent head-on collision, creating a massive gaseous bridge which is known to be highly turbulent. The bridge contains a complex web of narrow molecular filaments and clumps. The majority of the filaments are devoid of star formation, and fall significantly below the Kennicutt-Schmidt relationship for normal galaxies, especially for the numerous regions undetected in Pa$\alpha$ emission. Within the loosely connected filaments and clumps of gas we find regions of high velocity dispersion which appear gravitationally unbound for a wide range of likely values of $X_{\rm CO}$. Like the "Firecracker" region in the Antennae system, they would require extremely high external dynamical or thermal pressure to stop them dissipating rapidly on short crossing timescales of 2-5~Myrs. We suggest that the clouds may be transient structures within a highly turbulent multi-phase medium which is strongly suppressing star formation. Despite the overall turbulence in the system, stars seem to have formed in compact hotspots within a kpc-sized extragalactic HII region, where the molecular gas has a lower velocity dispersion than elsewhere, and shows evidence for a collision with an ionized gas cloud. Like the shocked gas in the Stephan's Quintet group, the conditions in the Taffy bridge shows how difficult it is to form stars within a turbulent, multi-phase, gas.
 \end{abstract}


\section{Introduction} 

Major mergers between massive gas rich galaxies are transformative events in galaxy evolution.  In many mergers, models suggest that tidal torques within host galaxy disks drives gas inwards, often forming an intense dust-enshrouded nuclear starburst (e. g. \citealt{Mihos1996}). Such a mechanism has been long accepted as the main explanation for the existence of rare, but powerful sources of far-IR emission (LIRGs and ULIRGs)\footnote{(U)LIRG=(Ultra)Luminous Infrared Galaxies in the local universe are defined as having high far-IR luminosities L$_{IR}$ $>$ (10$^{12}$~L$_{\odot}$) 10$^{11}$L$_{\odot}$.} in the local Universe (e.g.~\citealt{Soifer1984a,Soifer1984b,Sanders1986,Armus1987,Sanders1996,Armus2009, Saito2015,Armus2020}).  

While much of the early modeling of colliding and massive merging galaxies primarily was aimed at more general collision geometries, head-on collisions, thought to be responsible for collisional ring galaxies, have always been a special case \citep{Lynds1976,Toomre1978,Appleton1987,Struck-Marcell1990,Gerber1996}. Although several of these systems involved dissipative gas-rich collisons (e. g. \citealt{Appleton1996,Charm1996, Higdon1996, Braine2003,Braine2004}), there has been a resurgence of interest in the treatment of the dissipative effects of the gas \citep{Renaud2018}, especially the formation of a "splash" bridge \citep{Struck1997,Yeager2019,Yeager2020a,Yeager2020b}. 
Splash gas bridges are produced when two gas-rich disk systems collide nearly head-on. In such cases, the stellar components pass through each other, but leave behind a massive gas bridge \citep{Vollmer2012}. These kinds of collisions are challenging to models because a large fractions of the gas is strongly compressed and heated during the collision, and much of the multi-phase medium remains at various stages of cooling tens of millions of years after the collision (e. g. \citealt{Yeager2020a,Yeager2020b}). These latter models suggest that the gas between the galaxies continues to collide with high Mach-number, generating strong turbulent conditions in the bridge.  Gas in such bridge systems can be used to study the effects of turbulence and shocks on the formation of stars in a relatively "clean" environment, far from the complicating effects of nuclear starbursts or AGN found in many other major merger systems. 

\begin{figure*}
\includegraphics[width=0.9\textwidth]{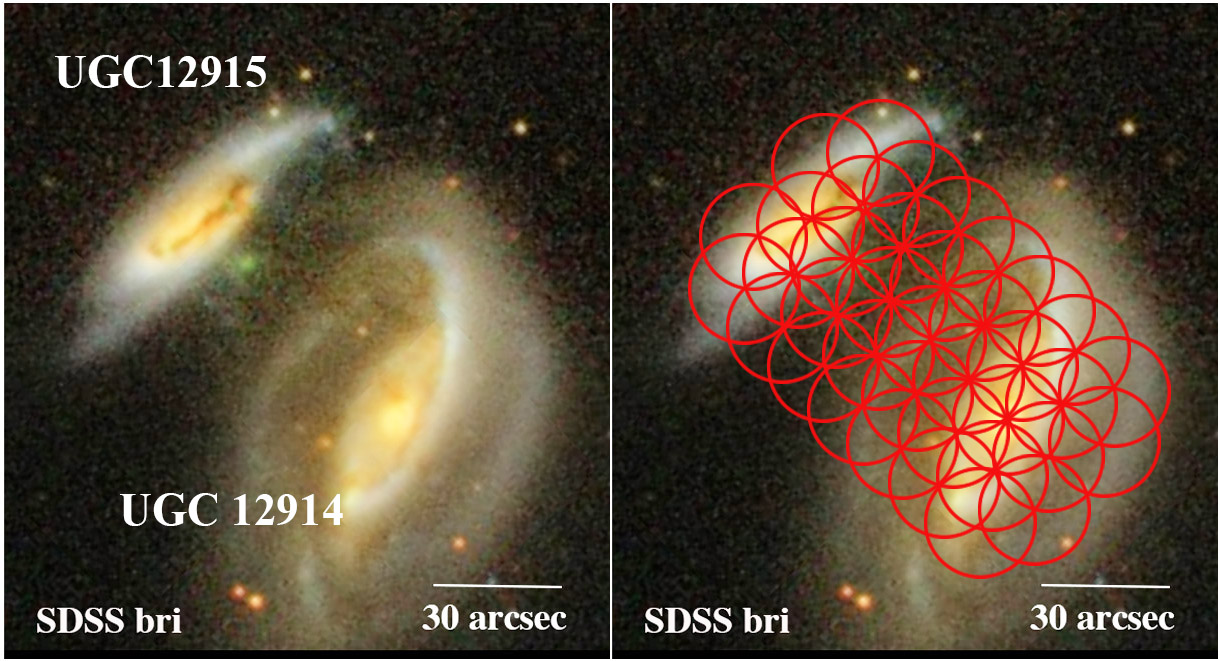}
\caption{The Taffy galaxy system, a) shown in a false color SDSS bri-band (blue = b, green = r, red = i) image designed to show the faintest emission, including a faint extragalactic HII region just south-west of UGC 12915, b) red circles show the primary beams of the 37 ALMA 12-m pointings (FWHM = 25.6 arcsec) for CO (2-1) observation described here. 1 arcsec corresponds to 300 pc at D = 62 Mpc} \label{fig:almafields}
\end{figure*}

We present  new high-resolution CO observations of one of the best studied splash bridge systems, known as the Taffy galaxies (UGC 12914/5, \citealt{Condon1993}; stellar masses of 7.4 and 4.2 $\times~10^{10}~M_{\odot}$ respectively;~\citealt{Appleton2015}). The two gas rich disks are believed to have collided almost face-on at high velocity (600-800 \kms) 25-30~Myr ago (see Figure~\ref{fig:almafields}a). The ionized gas in the galaxy disks appear to be counter-rotating \citep{Joshi2019}, suggesting the discs has oppositely oriented spins when the disks first collided. This geometry would create an even stronger affect on the gas collisions at the time of impact. We are now likely viewing the system almost edge-on after the stellar components have passed through each other \citep{Vollmer2012,Vollmer2021}, leaving behind a massive neutral and molecular gas bridge in the center of mass frame of the galaxies. 
 
In addition to the disturbed CO distribution studied with varying degrees of spatial resolution \citep{Gao2003,Zhu2007,Braine2003,Vollmer2021}, there is independent evidence that the gas in the bridge is highly disturbed. {\t Spitzer} IRS observations have shown the existence of large amounts of warm (T $\sim$ 100-300 K) emitting molecular hydrogen \citep{Peterson2012} with properties consistent with shock or turbulent heating \citep{Guillard2009}. The spectra showed powerful dominant emission lines of pure-rotational H$_2$, with large ratios of warm H$_2$/PAH and H$_2$/FIR, similar to those seen in the Stephan's Quintet intergalactic shock \citep{Appleton2006, Cluver2010}. {\it Herschel} PACS observations showed that the bridge also emits strong [CII]158$\mu$m and [OI]63$\mu$m emission, as well as emission from [CI], CO~(4-3) and CO (5-4) based SPIRE FTS observations \citep{Peterson2018}. The strength and unusual line ratios of the fine structure lines point towards heating by shocks and turbulence, as suggested by recent models \citep{Yeager2020b}. Direct evidence for fast atomic shocks (V $\sim$ 200 \kms) was found in the bridge \citep{Joshi2019}, as well as potentially faster shocks from {\it Chandra} observations of soft X-ray emission \citep{Appleton2015}. Radio continuum emission from the bridge \citep{Condon1993}, can also be explained as a result of Fermi accelerations of cosmic rays in shocks generated within the turbulent gas \citep{Lisenfeld2010}.

 Recently \citet{Vollmer2021} has reported the highest spatial resolution (2.7 arcsec) observations of CO (1-0) in the Taffy system with the IRAM Plateau de Bure Interferometer (PdBI), as well as detailed models of the structure and kinematics of the large scale gas distribution. The results provide very strong support for the collisional picture, and evidence for star formation suppression in the bridge. The results suggest that much of the gas in the bridge is not virialized on the 800 pc to few kpc scales, and they suggest turbulent adiabatic compression is responsible for the high velocity dispersion in the observed gas clumps. One exception is a luminous extragalactic HII region (hereafter X-HII region) which may have formed in the bridge, close to the northern-most Taffy galaxy UGC 12915.       
 
This paper  presents Atacama Large Millimeter Array (ALMA)\footnote{ALMA, an international astronomy facility, is a partnership of ESO, the U.S. National Science Foundation (NSF) and the National Institutes of Natural Sciences (NINS) of Japan in cooperation with the Republic of Chile.} observations of the CO (2-1) emission from the Taffy system on angular scales (0.24 x 0.18 arcsec$^2$), an order of magnitude higher than previous work. 

Because of the large quantity of data obtained in the ALMA mosaics, we split our discussion of the ALMA CO data into two papers. The current one concentrates on the CO (2-1) emission from the gas bridge from Cycle 4. A second paper will discuss more fully the condition of the molecular gas in the two Taffy galaxies themselves, and will present CO (3-2) data obtained both in Cycle 4, and in Cycle 7. 

The main goals of this paper are, 1) to explore the distribution of molecular gas in the Taffy bridge at 60-100 pc resolution and its relationship to Pa$\alpha$ emission observed with HST at similar resolution, 2) probe the kinematics of the CO gas emission and how that relates to shocks and star formation previously observed from ground and space based data, and 3) study in greater detail the one major area of star formation in the bridge (the extragalactic HII region). This region may provide further insights into the formation of stars in turbulent environments through comparison with new optical spectroscopy, new radio continuum, and archival {\it Spitzer} and X-ray observations. 

The paper is organized as follows. The observations and data calibrations are described in \S2. Results, including the large-scale molecular distribution and kinematics and gas surface densities estimates are presented in~\S3. \S4 describes the ionized gas emission in the bridge and in the X-HII regions.  \S5 is concerned with the relationship between molecular gas and star formation in the bridge, including testing the Kennicutt-Schmidt relationship for the clouds, the virial properties of the clouds, and quantifying the star formation rate in the X-HII region.  The Taffy bridge is compared with other similar intergalactic environments in \S6. The possible origin of large-scale star formation suppression in the bridge is described in \S7. The conclusions are given in \S8. An Appendix includes tabulated data and figures relating to the extracted regions discussed in the main body of the paper.

We assume a distance to the Taffy galaxies of 62~Mpc based on a  mean heliocentric
velocity for the system of 4350~\kms, and a Hubble constant of 70~\kms~Mpc$^{-1}$. At this distance, 1 arcsec corresponds to 300~pc.
\section{Observations} 

\subsection{CO Observations and reduction}
Table~\ref{tab:obs} is a complete list of the observations made with the ALMA 12-m  arrays in programs 2016.1.01037.S (40 antennas) and 2019.1.01050.S (41 antennas).  In 2016.1.01037.S, 37 full-sampled primary beam pointings with the 12-m array were made of the Taffy pair and bridge in $^{12}$CO$(2-1)$~($\nu_{rest}$=230.54~GHz; see Figure~\ref{fig:almafields}b). Two sets of seven pointings were also made in the $^{12}$CO$(3-2)$ ($\nu_{rest}$=345.795~GHz) centered on the brightest part of the bridge, and a bright region in UGC 12914. A further set of CO (3-2) observations were made of the bridge and UGC 12915 in program 2019.1.01050.S (41 antennas). Although we list all the observations in the table, we postpone discussion of the CO(3-2) observations until a second paper.  

The CO (2-1)  observations centered on a heliocentric (optically defined) velocity of 4450 \kms~were observed with a total bandwidth of 1.875 GHz (2472 \kms) and a channel separation of 1.953~MHz (2.6 \kms) in ALMA Band 6. A second continuum baseband was centered on $\nu$=228 GHz ($\lambda$1.3mm, and bandwidth 1.875GHz).   

Calibration of the CO(2-1) data was performed using ALMA flux and phase-reference calibration sources during the course of the observations. These data were processed with a standard ALMA calibration pipeline included in CASA v.5.5.1-5 
resulting in fully flux, phase and bandpass-calibrated visibility data. The quality of the calibration was carefully reviewed before performing exploratory Fourier transforms of the ALMA visibility data to produce initial "dirty channel maps" smoothed to a resolution of 10 \kms. The maps were made in each channel over a scale of 8000 x 8000 pixel$^{2}$, where the pixel scale was 0.018 arcsec. 

\begin{deluxetable*}{llllllccccl}
\tablecolumns{11}
\tablewidth{0pc}
\tablecaption{Detail of ALMA  Observations \label{tab:obs}}
\tablehead{
\colhead{Cycle} &
\colhead{Band} & 
\colhead{Array} &
\colhead{Date} &
\colhead{Line} &
\colhead{Frequency} &
\colhead{Mosaic} &
\colhead{Total} &
\colhead{Time} &
\colhead{Final Sythesized} &
\colhead{Refs.\tablenotemark{a} }
\\
\colhead{} & 
\colhead{} & 
\colhead{} &
\colhead{} &
\colhead{} &
\colhead{Sky (GHz)} &
\colhead{(number)} &
\colhead{Time (min) } &
\colhead{on-source (min)} &
\colhead{Beam size ($"$)} &
\colhead{}
}
\startdata
4 & 6    & 12m C40-6 & 2016-10-04 & CO(2-1) & 227.1 & 37\tablenotemark{b}  & 63.3 & 39.6 & 0.24x0.14 & 1 \\
4 & 6    & 12m C40-6 & 2016-10-05 & CO(2-1) & 227.1 & 37\tablenotemark{b}  & 64.2 & 39.6 & 0.24x0.14 & 1 \\
4 &6    & 12m C40-6 & 2016-10-06 & CO(2-1) & 227.1 & 37\tablenotemark{b}  & 72.97 & 39.6 & 0.24x0.14 & 1 \\
4& 6    & 12m C40-3 & 2016-12-03 & CO(2-1) & 227.1 & 37\tablenotemark{b}  & 59.6 & 39.6 & 0.24x0.14 & 1 \\
4 & 7    & 12m C40-5 & 2016-10-27 & CO(3-2) & 340.8 & 2-7\tablenotemark{c} & 96.5 & 53.4 & 0.23x0.18 & 2 \\
7 & 7 &  12m C43-5 & 2021-06-30 & CO (3-2) & 340.8 & 14\tablenotemark{d} & 197.4 & 108.6 & 0.21x0.18 & 2 
\\
\enddata
\tablenotetext{a}{1 = this paper, 2 =  Paper II-Appleton et al. (in preparation).}
\tablenotetext{b}{37 primary beam positions were observed, fully sampling the Taffy system; see Figure~\ref{fig:almafields}b.}
\tablenotetext{c}{Seven well-sampled primary beam positions were observed in 2 regions, one centered on the Taffy bridge, and a second in the S-E disk of UGC 12914 (to be described more fully in a second paper).}
\tablenotetext{d}{41 antennas, 14 pointings  covering bridge and inner regions of UGC 12915}
\end{deluxetable*}

\begin{figure*}
\includegraphics[width=0.95\textwidth]{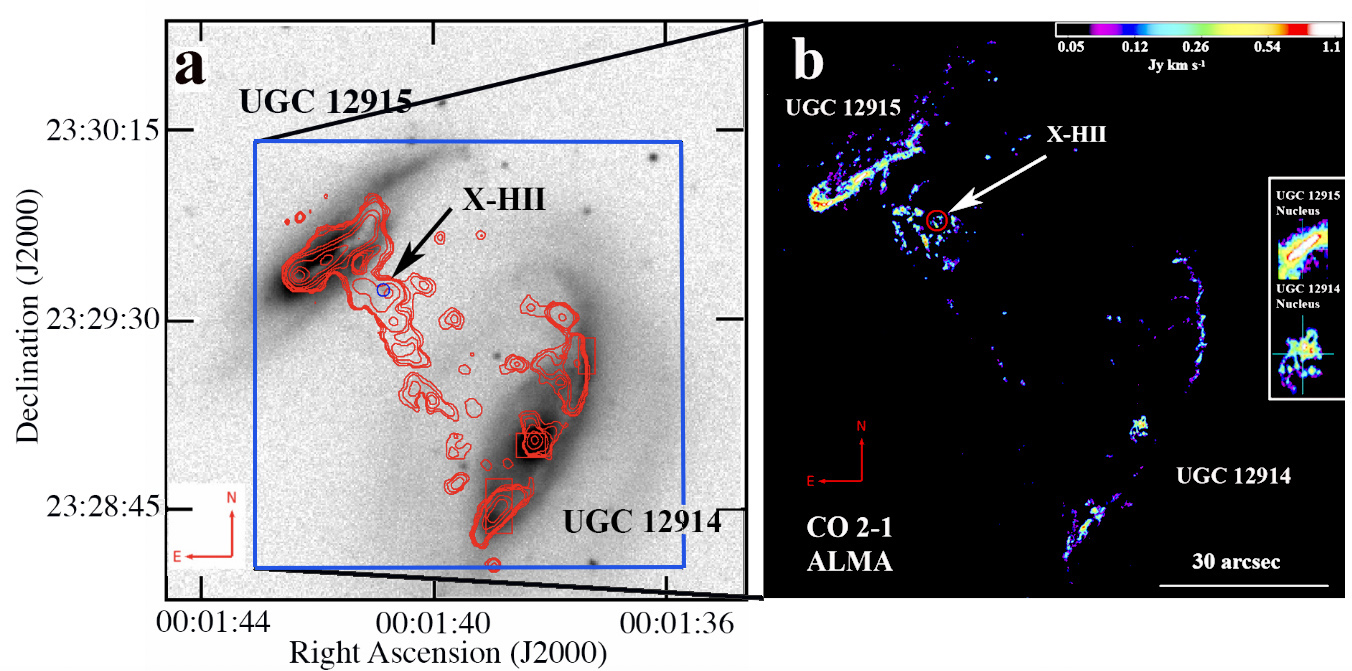}
\caption{(a) CO (1-0) emission contours \citep{Vollmer2021} taken with PdBI (beamsize 2.7 x 2.7 arcsec$^2$) superimposed on an SDSS i-band image of the Taffy system (UGC 12915/4) including the bridge emission, (b) A false color image of the ALMA total velocity integrated intensity  CO (2-1) (227GHz) mosaic at full-resolution (synthesized beam size 0.28x 0.14 arcsec$^2$),  built from 37 separate overlapping primary beam pointings. On this large scale view, the beam size is too small to display. The arrows show the position of the extragalactic HII region (circle 3 arcsec, marked as X-HII) on both figures, and the inset shows the position of the peak in 6 GHz nuclear radio emission (cross; Appleton et al. in prep.) in both galaxies superimposed on a zoomed-in image of the CO (2-1) emission. 1 arcsec at D = 62 Mpc corresponds to 300pc.}  \label{fig:radioCO21}
\end{figure*} 

Extended CO emission was suspected from the galaxies and weaker emission from the bridge, with emission being present not only at the smallest scales sampled by the ALMA observations (0.22"x 0.18") but also on slightly larger scales. This became clear when we initially tried a conventional CLEAN method (e. g. \citealt{Hogbom1974}) to deconvolve the "dirty maps" using the interferometric point-source response, or "dirty beam", 
for each channel where emission was detected. Because of the extended emission, this procedure always led to poor negative large-scale residual flux (bowls) in some of the residual maps.  Instead, we  used the multi-scale CLEAN algorithm  \citep[hereafter MSCLEAN;][]{Cornwell2008} as implemented in the CASA  task `tclean'. Unlike conventional CLEAN methods, which assume that the intrinsic brightness distribution of sources is made up of points sources (corresponding to a set of zero-scale delta-functions), the multi-scale clean method allows for both point-sources and larger scales to be present. Conventional CLEAN deconvolution methods iteratively build up the source distribution out of delta-functions, by subtracting the dirty beam from the observed dirty maps (the subtraction is usually done in the uv-plane).   MSCLEAN chooses from a set of smoothed dirty beams (the point source response convolved with the MSCLEAN scale), and progressively builds a model of the large-scale flux first, followed by flux on smaller and smaller scales, until it approximates that of a point source (zero-scale). This deconvolution of emission on different spatial scales decreases low-level artifacts caused by the PSF, allowing us to better recover the extended emission in the image. A full description of the method can be found in \citet{Cornwell2008}. Illustrative examples of its application to cases of nearby galaxies with extended HI emission are provided by \citet{Rich2008}. In our case, we tested, by trial and error, various scales to optimize the removal of the negative features seen previously in the residual maps. These final scales used corresponded to  scales of zero (delta function), 6, 12 and 24 pixels.  The method was then applied to all the channel maps (10 km s$^{-1}$ separation, covering a heliocentric optical velocity range from -419 to + 388 km s$^{-1}$ centered at 4350 km s$^{-1}$ ) containing emission, leading to a large data cube of the CO(2-1) emission covering a large part of the  Taffy field (see Figure~\ref{fig:almafields}b). As presented in Table~\ref{tab:obs}, the maps resulted in a synthesized beam with an angular resolution of  0.24 x 0.14 arcsec$^2$,  which corresponds to a projected physical scale of $\sim$60 pc for the Taffy system. The rms noise in each channel map was $\sim$0.7 mJy beam$^{-1}$.  

\subsection{HST Pa$\alpha$ Observations} 
Hydrogen recombination lines, like Pa$\alpha$, trace ionized gas associated with star formation, ionized shocks and other sources of diffuse ionized gas, and are commonly used to estimate star formation rates (e. g. \citealt{Kennicutt1998a,Calzetti2007}). In \S 4 we will discuss how the star formation rates estimated from the Pa$\alpha$ emission must also take into account strong contamination from shocked gas \citep{Joshi2019}. These authors did not find evidence for other sources of diffuse emission (e. g. DIG; see~\citealt{Haffner2009}) because of the low star formation rates in the system. 
We use archival Hubble Space Telescope (HST) Near Infrared Camera and Multi-Object Spectrometer (NICMOS) observations taken in the F187N and F190N filters, which were centered on UGC 12915 from the archive. These observations involved a small mosaic covering an area of 45.7 x 45.3 arcsec$^2$ centered on the galaxy.  We subtracted the continuum images to obtain a  Pa$\alpha$ image. The NICMOS-NIC3 image extends over part of the northern bridge, allowing us to compare the emission-line image with the CO map. Another NICMOS image of UGC 12914 is also available, but it does not cover any significant part of the bridge and is not presented here.  The NICMOS observations are obtained on a 0.2 arcsec pixel scale, which slightly under-sample the PSF at this wavelength (FWHM = 0.25 arcsec at 2$\mu$m).  This resolution is comparable to that of the CO (2-1) ALMA observations.  

Since the absolute astrometry of the NICMOS images is known to have significant uncertainty, we aligned the WCS coordinates of the well-defined nuclear peak of the galaxy in the F190N filter to that of a 6 GHz radio continuum image obtained with the Karl G. Jansky Very Large Array (VLA) (Appleton, in prep.) by re-registering the NICMOS image (a total shift of 0.9 arcsec). This offset was confirmed by finding good agreement between several compact 6 GHz radio sources, and corresponding bright compact Pa$\alpha$ knots in the western disk of UGC 12915. Similarly, we were also able to confirm the new coordinates by comparing the position of two radio hotspots embedded in the X-HII region with the corresponding knots of star formation at the same position in the Pa$\alpha$ image. From these tests we believe that the astrometry in the Pa$\alpha$ image is accurate to $\pm$0.1 arcsec (1/2 NICMOS pixel). 

Pa$\alpha$ flux densities were calculated using the conversion from counts s$^{-1}$ pixel$^{-1}$ to erg s$^{-1}$ cm$^{-2}$ arcsec$^{-2}$ evaluated using the relation F$_{line}$ = 1.054 $\times$ PHOTFLAM $\times$ FWHM (in erg s$^{-1}$ cm$^{-2}$) where PHOTFLAM is the photometric calibration parameter obtained from the observation FITS metadata, and FWHM is full-width half-maximum of the filter\footnote{Based on the metadata associated with the observations, PHOTFLAM = 3.33 $\times$ 10$^{-18}$ erg s$^{-1}$ cm$^{-2}$ \AA$^{-1}$ DN$^{-1}$, and the FWHM = 147.6 $\AA$}. The range of detected emission has a surface brightness of 1.5 $\times 10^{-16} < \Sigma(Pa\alpha) < 4.3 \times 10^{-15}$ erg s$^{-1}$ cm$^{-2}$ arcsec$^{-2}$, with a median SNR of 35.  Upper limits were calculated by 2.5 $\times$ the rms ADU/pixel over sample areas in the vicinity of the filaments. This upper limit, when converted to surface brightness units is $\Sigma(Pa\alpha) < 4.3 \times 10^{-17}$ erg s$^{-1}$ cm$^{-2}$ arcsec$^{-2}$.

\subsection{Radio Continuum Observations}
Radio continuum observations provide another means of estimating star formation through thermal and non-thermal processes associated with active star formation regions \citep{Condon1992,Murphy2011}.  
Deep observations were made as part of a radio polarization study of the Taffy  (project 19A-378) in the A-array of the VLA at 1.4~GHz (L-band) and 6~GHz (C-band) during a 12.6 hrs period in 2019 August 10.  Although the main aim of the project was to measure radio polarization, total intensity maps (Stokes I) were made in the two bands after processing with the task 'tclean' (using CASA v5.6.1) resulting in maps with restored synthesized beams of 0.36 x 0.30 arcsecs$^2$ (C-band) and 1.2 x 1.1 arcsec$^2$ (L-band). The rms noise in each map was 2$\mu$Jy beam$^{-1}$ and 7$\mu$Jy beam$^{-1}$ in C and L band, respectively. A more complete discussion of the radio observations will be provided in a future paper (Appleton, in preparation).

\subsection{Chandra X-Ray Observations}

We make use in this paper of a Chandra X-ray (0.5-8~kev) point-source image discussed in greater detail in \citep{Appleton2015}, and made available to us by those authors. The image was obtained in 2013 from a 39.5~ks exposure onto the back-illuminated S3 Advanced CCD Imaging Spectrometer \citep{Weisskopf2000}, and was smoothed with a Gaussian of FWHM 1.5 arcsec to emphasize the compact structure. In (\S 4.3) we discuss the relationship between the VLA radio, HST Pa$\alpha$ and Chandra X-ray observations of the extragalactic HII region . Because the Chandra X-ray observations had poor absolute astrometry, we used the VLA radio map of the nucleus of UGC~12914 (a point source as observed by both instruments) to carefully register the Chandra image 0.5-8~kev image, made available to us by those authors, to the radio position. The re-registration of the Chandra image to the VLA WCS frame resulted in a shift of  0.4 arcsec. This decisively shows that the brightest Ultra Luminous X-ray (ULX) source in Taffy system falls within the envelope of the high surface brightness Pa$\alpha$ and radio continuum emission from the extragalactic HII regions. A full discussion of the observations can be found in \citep{Appleton2015}.  

\subsection{{\it Spitzer} archival observations} 
IR emission is emitted from the dust grains heated by photons from young massive stars in star-forming regions. In $\S$~{\ref{sec:XHII}} we provide IR properties of the extragalactic HII region derived from archival {\it Spitzer} observations in the IRAC \citep{Fazio2004} 3.6, 4.5, 5.8 and 8$\mu$m bands, and in the MIPS \citep{Rieke2004} 24$\mu$m band. Images were dearchived from the Infrared Science Archive (IRSA) held at IPAC, Caltech. For the IRAC and MIPS images, the photometry reported in \S~{\ref{sec:XHII}} for the X-HII region was obtained by measuring flux densities in a fixed aperture of radius 4.8 arcsec centered on the clearly resolved emission from the source.  The same aperture was used for the 24$\mu$m data. However, because of the location of the X-HII region close to the disk of UGC~12915, care was taken to obtain local background estimates parallel to the extended disk of the galaxy. This effect was less than a few percent for the IRAC images, but represented a larger source of uncertainty for the MIPS 24$\mu$m emission. These uncertainties are reflected in the photometry reported in Table 2.

\subsection{Palomar 5 meter Spectroscopic Observations}
Observations of the Taffy System were made in moderate seeing conditions ($\sim$1~arcsec), and a 1~arcsec-wide slit with the Double Beam Spectrograph (DBSP; \citealt{Rahmer2012}) of the Palomar 5 meter telescope on 2021 January 9. The 600/1000 grating was used on the red arm of the spectrograph. At H$\alpha$ the spectral resolution was 85 \kms, and the scale along the slit was 0.29 arcsec/pixel. Flux calibration was performed using short observations of the white dwarf star G191B2B. The total on-source integration time was 3000 s. These  data were reduced using a Python-based pipeline PypeIt~\citep{Prochaska2020}, which performed bias, dark subtraction, flat field correction (using dome flats), flux and wavelength calibration using internal lamps.  

\section{Results} \label{sec:results}
\subsection{Large-scale molecular gas distribution}
Early CO (1-0) observations by \citet{Gao2003} of the Taffy system not only detected gas in the galaxies, but also measured large quantities of molecular gas in the bridge ($1.4 \times 10^{10}~ M_{\odot}$, assuming a standard Galactic N(H$_2$)/I$_{\rm CO}$ conversion factor\footnote{assumed Galactic conversion factor $ N(H_2)/I_{\rm CO} = 2 \times 10^{20}$ cm$^{-2}$ (K \kms)$^{-1}$. }). However, this mass is likely to be greatly overestimated \citep{Braine2003,Zhu2007} because there is evidence that a much smaller N(H$_2$)/I$_{\rm CO}$ conversion factor is appropriate in the bridge. Higher resolution observation of the system with a beam size 2.7x 2.7  arcsecs$^2$
were obtained by \citet{Vollmer2021} with the PdBI, and the integrated map is presented in Figure~\ref{fig:radioCO21}a superimposed over an SDSS image. The observations show the bridge is composed of partially resolved clumps scattered between the galaxies. 

In the current paper our ALMA CO (2-1) observations, which have a spatial resolution fifty times higher than the BIMA observations, and 12 times that of the PdBI, are shown in Figure~\ref{fig:radioCO21}b. This full resolution moment-0 map was constructed in the following way:  i) each channel map (of width 10 \kms) was spatially smoothed to an effective resolution of 0.4 x 0.4 arcsec$^2$, ii) a valid mask was constructed of all signal within the smoothed map that was $>$ 3.5 sigma above the noise per channel, iii) an integrated (moment-0) map was then made by applying the masks to the full-resolution channels, and summing the emission spatially in those channel maps (velocity space) where the mask indicated valid points above the masking threshold. Regions outside the valid regions were not summed.

The same mask was also applied to the smoothed version of the cube, resulting in a smoothed moment-0 map, which is presented in Figure~\ref{fig:smomom0-1}a. This figure shows some of the fainter features better than the full resolution image with 8000 x 8000 pixel$^2$ which is not well represented in a small figure.  
However, the full resolution map is used for the majority of the analysis.  The ALMA observations show that many of the features, seen in the lower-resolution PdBI map, appear as narrowly defined filaments and small clumps. 

In addition to the bridge regions, which is the main focus of this paper, most of the detected ALMA emission from UGC 12914 is seen in narrow structures, including peculiar narrow ripples of emission along its northwestern disk, with fainter gas filaments breaking off from the disk into narrow strands which point almost perpendicularly to the main arc of the emission in that arm. Narrow dense structure is also seen on the south-eastern part of the same disk. The region around the nucleus of UGC 12914 contains high surface brightness emission distributed in a series of loops and spiral filaments (see inset in Figure~\ref{fig:radioCO21}b).  On the other hand UGC~12915, which is more edge-on than its companion, is dominated by a  bright  south-eastern curved structure, or possible tightly wound spiral feature, and a high surface brightness inner edge-on nuclear disk (see upper inset in Figure~\ref{fig:radioCO21}b).  The main galaxy disk extends to the north-east following the inner optical dust lane where it breaks up into numerous clumps and extended filaments to the far north-east. Narrow filaments of gas are also seen extending away from many parts of UGC~12915's disk in strands to the north. 
Because of the complexity of the system, we will mainly concentrate on the Taffy bridge in the current paper,  and will return to a more complete description of the CO emission from the galaxy disks in Paper II. 

The structure of the bridge emission is striking. The region of the bridge closest to the northern Taffy galaxy is composed of a collection of filaments and bright clumps of emission, some of which give the appearance of a cone-shaped structure whose apex lies $\sim$~10 arcsec south-west (3kpc) from the center of UGC 12915. However, other fainter filaments cross the structure (Figure~\ref{fig:smomom0-1}), and we will show that the velocity structure of the bridge gas is quite complex, and the cone-shaped pattern does not form a single coherent kinematic structure.   
The faint X-HII region, seen in the optical image in Figure~\ref{fig:radioCO21}a, lies buried in the tangled north-western part of the main emission from the bridge. 

There are many other scattered CO complexes apparent in the observations, including gas to the far north-west of UGC~12915. Some of this gas may be part of a separate "north-western" bridge of emission identified in optical Integral Field Unit (IFU) observations of ionized gas in the Taffy bridge \citep{Joshi2019}.  Previous observations of both the neutral hydrogen \citep{Condon1993} and {\it Herschel} [CII]158$\mu$m emission \citep{Peterson2018} show neutral gas in this area.  

An important question to ask is what fraction of the CO emission in the Taffy bridge is detected on the small (60 - 100 pc) scale  compared with single-dish observations \citep{Braine2003,Zhu2007} or large-beam interferometric observations like BIMA \citep{Gao2003}. According to \citet{Zhu2007} and \citet{Gao2003}, the BIMA observations of the Taffy, made with a 9.8 x 9.7 arcsec$^2$ beam, capture most of the flux seen in the single dish observations. Based on a data cube provided by Y. Gao (personal communication), we centered a 20.5 arcsec diameter circular aperture on the main concentration in the northern bridge, and estimated the integrated flux over the measured profile to be 92.1 Jy\,km s$^{-1}$ in CO (1-0), which agrees with a statement made in \citet{Gao2003} of emission associated with the region they called the  "HII region", which actually includes most of the structures we have been discussing. To compare this with the ALMA observations requires converting
this flux to an equivalent CO (2-1) flux over the same area. Assuming r$_{21}$ = 0.79~\citep{Zhu2007} (where r$_{21}$ is the ratio of I$_{\rm CO(2-1)}$/I$_{\rm CO(1-0)}$), the equivalent CO (2-1) flux should be 291 Jy\,km s$^{-1}$ for the same aperture. From our ALMA observations, we derived, by integrating the CO emission over the same area, a total flux of 161.6 Jy-km s$^{-1}$.
From this we estimate that with ALMA we detect 55.5$\%$ of the emission seen in the BIMA observations. Thus a significant fraction of the CO emission in the northern bridge is in an extended component not detected by ALMA.  

\subsection{Large-scale molecular gas kinematics}

\begin{figure*}
\includegraphics[width=0.99\textwidth]{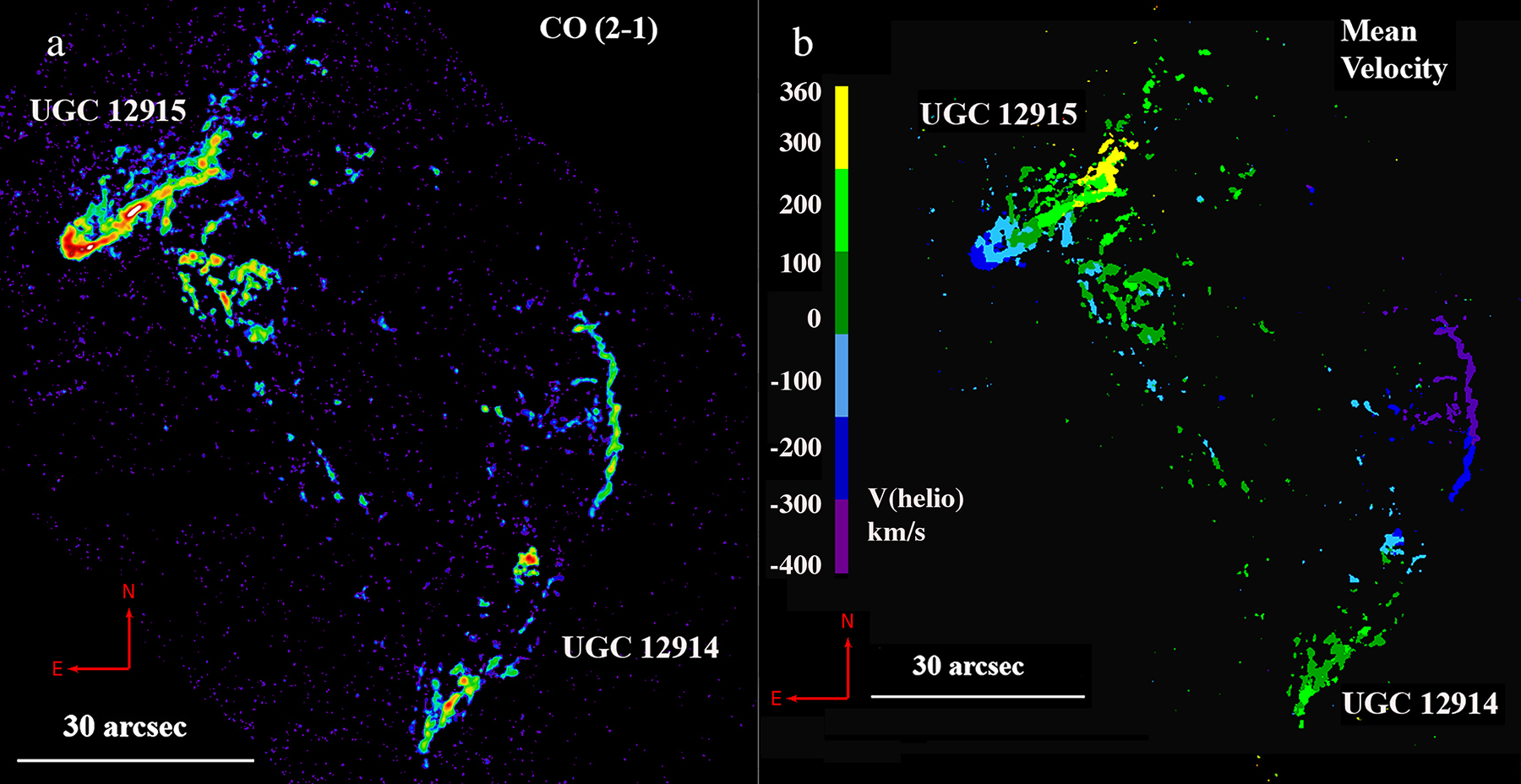}
\caption{(a) Smoothed  (to 0.4 x 0.4 arcsec$^2$) image of the CO (2-1) total velocity-integrated intensity map showing fainter features. The color stretch encompasses emission as faint as 0.05 ~Jy~km~s$^{-1}$beam$^{-1}$ (dark blue) to the brightest at 2 Jy~kms~s$^{-1}$beam$^{-1}$ (white), and (b) intensity weighted mean heliocentric velocity (optical velocity definition relative to 4350~\kms) of the full field.  } \label{fig:smomom0-1}
\end{figure*}

In Figure~\ref{fig:smomom0-1}b, we show an intensity-weighted (1st Moment) mean velocity map of the whole system (relative to the average heliocentric velocity of 4350 \kms  for the two galaxies)  with the same 0.4 x 0.4 arcsec$^2$ smoothing as Figure~\ref{fig:smomom0-1}a.  For reference, the systemic velocities of UGC 12915 and UGC 12914 are quite similar (-14 and +21 \kms~respectively, implying that most of the radial motion between the galaxies is in the plane of the sky \citep{Condon1993}. In this edge-on view of the collision, the counter-rotation of the two galaxies is particularly obvious. UGC 12915 appears to show blue-shifted emission in the SE and red-shifted emission in the NW, whereas  UGC 12914 shows the opposite behavior.  This counter-rotation may have contributed to an increased amount of cloud-cloud collisions when the two galaxies originally collided, almost face-on \citep{Vollmer2012,Yeager2019,Yeager2020a,Yeager2020b}. The extension of scattered gas clouds to the north-west of UGC 12915, discussed earlier, shows peculiar motions not consistent with regular rotation within UGC 12915. While the regular rotation of UGC 12915 is from blueshifted (-340 \kms; blue color) to redshifted gas  (+320 \kms; yellow color) in the figure, the clouds further north and to the west show systemic velocities between -50 to 100 \kms~(dark green and green).  Similarly, non-corrotating emission was noted in the velocity field of the ionized gas in this region, where even more discrepant velocities were observed \citep{Joshi2019}. This supports the idea that it may be part of a second, kinematically-distinct bridge between the two galaxies, or the remnants of a tidal tail from UGC 12914 \citep{Vollmer2021}.

\begin{figure*}
\includegraphics[width=1.0\textwidth]{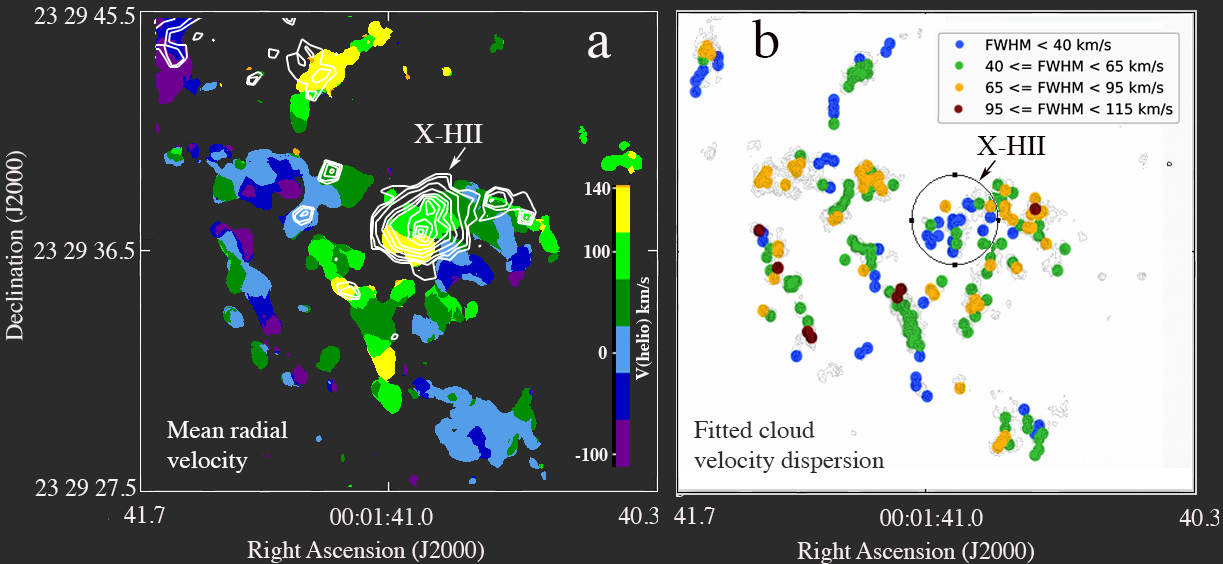}
\caption{(a) A zoom-in on the mean velocity field (heliocentric velocity relative to 4350 \kms) of the northern bridge to emphasize more finely the velocity structure. White contours show the Pa$\alpha$ emission from the X-HII region. (b) A color-coded representation of the fitted FWHM (km s$^{-1}$) of the CO-emitting clouds on 85pc scales superimposed on contours of CO surface density. Green and blue filled circles show clouds low dispersion, whereas orange and red show higher values. Only measurements with high quality profiles (Spectral Quality 1.0, see Table~\ref{tab:A1}) are shown.} \label{fig:bridgevel}
\end{figure*}

In the main CO bridge, the average kinematics is quite complex, as Figure~\ref{fig:bridgevel}a shows (displaying a velocity range from -100 to 140 km/s). The bridge as a whole does not show large-scale coherent motion, but is rather made up of many clumps and filaments with disparate average velocities. Some of the filaments in the cone-shaped structure do show weak systematic motions along parts of their length, but, except for the filaments near UGC 12915, appear kinematically distinct and do not seem obviously related to each other.  

Figure~\ref{fig:bridgevel}b shows a representation of the CO (2-1) line width as a function of position in the bridge. The color coding is based on the FWHM (in \kms) for the regions presented in Table~\ref{tab:A1}. The points are superimposed on a contour map of the total intensity of the CO (2-1) emission. Blue and green points show the gas with the lowest line-width, whereas many regions have FWHM greater than 65 \kms (orange and dark red filled circles), with values extending up to 115 \kms. Regions of high velocity dispersion are scattered throughout the filaments and clumps, with the quiescent gas (FWHM $<$ 40 \kms) being the minority.  Regions with the lowest velocity dispersion are mainly found in some of the filaments close to UGC~12915, and in the area near the X-HII region. Some example spectra are shown in Figure~\ref{fig:A1}. 

Although it is often traditional to show channel maps (intensity maps of the line emission as a function of velocity) of the full velocity cube of the observations, we will defer this to the second paper, since the emphasis of the current paper is the star formation properties of the bridge.
       
\subsection{Molecular surface density in selected regions}

\begin{figure}
\includegraphics[width=0.49\textwidth]{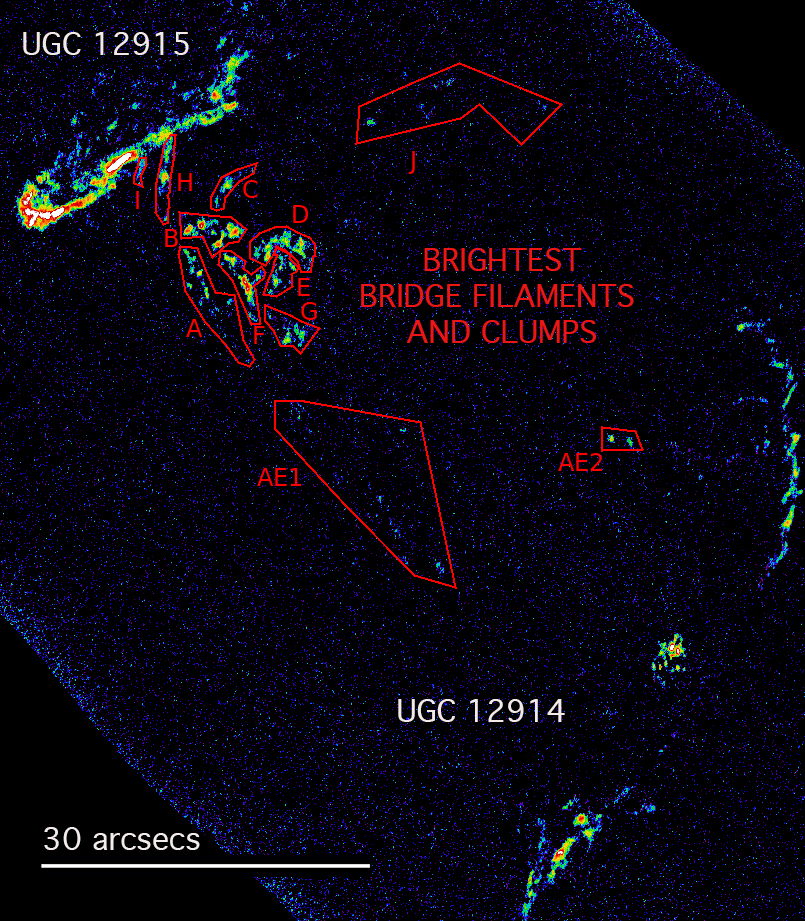}
\caption{The CO (2-1) integrated bridge emission divided into regions of bright emission. Regions A to G include the filaments and clumps associated with the main bridge (where D includes emission associated with the X-HII region) and H and I show filaments that may be part of the bridge but lie close to UGC 12915. Regions J and AE1/2 are scattered regions in the bridge. 
AE1 may be an extension of filament A.  Each region is divided further into many small extracted apertures, the properties of which are shown both graphically and in tabular form in the Appendix (Figures~\ref{fig:A2} and~\ref{fig:A3}, and Table~\ref{tab:A1}).} \label{fig:filaments}
\end{figure}

We extract the spectra of more than 239 small regions of the CO emission within the filaments and clumps, and compare their properties to star formation rate estimates derived from the NICMOS observations.  In Figure~\ref{fig:filaments}, the CO bridge filaments and cloud complexes are divided into 12 large regions which represent areas that seem spatially and kinematically related. Regions A to G cover emission complexes associated with the brighter part of the northern bridge. Region D includes emission associated with the X-HII region.   We also including two coherent filaments, H and I,  that fall close to the disk of  UGC 12915 and may be part of the bridge. Regions J, to the NW,  and the possible extension of filament A, called AE1, and two bright bridge clumps AE2 are also analyzed.  These bridge regions lie outside the area sampled by NICMOS, but are included for completeness.   

To study the details of the emission in each of the regions shown in Figure~\ref{fig:filaments}, we split each structure up into small extraction aperture sub-regions which are  described more fully in Appendix A, and Figures~\ref{fig:A2} and \ref{fig:A3}. Using the software package CASA and the CASA Viewer, we extracted spectra of each region (of dimension 0.23 x 0.28 arcsecs$^2$), slightly larger than the resolution of the ALMA data, and large enough to sample a significant part of the NICMOS PSF. Each of the extracted spectra were then fit with a Gaussian line profile.  In a few positions (Regions A1 through A4), we observed double-line profiles along the same line of sight. Here we fit two Gaussian components. For all the CO spectra, we estimate the mean radial velocity, the FWHM and peak flux density and finally the integrated CO  flux $S_{\rm CO(2-1)}\Delta V$ in Jy\kms. The line properties of all the extracted regions are presented in Table~\ref{tab:A1}.  Table~\ref{tab:A1} provides a flag of the quality of each spectrum. Of all the spectra extracted, 219 were deemed to be of sufficient quality (good baselines, signal to noise ratio) to be included in the analysis.

To estimate the molecular gas properties from the observed cloud line properties, it is necessary to make several assumptions, including a conversion to H$_2$ column density. We will use the standard conversion to molecular gas mass, including a 36$\%$ correction for Helium \citep{Bolatto2013}:   
\begin{equation}
\left(\frac{M_{\rm gas}}{M_{\odot}}\right)=1.05\times10^4X_{\rm CO,20}S_{\rm CO(1-0)}\Delta V D_L^2 (1+z)^{-1},
\end{equation}
where $D_L^2$ is the luminosity distance in Mpc and $X_{\rm CO,20}$ is the  conversion factor $N({\rm H_2})/I_{\rm CO}$ in units of 2$\times10^{20}$cm$^{-2}$ (K km s$^{-1})^{-1}$. Here N(H$_2$) is the H$_2$ molecular column density and I$_{\rm CO}$ is the velocity integrated intensity of the CO (1-0) transition in K \kms.  Therefore, to derive $M_{\rm gas}$,  we also need to make assumptions about both the value of $X_{\rm CO}$, and the ratio of  $S_{\rm CO(1-0)}/S_{\rm CO(2-1)}$.  

 \citet{Braine2003} suggested that $X_{\rm CO}$ was probably at least a factor of 4 times lower in the bridge  than the Galactic value, based on single dish observation of the ratio of the $^{13}$CO/$^{12}$CO (for the 1-0 transition), which implied the $^{12}$CO line was almost optically thin. A similar conclusion was reached by  \citealt{Zhu2007}, using the transitions CO (3-2), (2-1) and (1-0), and performing LVG modeling (Goldreich \& Kwan 1974). They estimated that in the bridge $X_{\rm CO}$ was between 2-3.6 $\times 10^{19}$ cm$^{-2}$ (K km s$^{-1})^{-1}$, which is 5 to 10 times lower than $X_{\rm CO,20}$. Both of these measurements were made with large filled apertures on scales of $\sim$11-12 arcsec. 
Recently, \citet{Vollmer2021} adopted an intermediate value of $X_{\rm CO}$ = 1/3 $X_{\rm CO,20}$ for the bridge in their PdBI $\sim$ 3 arcsec resolution beam.  Given the uncertainty in the LVG modeling, and the large difference in scale between the previous single dish observations and our ALMA observations, we adopt as an initial working hypothesis the \citet{Braine2003} value of $X_{\rm CO}$ = 5$\times 10^{19}$ cm$^{-2}$ (K km s$^{-1})^{-1}$, or 1/4 $X_{\rm CO,20}$. We will explore the implications of varying this value on the derived properties of the clouds and their line of sight extinction under different assumptions.  We finally assume $S_{\rm CO(1-0)}/S_{\rm CO(2-1)} = (\nu_{1-0}/\nu_{2-1})^2$(r$_{21})^{-1}$, and  r$_{21}$ = 0.79 (see \citealt{Zhu2007}). Gas masses from our extracted regions range from 0.1-1 $\times$ 10$^7$ ($X_{\rm CO}/X_{\rm CO,20}$) M$_{\odot}$.

\section{Ionized gas emission in the bridge and X-HII region} 

\subsection{The Pa$\alpha$ distribution in the bridge}
\begin{figure}
\includegraphics[width=0.49\textwidth]{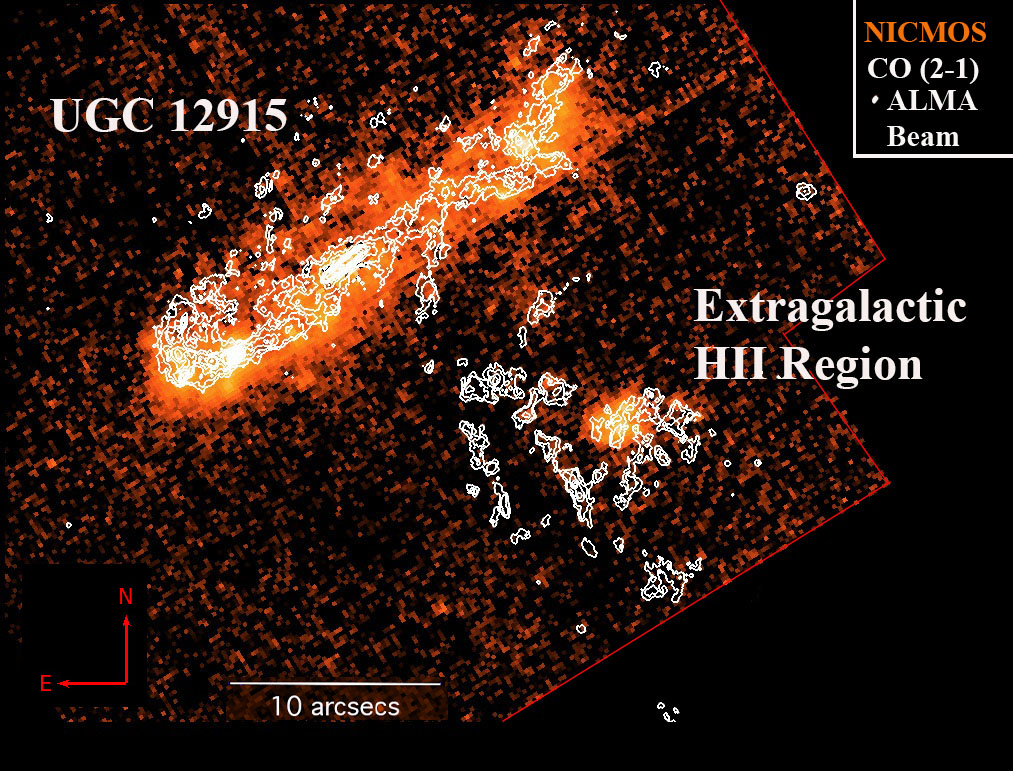}
\caption{Contours of the full-resolution ALMA CO (2-1) velocity integrated intensity superimposed on the false-color image of HST NICMOS image of Pa$\alpha$ around UGC 12915 including parts of the bridge. Contour levels are 0.1, 0.15, 0.25, 0.3, 0.35, 0.45, 0.55, 0.65,  0.75, 1.0, 1.2, 1.4, 1.75, 1.85 and 1.95 Jy\kms beam$^{-1}$. } \label{fig:NIC3CO2132}
\end{figure}

In Figure~\ref{fig:NIC3CO2132} we show  the complex filamentary structure of part of the  bridge in  CO (2-1)  superimposed on the Pa$\alpha$ NICMOS image of UGC 12915, which includes the brighter and most interesting parts of the bridge region. The figure shows that within UGC 12915, the CO is largely confined within the inner disk, and is well correlated with the bright Pa$\alpha$ emission. A narrow nuclear CO disk is detected,
and there are several CO filaments extending away from the disk to the north of UGC 12915, which, except for a few isolated cases, are devoid of obvious star formation. 

Concentrating on the bridge region, the most  prominent Pa$\alpha$ emission comes from the bright extended X-HII region \citep{Bushouse1987,Jarrett1999,Joshi2019}.  Several clumps of CO emission are seen projected against this star forming region, including an elongated finger of faint CO emission which crosses its center. Very little Pa$\alpha$ emission is seen from the other CO structures, except for one or two faint possible associations.
The lack of obvious star formation in these dense clumps is a characteristic of the CO emission observed in the bridge (see also \citealt{Vollmer2021}). 

For those regions with extracted CO spectra that fall within the area covered by NICMOS, we then proceeded to extract Pa$\alpha$ flux surface densities of the emission using the software package SAOImage-DS9\footnote{SAOImage~DS9 development has been made possible by funding from the Chandra X-ray Science Center (CXC), the High Energy Astrophysics Science Archive Center (HEASARC) and the JWST Mission office at Space Telescope Science Institute \citep{Joye2003}.}. Of the 219 high quality CO(2-1) extracted bridge spectra, 24 lay outside the NICMOS field of view either in the southern bridge or to the NW of UGC 12915. Of the remaining 194 spectra, only 35 (18$\%$) show a detectable Pa$\alpha$ emission in the bridge, and 5 are associated with Region H,  which is very close to the disk of UGC 12915. Except for the regions B18, 19, 20, 31, 32 and 33, all of the other detected regions are associated directly with the X-HII region.  Elsewhere, only upper limits were obtained for the surface brightness of the Pa$\alpha$. The flux surface densities and upper limits are tabulated in Table~\ref{tab:A1} in units of erg s$^{-1}$  cm$^{-2}$  arcsec$^{-2}$ for the same extraction regions as those measured for the CO emission line fluxes. 

\begin{figure*}[ht]
\includegraphics[width=0.9\textwidth]{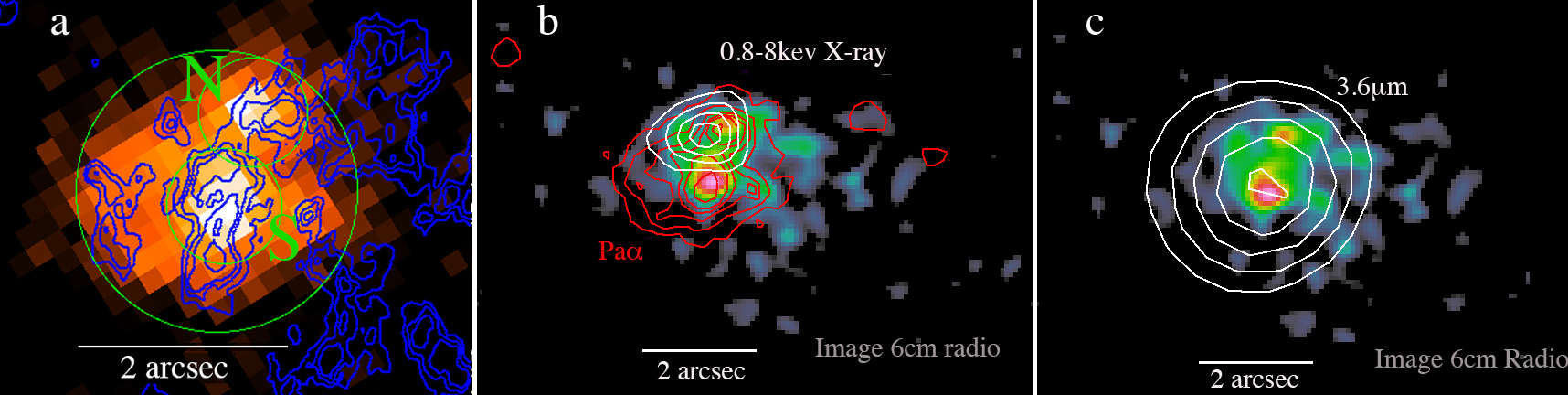}
\caption{Taffy X-HII region a) Pa$\alpha$ false color image with contours (blue) of CO (2-1) integrated emission overlaid. The green circles show photometric apertures (ALL, hotspot N and S) labelled in Table~\ref{tab:HII}. b) 6cm radio continuum false-color image from VLA A-array (Appleton et al. in preparation) with contours Pa$\alpha$ emission (red) and the 0.5-8keV X-ray ULX source CXOU J0001409+232938 (white contours; \citealt{Appleton2015}), c) same radio images as in b) but overlaying contours of the Spitzer IRAC band 1 (3.6$\mu$m emission (white).}\label{fig:multicolorHii}
\end{figure*} 

\begin{figure*}
\includegraphics[width=0.98\textwidth]{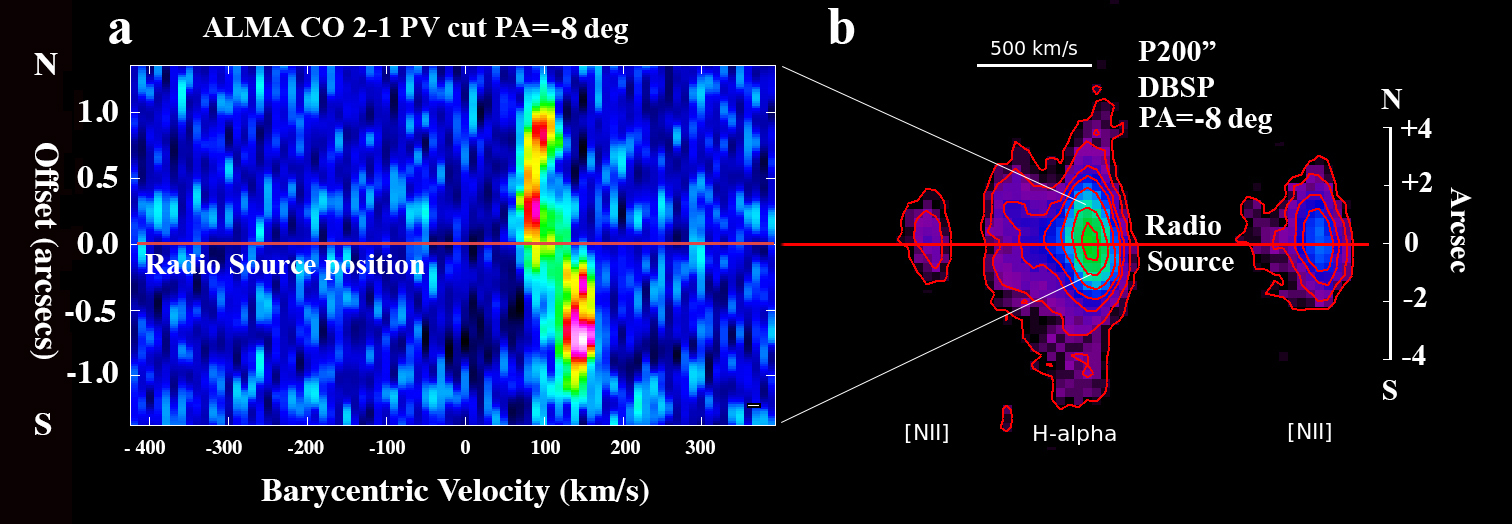}
\caption{(a) A position-velocity diagram taken along the main CO bar-like structure (PA = -8 deg) seen in the integrated map of the molecular gas. The diagram is reminiscent of two sides of a flat rotation curve displaced around the position of the brightest 6cm radio source (Hotspot-S of Figure~\ref{fig:multicolorHii}). The gas is extended over a scale of  2 arcsec (600pc), (b) Line emission taken along the same position angle with the Palomar 5-m DBSP (slit width = 1arcsec, seeing 1 arcsec) showing the distribution of [NII] and H$\alpha$ along the slit. The main emission shows a tilted distribution consistent with a rotating disk of $\pm$ 50-60 \kms in the inner part of the structure. The rotation and magnitude is consistent with the putative CO disk.  A blueshifted component is seen in H$\alpha$, but is not detected in CO.}\label{fig:COrot}
\end{figure*}

\subsection{The extragalactic HII (X-HII) region and its ionized gas}
\label{sec:XHII}

Figure~\ref{fig:multicolorHii} shows and compares the X-HII region at different wavelengths.
The overall scale of the Pa$\alpha$ emission from the X-HII region, shown in
Figure~\ref{fig:multicolorHii}a, is 2~arcsec (or 600~pc) and is composed of two compact regions of emission surrounded by more diffuse gas. 
The CO is dominated by a bar-like structure with a position angle of -8 degrees (north through east) crossing the face of the Pa$\alpha$ emission and clumpy structures surrounding it. In Figure~\ref{fig:multicolorHii}b we show 
the $\lambda$6~cm radio continuum image obtained with the VLA (Appleton et al., in preparation) which mimics the Pa$\alpha$ structure (red contours), again showing two dominant emission regions and diffuse emission. Also shown are (white) contours of X-ray emission defining the brightest ULX X-ray source CXOU J0001409+232938.  The ULX source falls close to the northern compact region. 
Figure~\ref{fig:multicolorHii}c shows the 6cm radio emission image with contours of {\it Spitzer} IRAC 3.6$\mu$m emission superimposed. The X-HII region is detected in all four IRAC bands and at 24$\mu$m with MIPS. The centroid of the 3.6$\mu$m image (which has  a much lower spatial resolution than the other images) falls close to the brighter southern compact radio source within the X-HII region. 

Returning to the relationship between the ionized gas and the molecular gas Figure~\ref{fig:COrot}a shows a position-velocity diagram constructed along a position-angle of the main CO-bar at PA = -8 degrees. This is also the angle projected onto the sky where the CO emission implies an approximate major axis of rotation. The resulting position-velocity diagram seems to show evidence of bulk motions from the north to the south, with an inflection point at the position of Hotspot-S (Figure~\ref{fig:multicolorHii}a), which is best defined by the radio position of the southern 6 GHz source. Relative to an assumed systemic velocity of ~106 \kms~ for the X-HII region\footnote{The velocities are all relative to the Taffy mean heliocentric velocity of 4350 \kms}, the gas shows a sudden jump from positive velocities (+60 \kms) to negative ones (-60 \kms) at the position of the radio source. This  supports the idea that Hotspot-S is the kinematic center about which the gas is rotating. The scale of this putative disk has a radius of 1-1.5 arcsec = 300-450 pc at D = 62 Mpc. 

To further support the idea that we are observing rotational motion, Figure~\ref{fig:COrot}b shows a part of the Palomar spectrum (slit width = 1 arcsec) positioned along PA = -8 degrees.  Bright H$\alpha$ emission is detected, as well as the satellite lines of [NII] which are well separated from H$\alpha$. The H$\alpha$ emission shows two features of note. Firstly the main core of the emission centroid is tilted, consistent with a rotating bright disk, with rotation of $\pm$~100~\kms across 1-1.5 arcsecs. Even though the Palomar spectral resolution at H$\alpha$ is much lower (85~\kms) than the CO data, its rotational motions appear to have the same sign and approximate magnitude as the CO emission. This suggests that both the CO and H$\alpha$ are part of the same disk.
We also observe a second, broad line-width blue-shifted  H$\alpha$ component observed at negative velocities,  which is centered at the same position as the main tilted disk-like structure, but shifted by 200-300 \kms\ from it. CO emission is clearly absent in Figure~\ref{fig:COrot}a at this velocity.
It is unlikely that the negative-velocity gas represents a powerful outflow from the X-HII region, since its overall star formation rate, as shown in this paper (\S 5.3), is quite low.  Discrepant velocity structures, similar to this one, are seen in the kinematics of molecular gas in the Antennae overlap region \citep{Tsuge2021a,Tsuge2021b}, where there is evidence of much stronger star formation. They may be evidence for shocks associated with cloud-collisions.  Similar examples of very faint star formation embedded in shocked intergalactic emission-line gas exhibiting multiple line profiles is observed in Stephan's Quintet (see for example \citealt{Xu2003,Konstantopoulos2014,Guillard2021}).

\section{ Relationship between molecular gas and star formation in the bridge}

\subsection{Deriving star formation rates from Pa$\alpha$ observations}
In normal galactic disks (e. g. \citealt{Calzetti2007,Calzetti2010}), it is generally assumed that a measure of the star formation rate can be inferred from the strength of a relatively unobscured hydrogen recombination line, like Pa$\alpha$, with the assumption that most of the emission is from gas heated by Lyman-continuum photons from hot stars. 

\subsubsection{Correcting for shocked gas}
In the Taffy bridge, we already have strong evidence from previous studies \citep{Joshi2019} that a significant fraction (up to approximately 45$\%$ of H$\alpha$ emission) comes from gas heated in shocks with characteristic shock velocities of 200-300 \kms. This result applies not only to the bright extragalactic HII region but also to fainter extended H$\alpha$ emission seen throughout the bridge. Ubiquitous shocks throughout the bridge are also supported by evidence for significant quantities of warm molecular gas in the bridge which presence cannot be explained by heating by star formation \citep{Peterson2012,Peterson2018}. Thus, it cannot be assumed that all the Pa$\alpha$ emission in the Taffy bridge is due to star formation alone. As a result of this uncertainty, we consider two limiting cases in our study. In the first, we assume all the Pa$\alpha$ is from star formation.  In the second, we assume a more realistic scenario where only 55$\%$ of the Pa$\alpha$ emission arises from star formation based on the observation of \citet{Joshi2019}. This range of possible values will be reflected in our plots of star formation surface density in the subsequent discussion.  
\subsubsection{Correcting for extinction}\label{sec:extinct}

A second source of uncertainty in measuring the star formation surface density from hydrogen recombination lines is extinction.  Given that the calculated H$_2$ surface densities approach 1000 $M_{\odot}$ ~pc$^{-2}$ under some assumptions, we cannot assume that extinction is negligible, even at the wavelength of the Pa$\alpha$ transition ($\sim$ 1.8$\mu$m). Assuming we can correct for the shocked fraction, and extinction, we can convert the corrected Pa$\alpha$ surface density fluxes into equivalent H$\alpha$ fluxes, assuming Case B recombination ($f(H\alpha)/f(Pa\alpha)$ = 8.59;~\citealt{Osterbrock1989}), and then convert the fluxes to star formation rates via the transformation of \citet{Kennicutt1998b}.   

To correct for  Pa$\alpha$ extinction both the detected fluxes and upper limits (which applies to the majority of the clouds) we adopt two different approaches: 
\begin{itemize}
\item{Case1: A minimal extinction assumption at Pa$\alpha$ based on the visible light measurements of the Balmer decrement.  Low extinction at Pa$\alpha$ can be inferred from measured values of A$_V$ using IFU spectroscopy to estimate the Balmer decrement across faint ionized gas filaments in the bridge from \citet{Joshi2019}. In the region of the bridge these authors estimated A$_V$ $\sim$ 1.5~mag. Assuming a \citet{Calzetti1994} extinction law, this would convert to A$_{Pa\alpha}$ of $\sim$~0.2~mag\footnote{Adopting other commonly used extinction laws does not affect this conclusion significant since most extinction curves deviate from one another much more in the UV,  and differ by only a few percent in the near IR \citep{Gordon2003}.} These results are also consistent with our Palomar 5-m spectroscopy obtained with a 1 arcsec slit size.}
\item{Case 2: Calculate A$_V$ from the measured molecular gas surface density: The  estimate of extinction from nebulae emission lines might be biased to lines of sight with lower extinction. We therefore consider a complementary method and derive the extinction from the total column density, N(H) (= 2N(H$_2$)) within each cloud. We assume that H$_2$ dominates the hydrogen column density on the scale of 0.2 arcsec (e. g. \citealt{Kahre2018}). This conversion is complicated by two factors. Firstly, the measurement of the total hydrogen column density N(H) (= 2N(H$_2$)) depends on our assumed X$_{\rm CO}$ factor.  Secondly, the assumed gas-to-dust mass ratio and the extinction law affect the extinction per total gas column density, $A_V/N(H)$. Neither of these factors are well constrained in the case of the Taffy CO bridge gas. We will perform a limited exploration of the effects of changing some of the assumptions.}  
\end{itemize}

A knowledge of the gas-to-dust mass ratio, which scales linearly with A$_V$, is particularly important in the Taffy bridge. Previous observations have suggested that dust is strongly depleted there. Dust was detected in emission at 450 and 850$\mu$m using the James Clark Maxwell Telescope (JCMT) with an angular resolution of 9.4 and 16 arcsec, respectively, by \citet{Zhu2007}.  These results suggest unusually large gas-to-dust mass ratios of 600-800 in the bridge, many factors ($>$5) higher   than values for the Galaxy or nearby galaxies($\sim$ 80-150). These large values are in contrast to the Taffy galaxies which show more normal  gas-to-dust ratios. Dust depletion might be a result of grain destruction in shocks in the bridge \citep{Jones1996}.  Based on these observational results, we consider a range of dust depletions from no-depletion, to depletions of up to a factor of 5. 

We take the relationship of \citet{Kahre2018} of N(H)/E(B-V) = 5.8 $\times 10^{21}$ H cm$^{-2}$ mag$^{-1}$, suitable for Galactic dust, and divide the extinction by an assumed dust depletion factor to account for the fewer dust particles per H-atom compared with normal Galactic gas.  We also consider other relationships between A$_V$ and and N(H), such as lowering the metallicity of the gas.

\begin{figure*}
\includegraphics[width=0.98\textwidth]{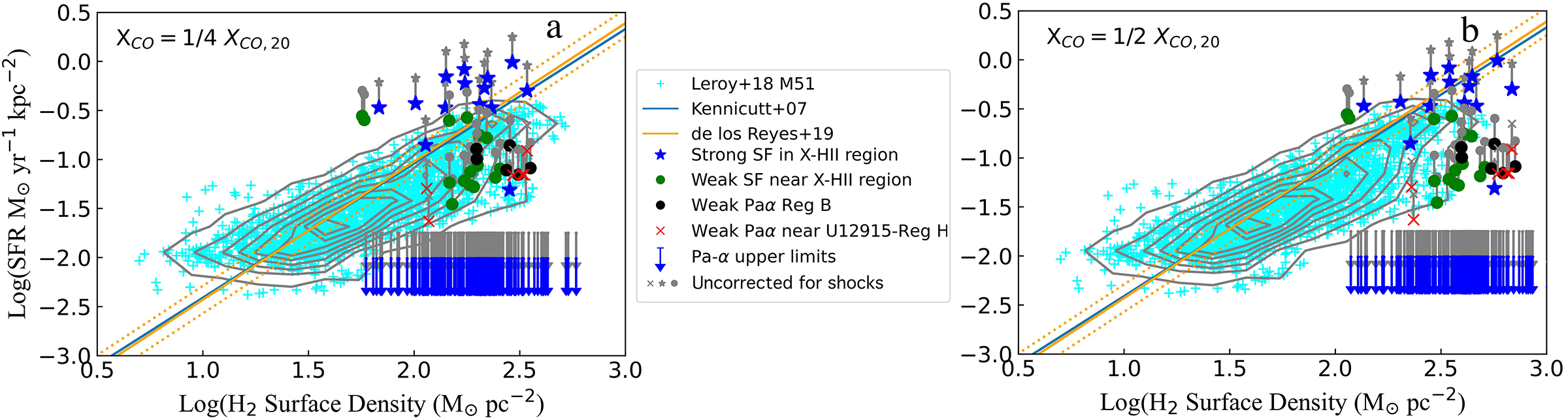}
\caption{Comparing the KS relation with the molecular and SF properties of the Taffy bridge under different assumed values for the CO conversion factor $X_{\rm CO}$ for minimal Pa$\alpha$ extinction (Case 1, see \S~\ref{sec:extinct}), a) assuming $X_{\rm CO}$ = 1/4 $X_{\rm CO,20}$,  and b) $X_{\rm CO}$ = 1/2 $X_{\rm CO,20}$. Grey symbols show the star formation rate density assuming no shock contribution to Pa$\alpha$, whereas the colored connected symbols are the more realistic case where shocks contribute up to 45$\%$ of the bridge Pa$\alpha$ emission \citep{Joshi2019}.   In each plot, the blue (or grey) downward arrows show SFR density limits derived from the undetected Pa$\alpha$ emission, with (and without) a shock contribution. The different detected regions are shown with symbols indicated in the central panel.  The blue and yellow lines shows two representations of the average KS relationship, for spiral and starburst galaxies  \citep{Kennicutt2007,Reyes2019}. The 1$\sigma$ spread is shown as dotted lines. The pale blue points (and density contours) are individual spatially-resolved CO measurements made with ALMA across the face of the nearby M51 for reference (for  $X_{\rm CO}$ = ($X_{\rm CO,20})$)  \citep{Leroy2017, Leroy2018}.
}
\label{fig:KSplot1}
\end{figure*} 

\begin{figure*}
\includegraphics[width=0.97\textwidth]{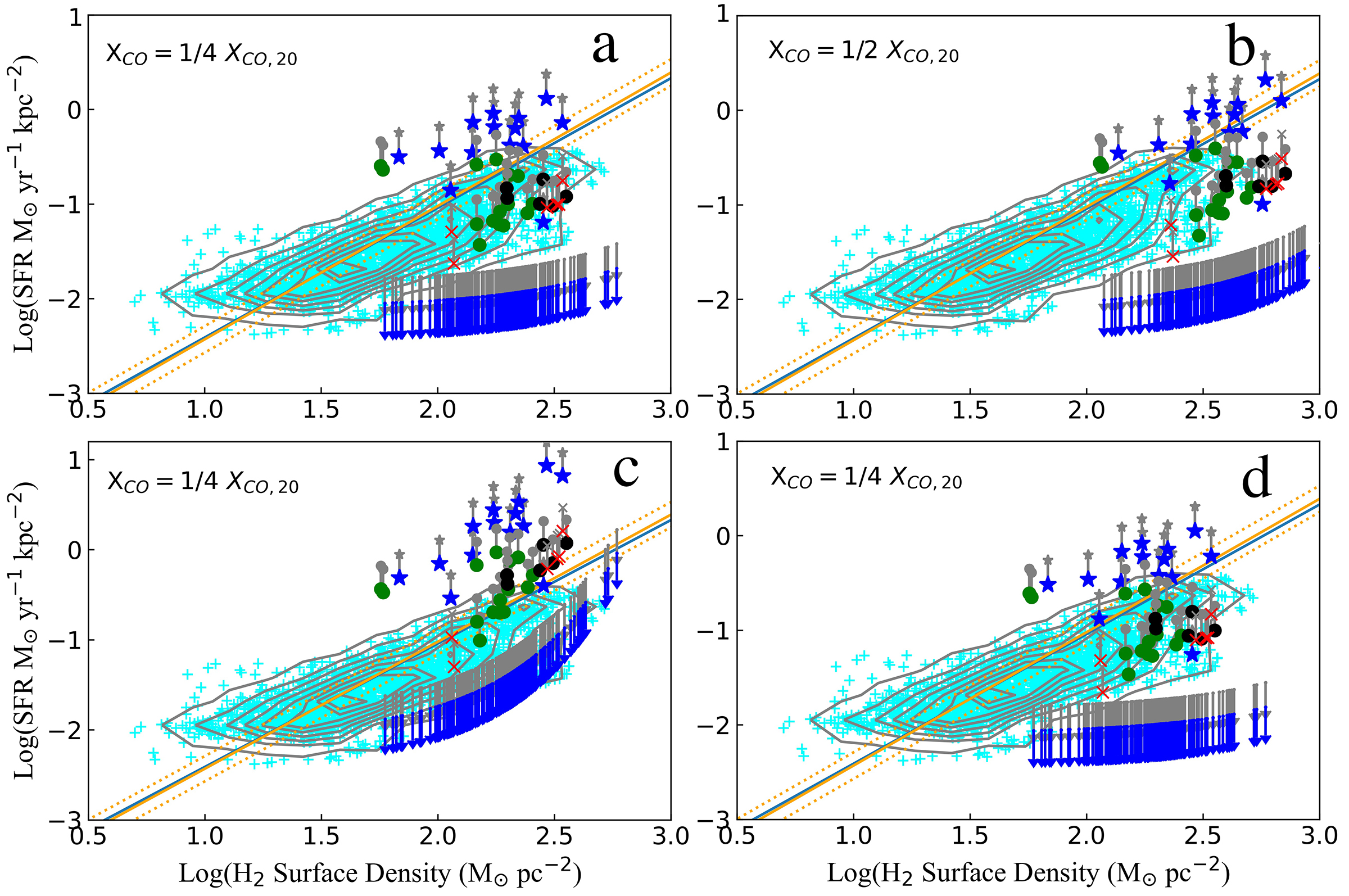}
\caption{The (logarithm of) SFR surface density in the bridge clouds versus molecular gas surface density, in which the star formation is corrected for extinction using the hydrogen column density method (Case 2, see \S~\ref{sec:extinct}), and for different assumed values of $X_{\rm CO}$,  (a) $X_{\rm CO}$ = 1/4 $X_{\rm CO,20}$,  assuming Milky Way (MW) extinction law, with a dust depletion of a factor of 5, (b) same as (a) but with $X_{\rm CO}$ = 1/2 $X_{\rm CO,20}$, (c) same as (a), but with no extra depletion and, (d) $X_{\rm CO}$ = 1/4 $X_{\rm CO,20}$, assuming an Small Magellanic Cloud(SMC)-like extinction law and no additional depletion. Significant bulk dust depletion of $\geq$ 5  was measured in the bridge by \citet{Zhu2007}.  The symbols and lines in the plot are the same as for Figure~\ref{fig:KSplot1}.} \label{fig:KSplot2}
\end{figure*}

Table~\ref{tab:A1} provides surface flux densities and logarithmic star formation rate surface densities for just one limiting case: that of minimal Pa$\alpha$ extinction and assuming 100$\%$ of the emission arises from star formation. In the figures and discussion that follows, we will explore a much wider range of possibilities.  The table also provides a Pa$\alpha$ flag that indicates whether Pa$\alpha$ is an upper limit, a detection, or is outside the NICMOS field of view. 
\vspace{0.2 cm}

\subsection{Testing the Kennicutt-Schmidt Relationship in the Taffy Bridge}

\begin{figure*}
\includegraphics[width=0.95\textwidth]{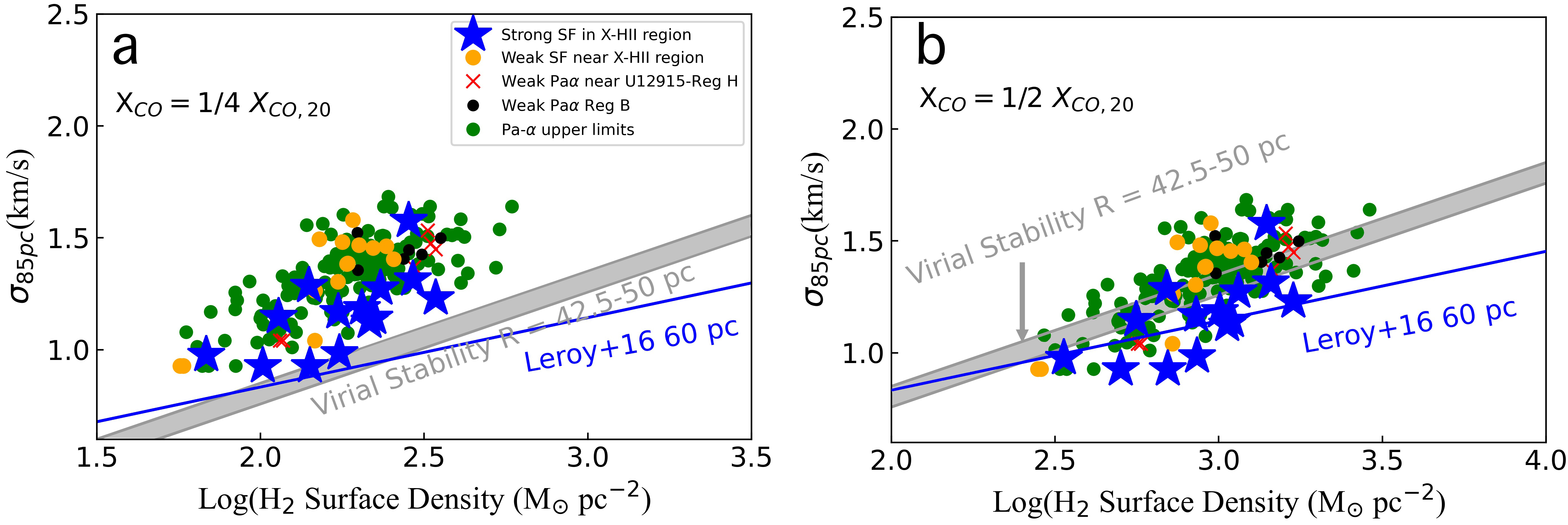}
\caption{The velocity dispersion of the CO-emitting clouds versus the gas surface density as a diagnostic of cloud stability for two CO/H$_2$ conversion factors, (a) $X_{\rm CO}$ = 1/4 $X_{\rm CO,20}$ and (b), $X_{\rm CO}$ = 1/2 $X_{\rm CO,20}$. The blue line shows the mean relationship for CO emitting clouds found by \citet{Leroy2016} from the PHANGS survey of normal galaxies and the grey shaded zone show the virial stability line for spherical constant-density clouds with diameters ranging from 85 to 100 pc (see text). Clouds below the grey shaded zone are gravitationally bound, whereas clouds above it are unbound. 
}
\label{fig:vsig_sig}
\end{figure*} 

Figure~\ref{fig:KSplot1}a-b shows a series of logarithmic plots of the star formation rate surface density versus the gas surface density for two plausible cases of X$_{\rm CO}$, and assuming Case 1 extinction (\S4.2). In normal galaxies, these two properties are related in what is referred to as the Kennicutt-Schmidt (hereafter KS) relationship \citep{Kennicutt1998a}.  
Each plot shows the range of possible star formation density associated with different regions (see key) for the case where there are no shocks contributing to the Pa$\alpha$ (grey symbols) connected to points with 45$\%$ shock contamination (colored symbols). A grey line joins the two extremes. 

The figures show that generally the points inside the X-HII region lie close to the expected KS relationship, but the majority of the other regions do not. Those with weaker assumed star formation are spread into and below the region occupied by clouds in M51 from \citet{ Leroy2017, Leroy2018}. 
The upper limits (with or without shock contamination), which include the majority of the more than 200 CO-emitting regions, 
lie significantly below the surface density they would have if they followed the KS relationship. Depending on the assumed X$_{\rm CO}$ value, the denser clouds lie at least an order of magnitude below the standard relationship. This strongly suggests these molecular regions show significant star formation suppression.  
 
Could this conclusion be driven by an underestimation of the actual extinction in the dense molecular gas? In Case 1 extinction, we assume that the  extinction measured for the X-HII region can be universally applied to the regions with Pa$\alpha$ upper limits. To explore a more general way of measuring extinction, independent of optical/near-IR emission line corrections, we apply  Case 2 extinction 
to all the points, even the upper limits, in  Figure~\ref{fig:KSplot2}~a-d. Figure~\ref{fig:KSplot2}a and b, both with a dust depletion factor of 5, show the effect of increasing the surface density of the gas from $X_{\rm CO}$ = 1/4 $X_{\rm CO,20}$ to $X_{\rm CO}$ = 1/2 $X_{\rm CO,20}$, respectively.  Figure~\ref{fig:KSplot2}c shows the (unlikely) case of no dust depletion, and a Milky Way-type extinction law, whereas Figure~\ref{fig:KSplot2}d shows an SMC-like extinction law, again with no extra dust depletion \citep{Bohlin1978,Gordon2003}). The symbols and meaning are the same as for Figure~\ref{fig:KSplot1}. 
As one might expect, the SMC extinction law looks very similar to the low MW gas-to-dust ratio case.  This latter example is provided to explore the unlikely possibility that the gas is of low-metallicity--not expected in a head-on collision of this kind with two massive galaxies.

The results of varying the assumed degree of extinction over a wide variety of reasonable parameter space (given the measured values for both lower-than-normal X$_{\rm CO}$ and large dust depletion inferred by  \citet{Zhu2007}) continue to support the idea that the majority of the CO-detected clouds in the Taffy bridge are strongly star-formation suppressed \citep{Vollmer2021}, even at the newly explored scale of 60-100 pc relation. The behavior in the diagrams is similar for all cases. Increasing the column density pushes the cloud points to both the right (increased H$_2$ surface density) and upwards (more dust and assumed extinction). The extinction "knob" can be turned by either keeping the dust depletion constant and increasing the gas surface density, or by keeping the surface density constant and increasing the dust content, or both. 

The plots in Figure~\ref{fig:KSplot1} and Figure~\ref{fig:KSplot2}  also demonstrate another interesting result. The sources with detected Pa$\alpha$ associated with the X-HII region lie generally at, or above the mean KS relationship, depending on the assumed star formation/shocked gas emission associated with the Pa$\alpha$.  A subset of the explored parameter space 
would lead the X-HII region exhibits "normal" star formation rates and high shock contamination with values of $ X_{\rm CO} \sim 1/2 X_{\rm CO,20}$ (e. g. Figure~\ref{fig:KSplot1}b;~Case 1 extinction, or Figure~\ref{fig:KSplot2}b ;~Case 2 extinction). 

In most of the realistic scenarios that fit most of the known properties of the Taffy bridge, we conclude that there is strong evidence for star formation suppression in the majority of the molecular gas in the bridge on scales of 60-100 pc. Only in the region where the CO emission is seen projected against the stronger Pa$\alpha$ emission do we begin to see much more normal relationship between the surface density of the gas and that of the young stars. 

\subsection{Small-scale molecular gas kinematics}

We next explore the line-of-sight velocity dispersion of the gas as a function of the gas surface density and its relationship to star formation. The condition for virial stability is given as $2T/U = 1$, where $T$ is the kinetic energy of the clouds (assumed to be dominated by internal motions), and $U$ is the internal energy.  For a spherically symmetric, constant density cloud, the condition for virial stability is a quadratic one between velocity dispersion and gas surface density, namely,  $\sigma^2 =(3/5) \pi G R_{cloud} \Sigma_{gas}$. Figure~\ref{fig:vsig_sig}a and Figure~\ref{fig:vsig_sig}b show the line of sight velocity dispersion of the gas as a function of the gas surface density for the Taffy bridge clouds, for two values of assumed X$_{\rm CO}$. As in Figure~\ref{fig:KSplot1}, the color coding for the symbols signifies the degree to which the observed molecular gas is exhibiting star formation.
The grey band in 
Figure~\ref{fig:vsig_sig} shows the range of loci of clouds in virial equilibrium for diameters of 85-100~pc (R$_{cloud}$ = 42.5-50~pc). Also shown is the average relationship found by \citet{Leroy2016} for clouds on a 60pc scale for a sample of nearby normal galaxies from PHANGS, derived with $X_{\rm CO,20}$.  This latter relationship is flatter in this representation than the quadratic one expected from the virial theorem. Thus the typical clouds in normal galaxies show a slower growth\footnote{\citet{Leroy2016} found a relationship of the form $log_{10}(\sigma) = 0.31~\times~log_{10}(\Sigma_{gas,50})+0.74$, where $\sigma$ is in km s$^{-1}$, and gas surface density $\Sigma_{gas,50}$ is in units of  50 $\times M_{\odot} pc^{-2}$. } in velocity dispersion $\sigma$ with increasing gas surface density $\Sigma_{gas}$, suggesting that the gas surface density is not the only factor  at play in controlling  properties in late-type disk galaxies (see \citealt{Leroy2016,Meidt2018,Meidt2020}). 

 We notice three aspects which are worthy of discussion in Figure~\ref{fig:vsig_sig}. Firstly, the blue stars associated with the brightest Pa$\alpha$ emission in the X-HII region lie along the inner edge of the distribution, showing systematically lower values of $\sigma$ for a given surface density of gas compared with the other points. Secondly, the slope of the distribution of Taffy bridge points in the figure is slightly steeper than both the relationship for normal galaxies, and that of virial equilbrium relationship. A similar steeper trend was found for gas clouds in the Antennae interacting system \citep[][not shown here]{Leroy2016} which shares some similarities with the Taffy. We will discuss a comparison between the Antennae system and Taffy in \S~\ref{sec:comparison-antennae}. Finally, all the points in Figure~\ref{fig:vsig_sig}a, and many of the points in Figure~\ref{fig:vsig_sig}b lie on the upper side of the virial equilibrium line, suggesting the clouds are not bound by self-gravity.  Indeed several observed profiles have high velocity dispersion with FWHM (2.36$\sigma$) $\geq$ 100 km/s on the scale of 85 pc. At X$_{\rm CO}$ = 1/2 X$_{\rm CO,20}$,  
 those points that lie on the bounded side of that plot tend to have the strongest Pa$\alpha$, perhaps suggesting that those clouds have overcome gravity to form stars. 
 
\begin{figure}
\includegraphics[width=0.45\textwidth]{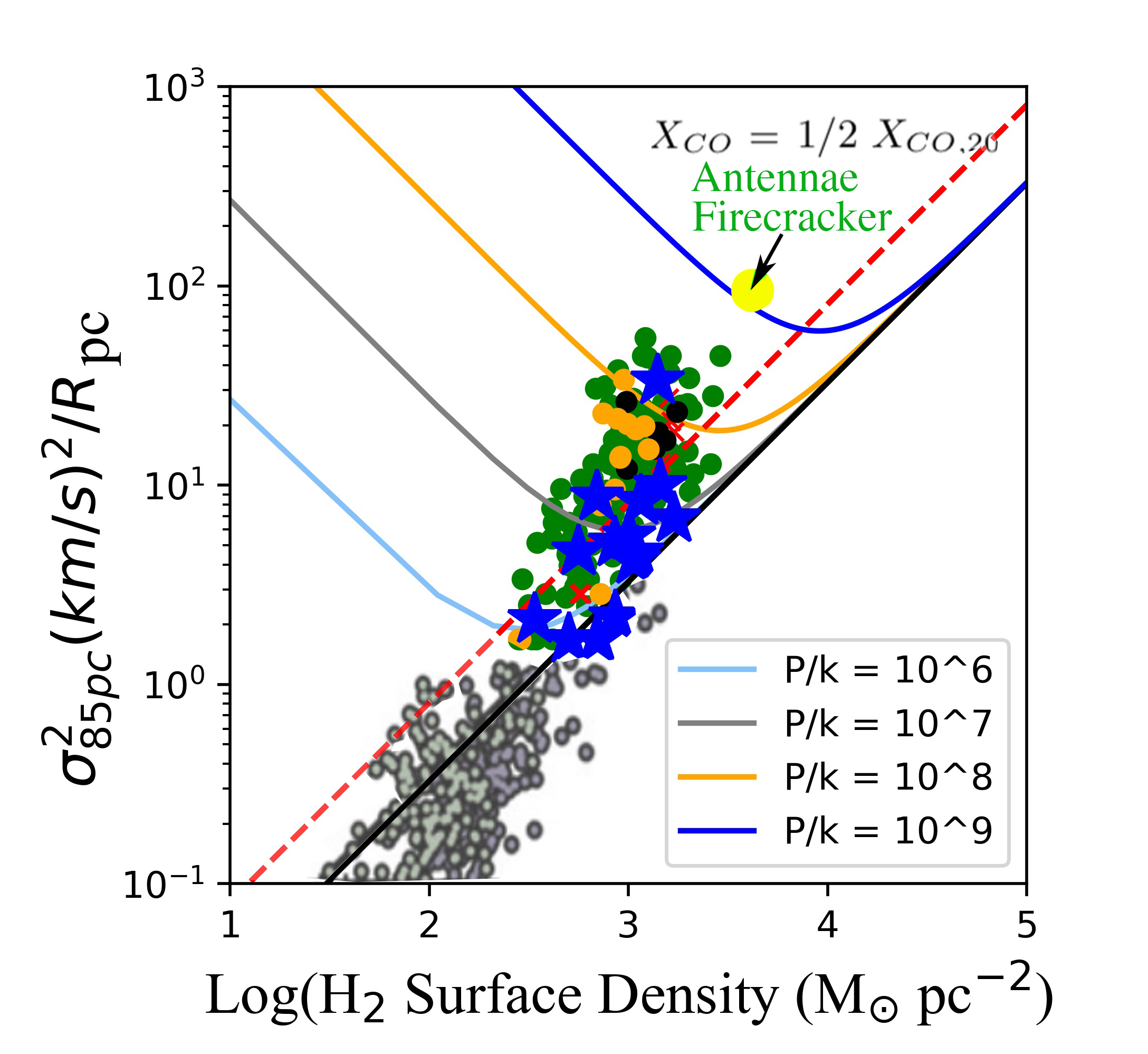}
\caption{The balance between bridge cloud internal motions $\sigma^2/R$  versus molecular surface density for the example case of $X_{\rm CO} = 1/2 X_{\rm CO,20}$ and diameter = 85~pc (R = 42.5~pc). The colored symbols for the points are the same as Figure~\ref{fig:vsig_sig}. Model lines for equilibrium with an external overpressure P$_e$ from a confining medium are shown for $10^6 <  P_e/k < 10^9$ in cgs units (from \citealt{Field2011}).  The black line and red dotted lines shows the condition for pure virial equilibrium for centrally concentrated and constant-density clouds, respectively.  The data for the protocluster candidate cloud in the Antennae ``overlap region'', sometimes called the ``Firecracker'' \citep{Johnson2015,Whitmore2014}, is shown as a large yellow filled circle, along with darker grey and grey-green filled points in the lower left corner, for Galactic and nearby galaxies clouds respectively, (adapted from Figure 6 of \citealt{Johnson2015}).}
 \label{fig:overpressure}
\end{figure} 

To explore what magnitude the overpressure would need to be to contain the Taffy bridge gas, we show in Figure~\ref{fig:overpressure} another common representation of the cloud kinematics (a logarithmic plot of $\sigma^2/R_{cloud}$ versus gas surface density) for the case of denser clouds (X$_{\rm CO} = 1/2 X_{\rm CO,20}$). The virial stability condition is shown for cloud diameter 85~pc (see caption). Also plotted are model curves for various over-pressures that would be required to provide stability to the clouds through an external medium\footnote{Increasing the assumed size of the clouds (to say 100 pc) would cause the points in Figure~\ref{fig:overpressure} to move down and to the left by a relatively small amount. More extended clouds would require less external pressure to bring them into equilibrium.}.  Adapted from Figure 6 of \citet{Johnson2015}, we also show the distribution of clouds found in nearby galaxies and Galactic clouds in the lower left corner of the plot. Those clouds tend to follow the line of virial stability within the observational scatter. The blue filled circles, which represent the hundreds of non-starforming Taffy bridge clouds, would require very high pressures, as high as P/k  of 10$^{6-8}$ (in cgs units), to stabilize or overpressure them. As discussed by \citet{Field2011} and \citet{Johnson2015} external pressures with P$_e$/k $>$ 10$^{5-6}$ are rarely found in normal galaxies.  

Could the direct high-speed collision of the Taffy galaxies create special conditions that might allow for regions of high gas overpressure in an otherwise relatively low-density turbulent medium?  \citet{Silk2019} has speculated that high-speed cloud-cloud collisions, produced in galaxies collisions,  can, on the one hand,  significantly suppress star formation in lower density regions because of supersonic turbulence and shear in the gas, but on the other hand, create rare over-pressured conditions (see \citealt{Padoan2012}). Here, dense self-gravitating clouds become over-pressured, collapsing from a slab-like compressed gas structure to form large numbers of "proto globular-like" star clusters with a relatively high star formation efficiency (see also \citealt{Madau2020}).  Regions of high density in a supersonic turbulent field is a natural result of the intermittency of turbulence, where highly non-linear behavior can occur for short periods of time within the turbulent system (e. g. \citealt{Anselmet2001,Hily-Blant2009}). Since intermittency occurs on scales close to the scale at which most of the dissipation is occurring (probably pc-scales, see \citealt{Guillard2009}), it is not clear how such high pressure regions within the multi-phase medium might affect the collapse of gas on the larger scales needed to form super star clusters.   

High overpressure from turbulence was suggested as an explanation for the dense CO emitting clouds found within a larger supergiant molecular cloud SGMC2 in the Antennae colliding galaxy system \citep{Wilson2000}, and  discussed in detail by \citet{Johnson2015}. This cloud (dubbed the "Firecracker") would require even higher external pressure than the Taffy clouds (see Figure~\ref{fig:overpressure} ).

\subsection{Star formation in the X-HII region}
Although the majority of the gas in the Taffy bridge is quite deficient in star formation, 
the X-HII region is an exception.  In Table~\ref{tab:HII} we quantify the overall star formation 
properties of the X-HII region using a variety of different measures, ranging from the Pa$\alpha$, 
H$\alpha$ and radio emission measures, as well as infrared measures of star formation, or a combination 
of both. Column 1 gives the observational method used, the radius (Column 2 and 3) of the circular aperture 
used in arcsec and area of the aperture in kpc$^2$ (assuming D = 62~Mpc).  Columns 4, 5 and 6  give the 
flux density, the measured line fluxes integrated over the apertures for Pa$\alpha$ (this work) and 
H$\alpha$ (from \citealt{Joshi2019}) and the total luminosity (for the IR this is $\nu F_{\nu}$). 
Column 7 and 8 give the estimated SFR (before and after extinction correction, where appropriate) 
based on methods shown in the table notes. Column 9 presented the equivalent SFR surface density.  
We also estimate the Pa$\alpha$ line and 6cm radio fluxes from the main extent of the X-HII region 
and the two hotspots (Hotspot N and S) within the X-HII region described earlier (Figure~\ref{fig:multicolorHii}). 
These are given the designation ALL, N and S respectively.  The notes to the table give the method and 
reference source for the various SFR measures used.  

\begin{deluxetable*}{lllclcccc}
\tablecolumns{9}
\tablewidth{0pc}
\tablecaption{Observed and derived properties of the Taffy extragalactic HII (X-HII) region\label{tab:HII}}
\tablehead{
\colhead{Obs/Method} & 
\colhead{R$_{ap}$} &    
\colhead{Area} & 
\colhead{Flux Density} & 
\colhead{Line Flux} &       
\colhead{LOG(L)} &        
\colhead{SFR} &      
\colhead{SFR$_{cor}$} &          
\colhead{LOG($\Sigma_{SFR}$)}
\\
\colhead{} & 
\colhead{($arcsec$)} &
\colhead{($kpc^2$)} & 
\colhead{($mJy$)} & 
\colhead{($erg~s^{-1}cm^{-2}$} & 
\colhead{($erg~s^{-1}$)} &   
\colhead{($M_{\odot} yr^{-1}$)} & 
\colhead{($M_{\odot} yr^{-1}$)} &  
\colhead{($M_{\odot} yr^{-1}kpc^{-2}$) } 
\\
\colhead{} & 
\colhead{} &
\colhead{} & 
\colhead{} & 
\colhead{$\times 10^{-15}$)} & 
\colhead{} &    
\colhead{} & 
\colhead{} &  
\colhead{} 
\\
\colhead{[1]} & 
\colhead{[2]} &
\colhead{[3]} & 
\colhead{[4]} & 
\colhead{[5]} & 
\colhead{[6]} &    
\colhead{[7]} & 
\colhead{[8]} &  
\colhead{[9]} 
}
\startdata
IRAC3.6$\mu$m                & 4.8  &  6.4 &  0.38$\pm$0.03 &          ---                  &     41.18$\pm$0.04\tablenotemark{b}  &          ---   &    ---        &          ---          \\
IRAC4.5$\mu$m                &  4.8 &  6.4 &  0.32$\pm$0.03 &          ---                  &    40.98$\pm$0.04\tablenotemark{b}   &          ---   &    ---        &          ---              \\
IRAC 5.8$\mu$m               & 4.8  &  6.4 &  1.46$\pm$0.15 &          ---                  &     41.54$\pm$0.04\tablenotemark{b}   &          ---   &    ---        &          ---                 \\
IRAC 8$\mu$m                  & 4.8  &  6.4 &  4.33$\pm$0.43 &          ---                  &     41.86$\pm$0.04\tablenotemark{b}   &          ---   &    ---        &          ---            \\
MIPS 24$\mu$m                & 4.8  &  6.4 & 6.1$\pm$1.9 &          ---                  &     41.55$\pm$0.13  &          0.08$\pm$0.04\tablenotemark{c} &    ---      &         -0.94\tablenotemark{c}               \\
NIC 3 Pa$\alpha$ & 1.56 & 0.69 & --- & 9.1$\pm$1.4           &     39.6$\pm$0.06   &    0.20 $\pm$0.03       &   0.24 $\pm$0.04\tablenotemark{i}        &   -0.5    \\
(ALL) & & & & & & & & \\
Pa$\alpha$$_{sf}$       & 1.56 &  0.69 &  ---  &         5.0$\pm$0.29\tablenotemark{e}                     &     39.36$\pm$0.06\tablenotemark{e}  &          0.11$\pm$0.03\tablenotemark{d,e} &    0.13$\pm$0.03\tablenotemark{i,d,e}     &          -0.72   \\
(ALL)\tablenotemark{e} & & & & & & & & \\
Pa$\alpha$ & 0.56 & 0.09 & --- & 2.1$\pm$0.2           &     38.97$\pm$0.06  &     0.04$\pm$0.01       &    0.05\tablenotemark{i}        &    -0.23    \\
(Hotspot N)& & & & & & & & \\
Pa$\alpha$ & 0.56 & 0.09 & --- & 3.0$\pm$0.45           &     39.14$\pm$0.06   &     0.07$\pm$0.02      &   0.08\tablenotemark{i}        &    -0.1    \\
(Hotspot S)& & & & & & & & \\
Pa$\alpha$+24$\mu$m      & 1.56\tablenotemark{f}  &   0.69\tablenotemark{f} &  ---  &          ---                   &     ---         &             0.15$\pm$0.05\tablenotemark{df}  &    ---       &          -0.66\tablenotemark{d,f}  \\
GCMS H$\alpha$   & 4.8  &  6.5 &  ---  &            16.6$\pm$3.3       &     39.88$\pm$0.08 &          ---   &    ---        &          --- \\
(ALL) & & & & & & & & \\
H$\alpha$$_{sf}$\tablenotemark{e}         & 4.8  &  6.5 &  ---  &          8.3$\pm$1.6\tablenotemark{e}                   &     39.58$\pm$0.08\tablenotemark{e}  &          0.14$\pm$0.03  &    0.50$\pm$0.1 \tablenotemark{a}      &        -1.11\tablenotemark{a}       \\
H$\alpha$$_{sf}$+24$\mu$m      & 4.8  &  6.5 &  ---  &          ---                   &     ---       &            0.09$\pm$0.02 \tablenotemark{f}  &    ---        &      -1.86        \\
Radio$_{(6GHz)}$                          & 1.56  &  0.69 & 0.33$\pm$0.03 & --- &[1.51$\pm$0.15 &            0.25$\pm$0.04\tablenotemark{g}  &   ---      &          -0.41 \\
(ALL) &    &     &                 &     & $\times$ 10$^{20} W $Hz$^{-1}$]\tablenotemark{g} & &  &. \\
Radio$_{(6GHz)}$                          & 0.56  &  0.08 & 0.052$\pm$0.005 & --- &[2.38$\pm$0.24 &            0.040$\pm$0.002\tablenotemark{g}  &   ---      &          -0.32  \\
(Hotspot N) &    &     &                 &     & $\times$ 10$^{19} W $Hz$^{-1}$]\tablenotemark{g} & &  &. \\
Radio$_{(6GHz)}$                          & 0.56  &  0.09 & 0.086$\pm$0.009 & --- &[3.94$\pm$0.40 &            0.07$\pm$0.01\tablenotemark{g}  &   ---      &          -0.11  \\
(Hotspot S) &    &     &                 &     & $\times$ 10$^{19} W $Hz$^{-1}$]\tablenotemark{g} & &  &. \\
Radio$_{(1.4GHz)}$                     & 1.56  & 0.69  & 1.2$\pm$0.1 & --- &[5.17$\pm$0.50 &             0.29$\pm$0.03\tablenotemark{h}  &    ---       &         -0.38     \\
 (ALL)                                  &    &     &                 &     & $\times$ 10$^{20} W $Hz$^{-1}$]\tablenotemark{h} & &  &. \\
\enddata
\tablenotetext{a}{SFR$_{cor}$:    extinction corrected for A$_v$ = 1.5 mag based on the Balmer decrement of \citet{Joshi2019}.}
\tablenotetext{b}{Luminosity $\nu$F($\nu)$, and  assuming a distance to Taffy of 62 Mpc. We used $\nu$ = 8.3, 6.7, 5.2, 3.8 and 1.25 $\times$ 10$^{13}$ Hz for the IRAC band 1, 2, 3 and 4 and MIPS 24$\mu$m respectively.}   
\tablenotetext{c}{Assuming a simple monochromatic relationship of \citet{Relano2007,Calzetti2010}, SFR$_{24\micron}[M_{\odot} yr^{-1}] = 5.66 \times 10^{-36}$ $\times L_{\nu}(24\micron)^{0.82}$, where $L_{\nu}(24\micron)$ is in erg s$^{-1}$. For the star formation surface density we assume the 24$\mu$m flux is emitted from the same area as the the P$\alpha$ flux, i.e. from an area of 0.69 kpc$^2$. }
\tablenotetext{d}{assume CASE B recombination, f(H$\alpha$) = 2.86/0.332 x f(Pa$\alpha$), and the SFR - H$\alpha$ relations of \citet{Calzetti2007,Calzetti2010}: SFR($[M_{\odot} yr^{-1})= 5.3 \times 10^{-42} L(H\alpha_{corr})$, where L(H$\alpha_{corr})$ is extinction corrected in erg s$^{-1}$.}
\tablenotetext{e}{assume both Pa$\alpha$ and H$\alpha$ fluxes have $\sim$ 45$\%$ contribution from shocks, so Pa$\alpha$$_{sf}$ (star formation) = 0.55 x Pa$\alpha$$_{tot}$ \citep{Joshi2019}}
\tablenotetext{f}{assume SFR (measured $H\alpha + 24\micron) = 5.3 \times 10^{-42}[L(H\alpha_{sf}) + 0.031 L(24\micron)]$, and that the 24$\mu$m flux come from the same area as the H$\alpha$ emission; \citep{Calzetti2007}.}
\tablenotetext{g}{from VLA 6 GHz flux densities and luminosities from Appleton et al. (in prep.) assuming the sum of thermal (T=10$^4$ K), non-thermal 6GHz contributions from \citet{Murphy2011}, and calculated spectral index of $\gamma$ = -0.76 (this work) within a common area of 2.3 arcsec$^2$ area.  }
\tablenotetext{h}{from VLA 1.48 GHz flux density and luminosity of Appleton et al (in prep.), and assuming the SFR prescription of \citep{Condon1992} and non-thermal fraction of flux = 0.8. The source is very extended at 20cm compared with the restoring beam = 1.27 x. 1.1 arcsec$^2$.}
\tablenotetext{i}{assumes 0.2 mag of extinction at Pa$\alpha$}
\end{deluxetable*}

The SFRs estimated in Table~\ref{tab:HII} provide a relatively consistent picture considering the different methods used. \citet{Joshi2019} showed that the H$\alpha$ line emission is contaminated by up to 45$\%$ by shocked gas in the bridge, and so the line fluxes for Pa$\alpha$ have been reduced accordingly before calculating the SFRs. For the entire X-HII  region, we can compare the shock corrected Pa$\alpha$ SFR, Pa$\alpha$~(ALL) = 0.13 $M_{\odot}$ yr$^{-1}$, with that derived from the extinction independent 6~GHz radio continuum, Radio$_{(6GHz)}$~(ALL) = 0.25 $M_{\odot}$~yr$^{-1}$. A low SFR is supported using a pure MIPS~24$\mu$m flux calculation\footnote{We note that 
the background-subtracted 24$\mu$m flux associated with the X-HII region is uncertain because of contamination by the disk of UGC~12915. Uncertainties in our estimated contamination are included in the uncertainty in Table~2.} of 0.08 $M_{\odot}$~yr$^{-1}$. \citet{Calzetti2007} provide a composite method of calculating the SFR using the 24$\mu$m flux in combination with the H$\alpha$ emission uncorrected for extinction. Using either the observed Pa$\alpha$ flux as a proxy for the H$\alpha$ flux, or the observed H$\alpha$ flux in combination with 24$\mu$m, yields 0.15 and 0.09 $M_{\odot}$~yr$^{-1}$ respectively. 
These values are in contrast to the extinction corrected H$\alpha$ flux measurements which yield  H$\alpha_{sf}$ = 0.5 $M_{\odot}$ yr$^{-1}$. The high value for H$\alpha_{sf}$ may imply that \citet{Joshi2019} overestimated the optical extinction at H$\alpha$. We conclude that the most likely SFR for the main Pa$\alpha$ emitting area of the X-HII region  is in the range 0.10-0.25 $M_{\odot}~yr^{-1}$. 

We estimate that the molecular gas mass associated with the entire X-HII region (4.8  arcsec$^2$) based on our CO observations is 1.29$\pm$0.01 $\times$ 10$^8$ [$X_{\rm CO}/X_{\rm CO,20}$] $M_{\odot}$. Formally, the depletion time for the molecular gas, $M_{mol}$/SFR, is therefore 645 Myrs and 160 Myrs, respectively for $X_{\rm CO} = X_{\rm CO,20}$, and $X_{\rm CO} = 1/4 X_{\rm CO,20}$. However this depletion time is likely much longer because the star formation is quite clumpy and inhomogeneous.  The KS-diagrams of Figure~\ref{fig:KSplot1}a+b, show that where many of the brighter SF regions lie close to the the average KS relationship, the implied depletion times are $\sim$ 1Gyr. 

The Pa$\alpha$ hotspots in the X-HII region could be examples of the formation of massive proto-clusters conceptually similar to fast gas-on-gas collisions between dwarf galaxies postulated as one formation mechanism for Ultra Diffuse Galaxies (\citealt{Silk2019}; UDFs).   Table~\ref{tab:HII} provides estimates of the SFR for the north (N) and south (S) hotspots of 0.05~$M_{\odot}~yr^{-1}$ and 0.08 $M_{\odot}~ yr^{-1}$ respectively, based on Pa$\alpha$. The radio continuum provides approximately similar values.  Over the next 10 Myrs, 10$^6 M_{\odot}$ of stars could form, making them future young massive star cluster candidates.


\section{Other potentially similar environments to the Taffy bridge gas}
\subsection{Comparison with Antennae "overlap" region}
\label{sec:comparison-antennae}

It is worthwhile contrasting the properties of the Taffy bridge with another well studied colliding galaxy system, the Antennae galaxy pair, NGC4038/9. The Antennae is known to be in the starburst phase of a major merger \citep{Whitmore1999,Schweizer2008}.  Unlike the Taffy system, which has experienced a head-on counter-rotating collision in the past (25-35 Myr ago), the Antennae is still in the compressive ``contact" stage of a relatively slow (100 Myr) prograde disk-disk merger \citep{Renaud2018}. In global properties, the Antennae and Taffy have very similar total far-infrared (FIR)  luminosities and total molecular gas masses. The Antennae has a FIR luminosity, L$_{\rm FIR}$= 5.6~$\times$~10$^{10} L_{\odot}$ \citep{Brandl2009} and total molecular mass $1.1~\times~10^{10} M_{\odot}$ \citep{Zhu2003}, whereas equivalent values for the Taffy are L$_{\rm FIR}$ = 6.5~$\times~10^{10} L_{\odot}$ \citep{Sanders2003}) and a total molecular mass $M_{\rm mol} = 0.97~\times~10^{10} M_{\odot}$ \citep{Zhu2007}.

The ``overlap'' region of the Antennae (which lies between the two component galaxies) also shares a
number of similarities with the Taffy bridge. They both
include significant quantities of molecular gas distributed in narrow filaments. Bearing in
mind the uncertainties in the X$_{\rm CO}$ factor for both galaxies, the Taffy bridge has roughly twice as much molecular gas as the ``overlap'' region. The Taffy bridge contains $\sim 1.3~\times~10^9 M_{\odot}$ (for X$_{\rm CO} = 1/5 ~X_{\rm CO,20}$;~\citealt{Zhu2007}), compared with $\sim 0.5~\times~10^9 M_{\odot}$ for the  Antennae ''overlap`` region (\citealt{Stanford1990,Wilson2000}, if we assume a common  X$_{\rm CO}$ factor. 

Unlike the Taffy bridge, which we have shown is largely devoid of star formation, the Antennae overlap region is bursting with activity. The total star formation rate in the overlap regions has been measured to be 5 $M_{\odot}$~yr$^{-1}$ \citep{Stanford1990} compared with 0.1-0.25 $M_{\odot}$~yr$^{-1}$ for the Taffy bridge (this paper). In the Antennae overlap region, the star formation activity is concentrated in a handful of very massive Super Giant Molecular Complexes \citep{Wilson2000}.
With the one exception of the ``Firecracker" discussed earlier, the majority of these giant molecular complexes in the Antennae dominate the star formation output of the overlap region \citep{Mirabel1998,Whitmore1995,Whitmore2010,Johnson2015}. 
In the Taffy, the molecular clouds are less massive, and only the X-HII region compares in molecular mass (10$^8$ $M_{\odot}$) with the lower end of the mass distribution of the Antennae SGMCs. Another difference is that, unlike the Taffy filaments, which contain clouds with locally high velocity dispersion (40-100 \kms), the Antennae CO filaments exhibit unusually low velocity dispersion ($\leq$ 10~\kms;~ \citealt{Whitmore2014}). Higher CO velocity dispersion are reported for the Antennae system in general \citep{Leroy2016}, which likely correlate with increased star formation activity. This seems the opposite of the Taffy, where the X-HII region exhibits the lowest velocity dispersion in the CO clouds, whereas the filaments and star formation-deficient clouds have the highest velocity dispersion.  In summary, except for some  regions of high star formation rate in the Taffy X-HII region, the filaments and clumps in the Taffy bridge are strongly suppressed in star formation compared with the Antennae. 
\subsection{Comparison with molecular gas in ram-pressure stripped systems} 
Another kind of environment that might have some similarities with the Taffy bridge are those involving the ram-pressure stripping of gas from  cluster galaxies, where stripped molecular clouds find themselves in the intergalactic medium. There is now significant evidence that the ISM of late-type cluster galaxies can be stripped \citep{Gunn1972} by their interaction with a hot gaseous cluster halo, including the formation of tails revealed in HI, H$\alpha$ emission, X-rays and warm and cold molecular hydrogen (e. g. \citealt{Gavazzi2001,Wang2004,Oosterloo2005,Sun2007,Yagi2007,Jachym2014}). Although the physical mechanism from stripping gas into the intergalactic medium (see \citealt{Schulz2001}) is rather different from the case of a major head-on collision, as in the Taffy case, there are some potential similarities. For example, \citet{Sivanandam2010} found evidence for warm molecular hydrogen in the stripped tail of ESO 137-001 that was likely shock-heated by turbulent interactions with its host ICM of Abell cluster 3627. Cold molecular gas has also been found in some likely ram-pressure stripped galaxies, primarily through single dish observations of low spatial resolution (e. g. \citealt{Jachym2014,Jachym2017,Lee2017}). Only recently have high resolution observations become available \citep{Jachym2019} including those studied by the GASP survey\footnote{GASP = Gas Stripping Phenomena with MUSE \citep{Poggianti2017}.}. In these cases, these intergalactic clouds, either stripped from their host galaxies or formed{\it~in situ} through turbulent compression, can be compared with the Taffy bridge. 

One particular gas-stripped system, JW 100, is a member of a class of similar "Jellyfish" galaxies ~\citep{Ebeling2014}, and was observed with ALMA in the CO (1-0) and (2-1) transitions \citep{Moretti2018,Moretti2020}. This massive cluster galaxy (with a diameter of at least 45 kpc) exhibits a clumpy tail of molecular hydrogen extending 35 kpc away from its disk. Approximately $ 7~\times~10^9 M_{\odot}$ of molecular gas is present in the tail, which is comparable with the Taffy bridge, representing about 30$\%$ of the gas in the galaxy disk. Like the Taffy, several clumps in the tail exhibit high velocity dispersion (~70-80 \kms), although much of the gas exhibits lower linewidths. 
Although the mechanism for stripping the gas in such tails may be markedly different from the Taffy~\citep{Pedrini2022}, it is possible that gas may be formed in shocks or eddies in the wake of the main shocked region around the galaxy. Is there evidence for star formation suppression here too? 

\begin{figure}
\includegraphics[width=0.49\textwidth]{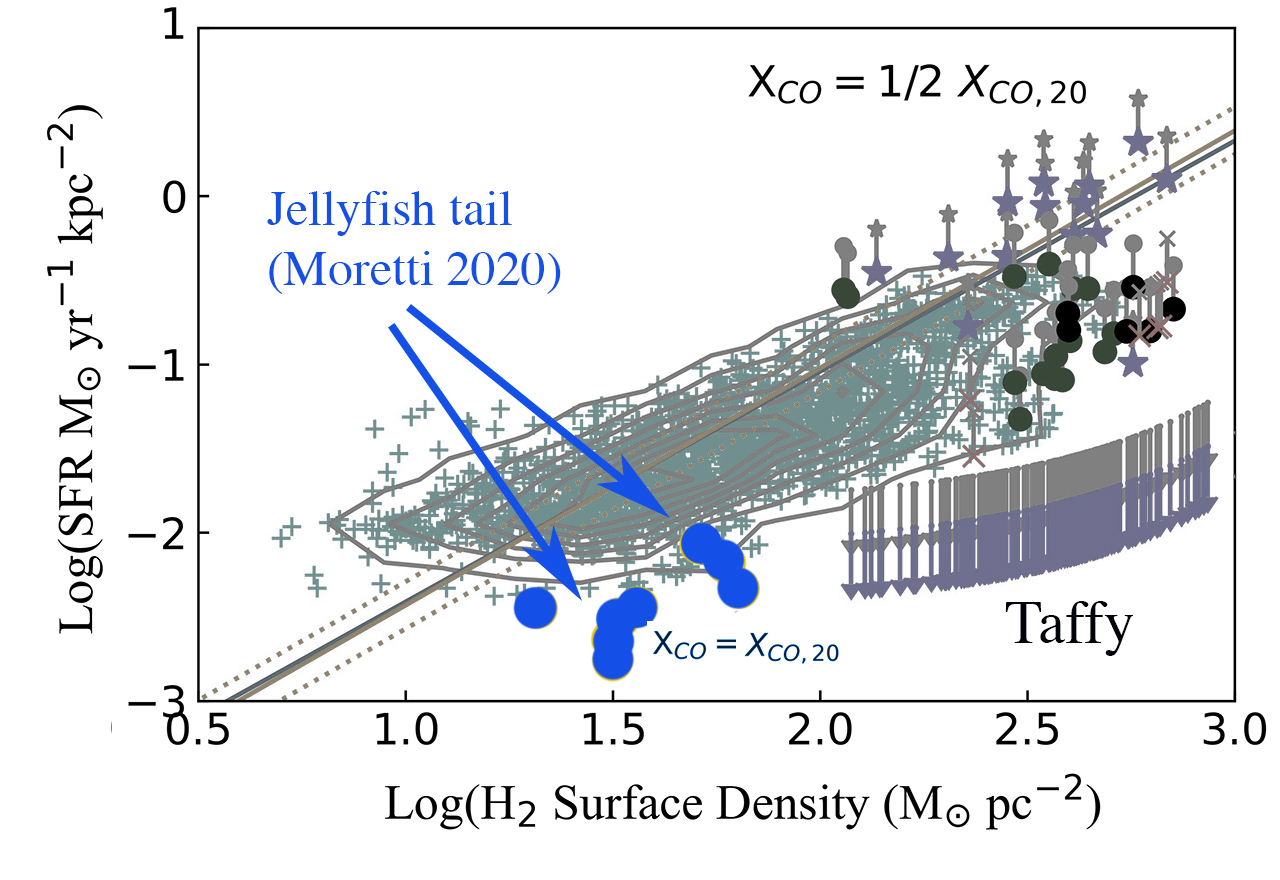}
\caption{A comparison between the surface formation rate density and molecular hydrogen surface density for clouds in the ram-pressure stripped tail of the jellyfish galaxy JW100 (blue filled circles~\citealt{Moretti2020}, where X$_{\rm CO}$ was assumed to be Galactic), and a partially greyed-out version of Figure~\ref{fig:KSplot2}b of the current paper (symbols are the same as that figure). Like the majority of the Taffy clouds, which have upper limits to star formation, the JW100 clouds also show significant star formation suppression.}
 \label{fig:jelly}
\end{figure}

In Figure~\ref{fig:jelly} we show examples taken from \citet{Moretti2020}, of clouds in the stripped tail of the JW100 (blue circles) overplotted on a partially greyed-out version of one characteristic KS-relationship plots of the  Taffy from this paper (Figure 10b). As is discussed in more detail by \citet{Moretti2020}, many of the stripped clouds fall below the standard KS relationship. Thus, like the majority of non-detected Pa$\alpha$ upper limits in Taffy, the stripped clouds seem to show very low star formation rates, as measured by optical spectroscopy. One difference between those stripped clouds and Taffy is that the former appear to have significantly lower overall H$_2$ surface densities compared with the Taffy bridge, even if (as was assumed by the authors) X$_{\rm CO}$ has a galactic value. However, one should caution that the ALMA data discussed by \citet{Moretti2020} was taken with 1 arcsec resolution, which corresponds to 1kpc at the distance of JW100. Thus the surface densities of molecular gas could well be much higher than observed, if the system had been observed at the same linear resolution as the Taffy. Nevertheless, it is clear that there are some similarities between the properties of the gas in the stripped clouds in the JW 100 and those suppressed star formation clouds in the Taffy bridge. 

The careful comparison of the multi-wavelength observations of JW100 by \citet{Poggianti2019} further strengthens those similarities. Like the Taffy X-HII region, an ULX source was found associated with one of the brighter HII regions. These authors also showed that although parts of the stripped gas show evidence of star formation, there are regions of the tail that show LINER-type optical emission (similar to regions of the Taffy bridge \citealt{Joshi2019}), and X-ray emission not consistent with star formation.  A variety of possible heating mechanisms for such systems, including strong turbulent mixing, plasma interactions, thermal conduction, and shocks have been put forward as possible explanations~\citep{Poggianti2019,Campitiello2021,Pedrini2022}.

\section{Origin of the Star Formation Suppression in the Bridge}\label{sec:origin-sf-supression}

\begin{figure}
\includegraphics[width=0.48\textwidth]{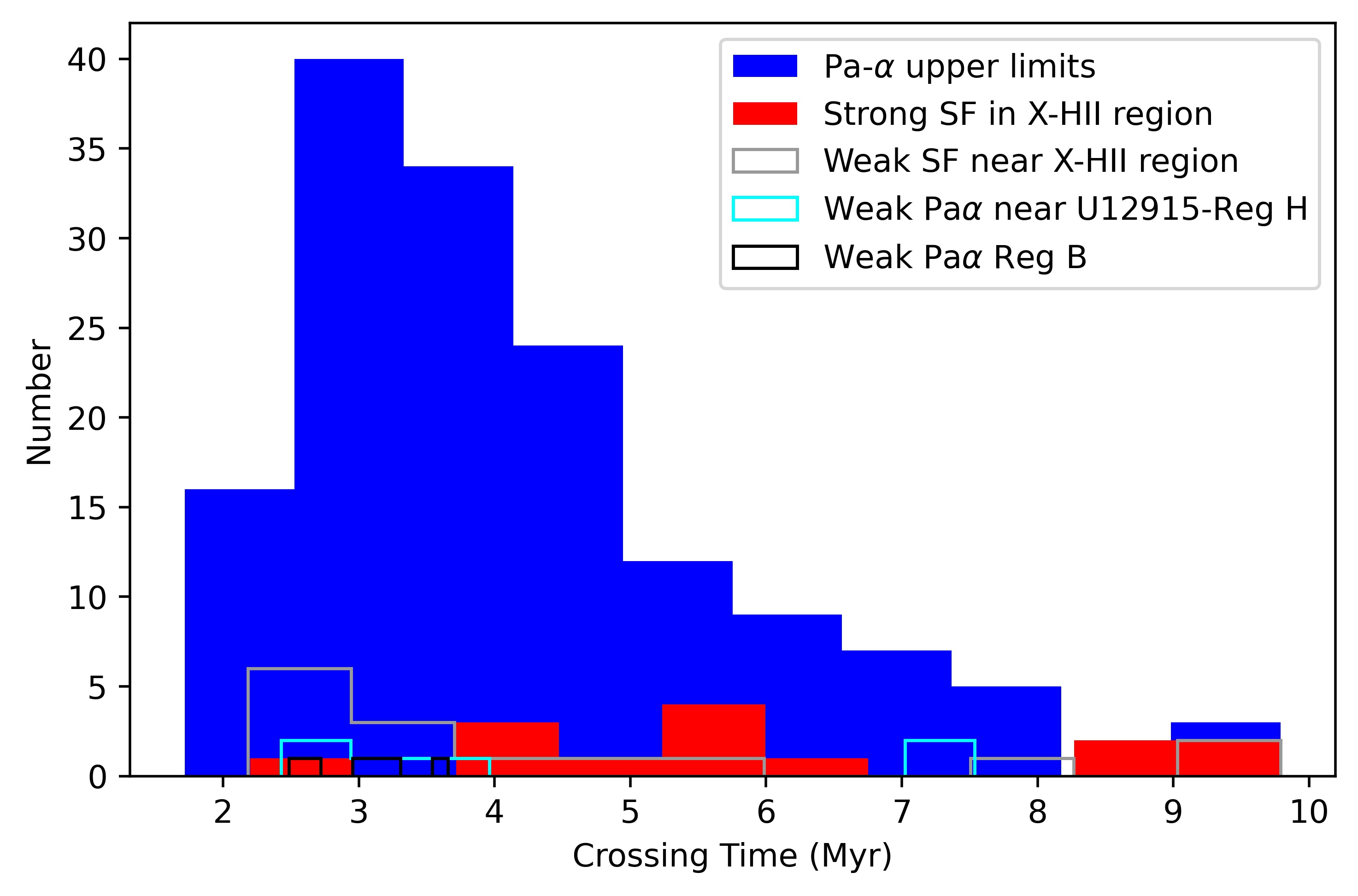}
\caption{Cloud crossing times in the bridge in Myr for an assumed diameter of 85~pc. If the clouds are smaller the crossing times will be even shorter. The clouds devoid of star formation (solid blue) have generally shorter crossing times than those containing star formation (solid red). The other samples are the same as those described in Figure~\ref{fig:KSplot1}.}

\label{fig:crosstime}
\end{figure}

The majority of the clouds in the Taffy bridge are likely unbound if self-gravity is the only restoring force for a wide range of reasonable values of assumed X$_{\rm CO}$. Such clouds would evaporate on a crossing time ($2R/\sigma$), which is typically 2-8~Myr (see Figure~\ref{fig:crosstime}) for the average Taffy bridge cloud. This is much shorter then the nominal time since the Taffy galaxies collided of 25-35~Myrs \citep{Condon1993}. This short timescale seems to rule out the possibility that the clouds in the bridge were originally normal molecular clouds confined to their host disks before the collision. In such a simple picture, such clouds would suddenly no longer feel the gravitational forces of the original disk, and maybe expected to freely expand when ejected into intergalactic space. Although it is possible that a sub-set of clouds (the long tail of the distribution in Figure~\ref{fig:crosstime}) may have crossing times long enough to have survived without an external medium since the collision (we cannot resolve the linewidths of clouds 
with  FWHM $<$ 15 \kms), it is clear that the majority of the gas, including those associated with the X-HII region, have crossing times which suggest they are relatively recently formed. Furthermore, models of the head-on collision like the Taffy system show significant ionization of the ISMs in both galaxies at and shortly after the collision suggesting that only a relatively small number of clouds punch through unscathed \citep{Yeager2020a,Yeager2020b}.

To understand why the CO clouds are not virialized, we are reminded that the Taffy bridge CO clouds are embedded in an highly turbulent multi-phase medium. There is evidence for the existence of copious quantities of shock-heated warm H$_2$ through direct mid-IR emission lines, and indirectly through the detection of [CI] and [CII]  emission~\citep{Peterson2012,Peterson2018}.  Additionally, there is evidence for shock-excited atomic emission line gas in the bridge \citep{Joshi2019}. Thus the CO-emitting clouds are likely immersed in a highly turbulent medium, similar to the intergalactic shocked filament in Stephan's Quintet \citep{Appleton2017,Guillard2021}. \citet{Vollmer2021} has discussed this in the context of their dynamical model of the Taffy system which was tuned to provide a good match with the lower-resolution PdBI CO observations of the Taffy. In that paper, the authors use their model to show that turbulence driven by cloud-cloud collisions on large-scales in the bridge would lead to clouds which are not virialized.  Our observation, which push down by an order of magnitude in scale compared with the PdBI observations, fully support this conclusion. 

However the actual details of the mechanism by which cold CO clouds are dynamically heated are likely to be quite complex. For example, the cascade of energy from the large driving scales of the collision, and subsequent continued high Mach-number cloud-cloud collisions can only be properly modeled with  multi-phase model which takes into account atomic and molecular cooling phases, phase transitions, shocks, turbulence and gas phase-mixing in different media \citep{Guillard2009}.  For example, in the analogous Stephan's Quintet shocked filament, recent observations of Ly$\alpha$ emission \citep{Guillard2021} combined with previous observations of molecular hydrogen and X-rays \citep{Appleton2017,Osullivan2009} show strong evidence that the dissipation of kinetic energy involves all gas phases. These are likely connected through a turbulent cascade involving atomic and molecular magneto-hydrodynamic shocks \citep{Lesaffre2020,Durrive2021}, and turbulent mixing layers \citep{Slavin1993}. The models of \citet{Yeager2020a,Yeager2020b} are a step in the right direction in attempting to treat multi-phase cooling of the gas in Taffy-like bridge systems. Those models predict rapid evolution in the gas phases as the collision progresses as shocks permeate the medium, with the appearance of waves of cooling, and periods of re-excitation by cloud-cloud collisions. The inclusion of mid-IR H$_2$ and [CII] cooling lines is another important step (Yeager et al, in preparation). These models, however, cannot yet capture the turbulent mixing that must occur at the boundaries between the hot, warm and cold components. Such models only partially deal with the possibility that the cold molecular clouds are transient material that is constantly being recycled through different thermal phases. In such cases, clouds must cool and collapse rapidly to have a chance of forming stars. 

The existence of high Mach-number gas collisions may stochastically generate regions like the X-HII region \citep{Yeager2020b}. The Pa$\alpha$ hotspots in the X-HII region could be examples of the formation of massive proto-clusters conceptually similar to fast gas-on-gas collisions between dwarf galaxies postulated as one formation mechanism for Ultra Diffuse Galaxies (\citealt{Silk2019}; UDFs).  The hotspots have locally high star formation surface densities, and over the next 10 Myrs the extragalactic HII region could form $\sim$ 1.3 $\times 10^6~M_{\odot}$ of stars,  making them future young massive star cluster candidates. Although in the case of the Taffy, where the size of the colliding systems is larger than \citet{Silk2019} envisaged, high Mach-number gas collisions will likely continue to generate regions like the X-HII region across the full extent of the bridge over time \citep{Yeager2020b}. 

An added complexity is the possible influence of magnetic fields. The Taffy was discovered because of the detection of radio continuum between the galaxies resulting from synchrotron radiation from cosmic rays spiraling in magnetic fields \citep{Condon1993}. \citet{Lisenfeld2010} explained the radio emission as the result of cosmic rays accelerated in shocks produced {\it in situ} in the Taffy bridge. The powerful emission from the lowest pure rotational transitions of H$_2$ in Stephan's Quintet, which exhibits very similar Mid-IR H$_2$ spectra to the Taffy bridge, is most easily explained \citep{Lesaffre2013} by magnetic shocks which moderate the temperature of the shock into the temperature range where the 0-0S(0) and 0-0S(1) line dominate (100 $<$ T $<$ 300 K). New deep VLA radio continuum observations with much higher resolution than the original Condon et al. observation have been recently made (Appleton, in preparation) which will provide a more complete picture of the distribution of thermal and non-thermal emission in the Taffy bridge, and may help determine the degree of importance of magnetic fields in the process of energy dissipation and cloud dynamics.

\section{Conclusions} 

This is the first of two papers describing ALMA observations of CO in the Taffy galaxy system,~UGC 12914/5.  In this paper we present a large mosaic of primary beam pointings made in the CO (2-1) transition at a spatial resolution more than an order of magnitude greater than previous observation (0.24 x 0.18 arcsec$^2$). The paper concentrates on the molecular bridge extending between the galaxies.  When compared with archival HST NICMOS Pa$\alpha$ and {\it Spitzer} observations, as well as new VLA radio continuum, Palomar 5-m optical spectroscopy and previously published {\it Chandra} X-ray observations,  we conclude the following:

\begin{itemize}
\item{The observations provide evidence that gas on scales of 60-100~pc in the Taffy galaxies is significantly disturbed in the aftermath of the head-on collision between the galaxies. The gas in UGC~12915 is particularly disturbed, with streamers of likely tidal debris  extending northwards far from the inner disk. UGC 12914, although less obviously disturbed that its companion, exhibits a singular powerful narrow structure of CO which follows the southern edge of the inner disk into a hook-like arm.  }
\item{The molecular gas bridge between UGC~12914/5 is resolved into a myriad of narrow filaments (60-100pc~$\times$~1kpc) and clumpy structures extending between the two galaxies which have some similarities with the filaments seen in the "overlap" region of the Antennae galaxies, except that almost all of them are devoid of star formation. By analyzing 234 regions in the bridge, we compare the star formation rate surface density derived for Pa$\alpha$ HST observations with the ALMA data at 85~pc resolution.  We show that the majority of the filaments are devoid of star formation, and fall significantly below the Kennicutt-Schmidt relationship for normal galaxies,~especially for the numerous regions undetected in Pa$\alpha$. This statement is shown to be true over a wide range of assumed extinction, X$_{\rm CO}$ values and dust to gas ratios. The result strongly supports the idea that, except in the X-HII region, the gas in the bridge is experiencing strong star formation suppression, even at the high surface densities probed by the new ALMA observations.}
\item{The kinematics of the majority of the bridge molecular gas shows unusually high velocity dispersion on the scale of 85~pc, with some regions within the filaments showing Gaussian line profiles with FWHM $\geq$ 100 \kms. Gas associated with the X-HII region tends to have lower velocity dispersion, suggesting it is less turbulent there. Like clouds seen in the Antennae galaxies, the Taffy clouds show a steeper slope in log($\sigma$) versus log($\Sigma_{gas}$) than normal galaxies. In addition, most of the Taffy bridge regions are gravitationally unbound for most reasonable values of assumed $X_{\rm CO}~(1/4 < (X_{\rm CO}/X_{\rm CO,20}) < 1/2$), and, like the "Firecracker" region in the Antennae system, would require an extremely high external pressure ($10^6 < P_e/k < 10^8$) to pressure-support them. If they are not externally supported, many of the clouds would disperse on a crossing timescale of $\sim$2-5~Myrs and are therefore likely transient. Such clouds may be continuously created and destroyed in such a highly turbulent medium.} 
\item{Despite the highly turbulent medium, stars are able to sometimes form there, albeit at a low rate.  In the X-HII region we derive a star formation rate of~0.1-0.25 $M_{\odot}$~yr$^{-1}$. Gas associated with this region exhibits high local star formation rates, on or above the KS relationship for normal systems depending on the assumed X$_{\rm CO}$ value. These higher star formation rate densities of~0.6-0.8~$M_{\odot}$~yr$^{-1}$~ kpc$^{-2}$ in two regions (containing faint compact radio sources and a ULX source) on scales of $<$ 100 pc might be evidence of the formation of massive star clusters. We also present evidence of potential rotational motions in the molecular gas and ionized gas associated with the X-HII region. }
\end{itemize}

The discovery that the majority of the dense molecular clouds in the bridge are not able to form stars suggests that turbulence, driven by the collision of the two Taffy galaxies more than 25-30~Myr previously, is still actively suppressing star formation down the the scale of 60-100~pc across the majority of the bridge.  Recent gas-dynamical models of Taffy-like head-on collisions support the idea that even after 30 Myr, cloud-cloud collisions are still creating high Mach-number collisions in the bridge. Such collisions may also explain the large quantities of warm molecular hydrogen discovered in the bridge by the {\it Spitzer} IRS, as well as the existence of extensive shock-excited ionized gas throughout the bridge, including strong H$\alpha$ emission at a discrepant velocity near the X-HII region in the bridge. The process in which large-scale driven turbulence caused by the galaxy collision can influence dense cold molecular clouds on 60-100~pc scales is likely to involve many different processes, including atomic and  molecular magnetic shocks and turbulent mixing of gas, similar to that suspected in the Stephan's Quintet system. 

\vspace{0.2 cm}

This paper made use of data from the following observatories/instruments: The Atacama Large Millimeter/submillimeter Array (ALMA), the Hubble Space Telescope (HST; NICMOS), the Chandra X-ray telescope S3 Advanced CCD Imaging Spectrometer, the Very Large Array (VLA), the {\it Spitzer} Space telescope; IRAC and MIPS and the Double-beam Spectrograph on the 5-m Hale telescope at the Palomar Observatory.
The paper made use of the following software: Community Astronomy Software Applications (CASA; NRAO), Interactive Data Language (IDL; L3Harris Geospatial Solutions), Python 3 open source software, PypeIt \citep{Prochaska2020}, SAOImage DS9 (Chandra X-ray Science Center, HEASARC and JWST Mission office at STSCI.)

\acknowledgments  
This paper makes use of the following ALMA data: ADS-JAO.ALMA.2016.1.01037.S. ALMA is a partnership of ESO (representing its member states), NSF (USA) and NINS (Japan), together with NRC (Canada), MOST and ASIAA (Taiwan), and KASI (Republic of Korea), in cooperation with the Republic of Chile. The Joint ALMA Observatory is operated by ESO, AUI/NRAO and NAOJ.
The DBSP spectra are based on observations obtained at the Hale Telescope, Palomar Observatory as part of a continuing collaboration between California Institute of Technology, NASA/JPL, Yale University, and the National Astronomical Observatories of China. This work also contains archival data obtained with the {\it Spitzer} Space Telescope, which was operated by the Jet Propulsion Laboratory, California Institute of Technology under a contract with NASA. The National Radio Astronomy Observatory is a facility of the National Science Foundation operated under cooperative agreement by Associated Universities, Inc.. This research has made use of the NASA/IPAC Infrared Science Archive, which is funded by the National Aeronautics and Space Administration and operated by the California Institute of Technology.  PA would like to acknowledge the NRAO visitor support during a visit to NRAO-Charlottesville early in this project. PA would also like to thank the Institut d'Astrophysique de Paris for support as a Visiting Scientist in October 2019 in connection with this work.  UL acknowledges support from project PID2020-114414GB-100, financed by MCIN/AEI/10.13039/501100011033, from projects P20\_00334  and FQM108, financed by the Junta de Andalucia and from FEDER/Junta de Andaluc\'ia-Consejer\'ia de Transformaci\'on Econ\'omica, Industria, Conocimiento y Universidades/Proyecto A-FQM-510-UGR20.
The authors wish to thank an anonymous referee for many important suggestions for improving this paper. 
\clearpage
\appendix
\restartappendixnumbering
\section{Extracted Regions}
In this Appendix we provide sample Tables (in machine readable format) of the regions extracted in the bridge and the measured and derived properties. The full contents of the tables will be made available online. We provide here only a sample of the full table. For the refereed version we include the full tables. 
Column 1 provides the Region which corresponds to the regions marked in Figures \ref{fig:A2} and \ref{fig:A3}, and were chosen to be representative regions in the bridge structures. Column 2 and 3 are the RA (J2000) and Dec (J2000) of the regions. Column 4 and 5 are the best fitting systemic heliocentric velocity (optical definition) and full-width half-maximum (FWHM), with uncertainties, of the CO profiles after fitting with a Gaussian in \kms. Column 6 is the integrated flux of the CO (2-1) profile in Jy \kms and uncertainties. Column 7 is a spectral quality flag based. Flags of unity are the best quality spectra, and the table notes explain the other flags. Only Quality 1 flagged data are used in the figures presented in the paper.  Column 8 provides the molecular mass surface density, in units of log$_{10}$($M_{\odot}$~pc$^{-2}$), for the nominal case of X$_{\rm CO}$ = 1/4 X$_{\rm CO,20}$. In the paper we consider potentially realistic cases of X$_{\rm CO}$ in the range 1/4 $<$ X$_{\rm CO,20}$ $ <$ 1/2, where X${\rm CO,20}$ is the standard Galactic value. Column 9 tabulates the surface density of Pa$\alpha$ emission in units of log$_{10}$(erg s$^{-1}$ cm$^{-2}$ kpc$^{-2}$) for the Case 1 extinction example given in \S 5.1. Column 10 gives the star formation surface density in units of log$_{10}$($M_{\odot}$~yr$^{-1}$~kpc$^{-2}$) under the same assumptions. Column 11 provides a Pa$\alpha$ detection flag (see table note).   

\begin{figure*}
\includegraphics[width=0.9\textwidth]{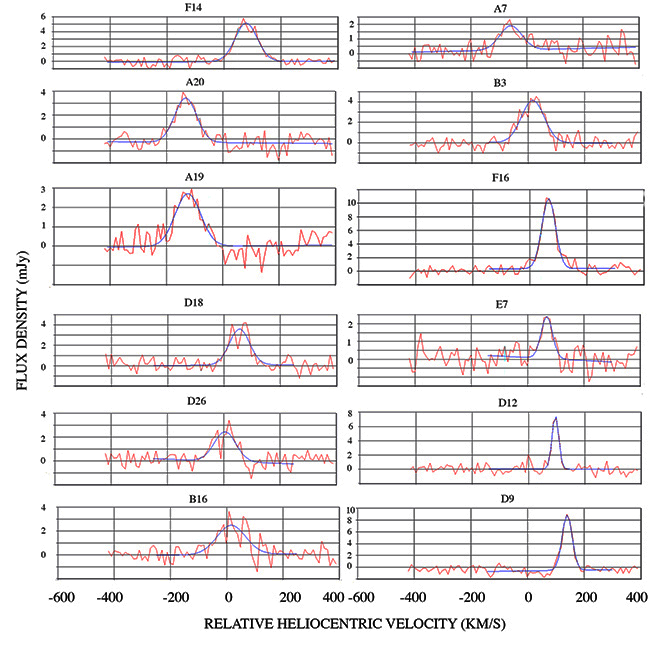}
\caption{Examples of spectra described in Table~\ref{tab:A1} and in Figure~\ref{fig:A2} and \ref{fig:A3}. The designation refers to the entry in the Table and the blue lines show Gaussian fits to the spectra. Narrow lines are seen in the region associated with the X-HII region (e. g. D9 and D12).} 
\label{fig:A1}
\end{figure*}

\begin{longrotatetable}
\begin{deluxetable*}{clllllllcll}
\tabletypesize{}
\tablecolumns{11}
\tablewidth{0pt}
\tablecaption{Properties of the Extracted Regions} \label{tab:A1}
\tablehead{
\colhead{Reg.} & 
\colhead{RA} & 
\colhead{Dec} & 
\colhead{$V_{sys}$\tablenotemark{b}} &       
\colhead{FWHM\tablenotemark{c}}  &        
\colhead{$\Sigma S_V\Delta V$} &          
\colhead{Spec\tablenotemark{d}} &
\colhead{Log($\Sigma_{mol}$)\tablenotemark{e}} &
\colhead{$\Sigma$(Pa$\alpha$)\tablenotemark{f}} &
\colhead{Log($\Sigma_{SFR})$\tablenotemark{f}} &
\colhead{Pa$\alpha$\tablenotemark{g}}
\\
\colhead{$\#$} & 
\colhead{(J2000)} & 
\colhead{(J2000)} &
\colhead{CO~(2-1)} &       
\colhead{CO~(2-1)} &        
\colhead{CO~(2-1)} &          
\colhead{Qual.} &
\colhead{1/4 X$_{\rm CO,20}$} &
\colhead{($10^{-16}\times erg~s^{-1}$} &
\colhead{($M_{\odot} yr^{-1}$} &
\colhead{Det.}
\\
\colhead{} & 
\colhead{({\it hh:mm:ss})} & 
\colhead{({\it dd:mm:ss})} & 
\colhead{($km~s^{-1}$)} &       
\colhead{($km~s^{-1}$)} &        
\colhead{($Jy$~\kms)} &          
\colhead{} &
\colhead{($M_{\odot}~pc^{-2}$)} &
\colhead{$cm^{-2}~{arcsec}^{-2}$)} &
\colhead{$kpc^{-2}$)} &
\colhead{Flag}
\\
\colhead{[1]} & 
\colhead{[2]} &    
\colhead{[3]} & 
\colhead{[4]} & 
\colhead{[5]} &       
\colhead{[6]} &        
\colhead{[7]} &          
\colhead{[8]} &
\colhead{([9]} &
\colhead{[10]} &
\colhead{[11]} 
}
\startdata
A1a & 00:01:41.479 & +23:29:37.1 &   4380 &  96$\pm$14 &   0.29$\pm$0.06 &    1.0 &       2.46$\pm$0.08\tablenotemark{a} &    0.43\tablenotemark{a} &      -1.75\tablenotemark{a} & 0  \\
A1b & 00:01:41.479 & +23:29:37.1 &   4510 &  28$\pm$4 &   0.08$\pm$0.02 &    1.0 &        &     &       & 0  \\
A2a & 00:01:41.466 & +23:29:37.0 &   4376 &  32$\pm$5 &   0.09$\pm$0.02 &    1.0 &       2.27$\pm$0.08\tablenotemark{a} &    0.43\tablenotemark{a} &      -1.75\tablenotemark{a} & 0  \\
A2b & 00:01:41.466 & +23:29:37.0 &   4519 &  28$\pm$3 &   0.14$\pm$0.02 &    1.0 &        &     &       & 0  \\
A3a & 00:01:41.463 & +23:29:36.7 &   4375 &  36$\pm$7 &   0.09$\pm$0.02 &    1.0 &       2.23$\pm$0.11\tablenotemark{a} &    0.43\tablenotemark{a} &      -1.75\tablenotemark{a} & 0  \\
A3b & 00:01:41.463 & +23:29:36.7 &   4522 &  30$\pm$3 &   0.13$\pm$0.01 &    1.0 &        &     &       & 0  \\
A4a & 00:01:41.448 & +23:29:36.4 &   4376 &  41$\pm$6 &   0.11$\pm$0.02 &    1.0 &       2.33$\pm$0.08\tablenotemark{a} &    0.43\tablenotemark{a} &      -1.75\tablenotemark{a} & 0  \\
A4b & 00:01:41.448 & +23:29:36.4 &   4508 &  47$\pm$3 &   0.16$\pm$0.02 &    1.0 &        &     &       & 0  \\
A5 & 00:01:41.451 & +23:29:36.1 & 4481 &  57.8$\pm$5 &   0.20$\pm$0.02 &    1.0 &       2.20$\pm$0.05 &        0.43  &      -1.75 & 0\\
A6 & 00:01:41.423 & +23:29:36.5 &   4376 &  67$\pm$5 &   0.23$\pm$0.03 &    1.0 &       2.26$\pm$0.05 &        0.43  &      -1.75 & 0\\
A7 & 00:01:41.424 & +23:29:36.3 &   4390 &  86$\pm$17 &   0.17$\pm$0.05 &    1.0 &       2.14$\pm$0.11 &       0.43  &      -1.75 & 0\\
A8 & 00:01:41.429 & +23:29:35.7 & 4496 &  95.8$\pm$24 &   0.23$\pm$0.08 &    1.0 &       2.25$\pm$0.13 &       0.43  &      -1.75 & 0\\
A9 & 00:01:41.445 & +23:29:35.6 & 4496 &  63.8$\pm$5 &   0.22$\pm$0.02 &    1.0 &       2.24$\pm$0.05 &        0.43  &      -1.75 & 0\\
A10 & 00:01:41.435 & +23:29:35.3 & 4484 &  70.5$\pm$6 &   0.24$\pm$0.03 &    1.0 &       2.27$\pm$0.05 &       0.43  &      -1.75 & 0\\
A11 & 00:01:41.393 & +23:29:35.2 &   4446 &  52$\pm$4 &   0.27$\pm$0.03 &    1.0 &       2.34$\pm$0.05 &       0.43  &      -1.75 & 0\\
A12 & 00:01:41.377 & +23:29:35.2 &   4450 &  62$\pm$5 &   0.28$\pm$0.03 &    1.0 &       2.34$\pm$0.05 &       0.43  &      -1.75 & 0\\
A13 & 00:01:41.396 & +23:29:35.0 &   4441 &  52$\pm$4 &   0.27$\pm$0.03 &    1.0 &       2.32$\pm$0.05 &       0.43  &      -1.75 & 0\\
A14 & 00:01:41.380 & +23:29:35.0 &   4443 &  57$\pm$5 &   0.26$\pm$0.03 &    1.0 &       2.32$\pm$0.05 &       0.43  &      -1.75 & 0\\
A15 & 00:01:41.365 & +23:29:34.2 &   4409 &  45$\pm$4 &   0.16$\pm$0.02 &    1.0 &       2.10$\pm$0.05 &       0.43  &      -1.75 & 0\\
A16 & 00:01:41.446 & +23:29:33.9 &   4459 &  64$\pm$5 &   0.25$\pm$0.03 &    1.0 &       2.30$\pm$0.05 &       0.43  &      -1.75 & 0\\
A17 & 00:01:41.463 & +23:29:34.1 &   4492 &  68$\pm$5 &   0.26$\pm$0.03 &    1.0 &       2.32$\pm$0.05 &       0.43  &      -1.75 & 0\\
A18 & 00:01:41.342 & +23:29:33.7 &   4373 &  54$\pm$4 &   0.25$\pm$0.03 &    1.0 &       2.29$\pm$0.05 &       0.43  &      -1.75 & 0\\ 
A19 & 00:01:41.347 & +23:29:33.3 &   4330 & 104$\pm$16 &   0.30$\pm$0.06 &    1.0 &       2.38$\pm$0.08 &      0.43  &      -1.75 & 0\\
A20 & 00:01:41.337 & +23:29:33.1 &   4320 &  97$\pm$8 &   0.39$\pm$0.04 &    1.0 &       2.49$\pm$0.05 &       0.43  &      -1.75 & 0\\ 
A21 & 00:01:41.370 & +23:29:32.8 & 4462 &  48.3$\pm$12 &   0.10$\pm$0.04 &    0.5 &       1.91$\pm$0.13 &      0.43  &        -1.75 & 0\\
A22 & 00:01:41.378 & +23:29:32.4 & 4505 &  34.2$\pm$3 &   0.16$\pm$0.02 &    1.0 &       2.09$\pm$0.05 &       0.43  &        -1.75 & 0\\
A23 & 00:01:41.188 & +23:29:32.7 & 4505 &  30.0$\pm$2 &   0.10$\pm$0.01 &    1.0 &       1.89$\pm$0.05 &       0.43  &        -1.75 & 0\\
A24 & 00:01:41.200 & +23:29:32.5 & 4520 &  33.5$\pm$3 &   0.14$\pm$0.02 &    1.0 &       2.05$\pm$0.05 &       0.43  &        -1.75 & 0\\
A25 & 00:01:41.281 & +23:29:32.0 &   4342 &  71$\pm$14 &   0.16$\pm$0.05 &    0.7 &       2.11$\pm$0.11 &      0.43  &      -1.75 & 0\\
A26 & 00:01:41.264 & +23:29:31.8 &   4348 & 146$\pm$29 &   0.30$\pm$0.08 &    0.7 &       2.37$\pm$0.11 &      0.43  &      -1.75 & 0\\
A27 & 00:01:41.340 & +23:29:31.2 &   4437 &  73$\pm$15 &   0.07$\pm$0.02 &    0.4 &       1.73$\pm$0.11 &      0.43  &      -1.75 & 0\\
A28 & 00:01:41.219 & +23:29:29.5 &   4469 &  26$\pm$4 &   0.08$\pm$0.02 &    0.5 &       1.78$\pm$0.08 &       0.43  &      -1.75 & 0\\ 
A29 & 00:01:41.155 & +23:29:28.5 &   4475 & 144$\pm$29 &   0.22$\pm$0.06 &    0.4 &       2.24$\pm$0.11 &      0.43  &      -1.75 & 0\\
A30 & 00:01:41.070 & +23:29:27.2 &   4300 &  68$\pm$14 &   0.14$\pm$0.04 &    0.4 &       2.04$\pm$0.11 &      0.43  &      -1.75 & 0\\
B1 & 00:01:41.476 & +23:29:38.9 & 4500 &  76.9$\pm$6 &   0.53$\pm$0.06 &    1.0 &       2.62$\pm$0.05 &        0.43  &       -1.75 & 0\\ 
B2 & 00:01:41.466 & +23:29:39.1 & 4477 &  78.7$\pm$6 &   0.47$\pm$0.05 &    1.0 &       2.57$\pm$0.05 &        0.43  &       -1.75 & 0\\ 
B3 & 00:01:41.450 & +23:29:39.3 & 4466 &  94.4$\pm$8 &   0.41$\pm$0.05 &    1.0 &       2.51$\pm$0.05 &        0.43  &       -1.75 & 0\\ 
B4 & 00:01:41.455 & +23:29:38.9 & 4481 &  73.0$\pm$6 &   0.38$\pm$0.04 &    1.0 &       2.47$\pm$0.05 &        0.43  &       -1.75 & 0\\ 
B5 & 00:01:41.421 & +23:29:38.6 &   4380 & 115$\pm$17 &   0.31$\pm$0.06 &    1.0 &       2.39$\pm$0.08 &       0.43  &      -1.75 & 0\\
B6 & 00:01:41.405 & +23:29:38.8 &   4369 &  72$\pm$6 &   0.25$\pm$0.03 &    1.0 &       2.29$\pm$0.05 &        0.43  &      -1.75 & 0\\ 
B7 & 00:01:41.387 & +23:29:39.5 &   4454 &  78$\pm$6 &   0.49$\pm$0.06 &    1.0 &       2.59$\pm$0.05 &        0.43  &      -1.75 & 0\\ 
B8 & 00:01:41.373 & +23:29:39.3 &   4449 &  83$\pm$7 &   0.68$\pm$0.08 &    1.0 &       2.73$\pm$0.05 &        0.43  &      -1.75 & 0\\ 
B9 & 00:01:41.369 & +23:29:39.6 &   4441 &  79$\pm$6 &   0.51$\pm$0.06 &    1.0 &       2.60$\pm$0.05 &        0.43  &      -1.75 & 0\\ 
B10 & 00:01:41.361 & +23:29:39.9 &   4427 &  61$\pm$5 &   0.26$\pm$0.03 &    1.0 &       2.31$\pm$0.05 &       0.43  &      -1.75 & 0\\ 
B11 & 00:01:41.310 & +23:29:39.6 &   4428 &  34$\pm$3 &   0.13$\pm$0.01 &    1.0 &       2.01$\pm$0.05 &       0.43  &      -1.75 & 0\\ 
B12 & 00:01:41.294 & +23:29:39.8 &   4436 &  33$\pm$3 &   0.14$\pm$0.02 &    1.0 &       2.06$\pm$0.05 &       0.43  &      -1.75 & 0\\ 
B13 & 00:01:41.275 & +23:29:39.7 &   4435 &  31$\pm$2 &   0.14$\pm$0.02 &    1.0 &       2.04$\pm$0.05 &       0.43  &      -1.75 & 0\\ 
B14 & 00:01:41.304 & +23:29:38.8 &   4354 &  56$\pm$4 &   0.30$\pm$0.03 &    1.0 &       2.38$\pm$0.05 &       0.43  &      -1.75 & 0\\ 
B15 & 00:01:41.283 & +23:29:38.7 &   4301 &  31$\pm$2 &   0.14$\pm$0.02 &    1.0 &       2.04$\pm$0.05 &       0.43  &      -1.75 & 0\\ 
B16 & 00:01:41.296 & +23:29:37.7 &   4468 & 116$\pm$23 &   0.30$\pm$0.08 &   0.5 &       2.37$\pm$0.11 &       0.43  &      -1.75 & 0\\
B17 & 00:01:41.279 & +23:29:37.5 &   4462 &  80$\pm$6 &   0.38$\pm$0.04 &    1.0 &       2.48$\pm$0.05 &       0.43  &      -1.75 & 0\\ 
B18 & 00:01:41.277 & +23:29:37.8 &   4454 &  65$\pm$5 &   0.39$\pm$0.04 &    1.0 &       2.49$\pm$0.05 &       3.07$\pm$0.36 &      -0.89 & 1 \\
B19 & 00:01:41.260 & +23:29:37.6 &   4433 &  76$\pm$6 &   0.45$\pm$0.05 &    1.0 &       2.55$\pm$0.05 &       3.58$\pm$0.38 &      -0.83 & 1 \\
B20 & 00:01:41.257 & +23:29:38.0 &   4454 &  62$\pm$5 &   0.35$\pm$0.04 &    1.0 &       2.44$\pm$0.05 &       3.43$\pm$0.37 &      -0.85 & 1 \\
B21 & 00:01:41.244 & +23:29:38.2 &   4472 &  49$\pm$4 &   0.33$\pm$0.04 &    1.0 &       2.42$\pm$0.05 &       0.43  &      -1.75 & 0\\ 
B22 & 00:01:41.237 & +23:29:38.5 &   4469 &  47$\pm$4 &   0.33$\pm$0.04 &    1.0 &       2.42$\pm$0.05 &       0.43  &      -1.75 & 0\\ 
B23 & 00:01:41.236 & +23:29:38.8 &   4454 &  46$\pm$4 &   0.22$\pm$0.03 &    1.0 &       2.25$\pm$0.05 &       0.43  &      -1.75 & 0\\ 
B24 & 00:01:41.185 & +23:29:38.4 &   4456 &  51$\pm$4 &   0.29$\pm$0.03 &    1.0 &       2.36$\pm$0.05 &       0.43  &      -1.75 & 0\\ 
B25 & 00:01:41.193 & +23:29:38.7 &   4464 &  67$\pm$5 &   0.24$\pm$0.03 &    1.0 &       2.28$\pm$0.05 &       0.43  &      -1.75 & 0\\ 
B26 & 00:01:41.174 & +23:29:38.7 &   4466 &  61$\pm$5 &   0.41$\pm$0.05 &    1.0 &       2.51$\pm$0.05 &       0.43  &      -1.75 & 0\\ 
B27 & 00:01:41.192 & +23:29:39.0 &   4476 &  78$\pm$6 &   0.30$\pm$0.03 &    1.0 &       2.38$\pm$0.05 &       0.43  &      -1.75 & 0\\ 
B28 & 00:01:41.172 & +23:29:39.0 &   4490 &  65$\pm$5 &   0.36$\pm$0.04 &    1.0 &       2.45$\pm$0.05 &       0.43  &      -1.75 & 0\\ 
B29 & 00:01:41.158 & +23:29:38.8 &   4496 &  92$\pm$7 &   0.52$\pm$0.06 &    1.0 &       2.61$\pm$0.05 &       0.43  &      -1.75 & 0\\ 
B30 & 00:01:41.141 & +23:29:38.6 &   4507 &  78$\pm$6 &   0.41$\pm$0.05 &    1.0 &       2.51$\pm$0.05 &       0.43  &      -1.75 & 0\\ 
B31 & 00:01:41.201 & +23:29:39.3 &   4497 &  68$\pm$5 &   0.36$\pm$0.04 &    1.0 &       2.45$\pm$0.05 &       6.09$\pm$0.47 &      -0.60 & 1 \\
B32 & 00:01:41.189 & +23:29:39.5 &   4510 &  80$\pm$12 &   0.25$\pm$0.05 &    1.0 &      2.30$\pm$0.08 &       5.63$\pm$0.46 &      -0.63 & 1 \\
B33 & 00:01:41.177 & +23:29:39.3 &   4499 &  56$\pm$4 &   0.25$\pm$0.03 &    1.0 &       2.30$\pm$0.05 &       4.45$\pm$0.41 &      -0.73 & 1 \\
B34 & 00:01:41.154 & +23:29:39.1 &   4500 &  73$\pm$6 &   0.36$\pm$0.04 &    1.0 &       2.46$\pm$0.05 &       0.43  &      -1.75 & 0\\ 
B35 & 00:01:41.139 & +23:29:39.0 &   4480 &  60$\pm$9 &   0.19$\pm$0.04 &    0.5 &       2.18$\pm$0.08 &       0.43  &      -1.75 & 0\\ 
B36 & 00:01:41.117 & +23:29:38.8 &   4482 &  47$\pm$4 &   0.25$\pm$0.03 &    1.0 &       2.30$\pm$0.05 &       0.43  &      -1.75 & 0\\ 
C1 & 00:01:41.277 & +23:29:41.1 &   4514 &  42$\pm$6 &   0.13$\pm$0.03 &    1.0 &       2.01$\pm$0.08 &        0.43  &      -1.75 & 0\\ 
C2 & 00:01:41.276 & +23:29:41.5 &   4521 &  35$\pm$3 &   0.15$\pm$0.02 &    1.0 &       2.06$\pm$0.05 &        0.43  &      -1.75 & 0\\ 
C3 & 00:01:41.274 & +23:29:41.7 &   4526 &  32$\pm$3 &   0.16$\pm$0.02 &    1.0 &       2.11$\pm$0.05 &        0.43  &      -1.75 & 0\\ 
C4 & 00:01:41.275 & +23:29:42.0 &   4528 &  25$\pm$2 &   0.08$\pm$0.01 &    1.0 &       1.82$\pm$0.05 &        0.43  &      -1.75 & 0\\ 
C5 & 00:01:41.234 & +23:29:42.5 &   4568 &  61$\pm$5 &   0.20$\pm$0.02 &    1.0 &       2.21$\pm$0.05 &        0.43  &      -1.75 & 0\\ 
C6 & 00:01:41.228 & +23:29:42.8 &   4586 &  59$\pm$5 &   0.23$\pm$0.03 &    1.0 &       2.26$\pm$0.05 &        0.43  &      -1.75 & 0\\ 
C7 & 00:01:41.224 & +23:29:43.0 &   4591 &  42$\pm$3 &   0.21$\pm$0.02 &    1.0 &       2.22$\pm$0.05 &        0.43  &      -1.75 & 0\\ 
C8 & 00:01:41.207 & +23:29:42.9 &   4584 &  57$\pm$5 &   0.30$\pm$0.03 &    1.0 &       2.38$\pm$0.05 &        0.43  &      -1.75 & 0\\ 
C9 & 00:01:41.207 & +23:29:43.2 &   4576 &  52$\pm$4 &   0.23$\pm$0.03 &    1.0 &       2.26$\pm$0.05 &        0.43  &      -1.75 & 0\\ 
C10 & 00:01:41.188 & +23:29:43.0 &   4572 &  44$\pm$4 &   0.21$\pm$0.02 &    1.0 &       2.23$\pm$0.05 &       0.43  &      -1.75 & 0\\ 
C11 & 00:01:41.208 & +23:29:43.5 &   4583 &  37$\pm$3 &   0.13$\pm$0.01 &    1.0 &       2.00$\pm$0.05 &       0.43  &      -1.75 & 0\\ 
C12 & 00:01:41.187 & +23:29:43.3 &   4573 &  50$\pm$4 &   0.20$\pm$0.02 &    1.0 &       2.20$\pm$0.05 &       0.43  &      -1.75 & 0\\ 
C13 & 00:01:41.183 & +23:29:43.6 &   4572 &  42$\pm$6 &   0.12$\pm$0.02 &    0.5 &       1.98$\pm$0.08 &       0.43  &      -1.75 & 0\\ 
C14 & 00:01:41.136 & +23:29:43.8 &   4578 &  29$\pm$4 &   0.08$\pm$0.02 &    1.0 &       1.81$\pm$0.08 &       0.43  &      -1.75 & 0\\ 
C15 & 00:01:41.121 & +23:29:44.1 &   4574 &  39$\pm$6 &   0.11$\pm$0.02 &    1.0 &       1.92$\pm$0.08 &       0.43  &      -1.75 & 0\\ 
C15 & 00:01:41.121 & +23:29:44.1 &   4574 &  39$\pm$6 &   0.11$\pm$0.02 &    1.0 &       1.92$\pm$0.08 &       0.43  &      -1.75 & 0\\ 
C16 & 00:01:41.087 & +23:29:44.3 &   ---  &  ---      &   ---           &    0.0 &       ---          &        0.43  &      -1.75 & 0\\       
C17 & 00:01:41.046 & +23:29:44.2 &   ---  &  ---      &   ---           &    0.0 &       ---          &        0.43  &      -1.75 & 0\\       
C18 & 00:01:41.043 & +23:29:44.5 &   4587 &  45$\pm$9 &   0.09$\pm$0.03 &    0.5 &       1.87$\pm$0.11 &       0.43  &      -1.75 & 0\\ 
D1 & 00:01:41.010 & +23:29:36.6 &   4584 &  25$\pm$2 &   0.07$\pm$0.01 &    1.0 &       1.76$\pm$0.05 &       11.10$\pm$0.62 &      -0.34 & 1 \\
D2 & 00:01:41.022 & +23:29:36.9 &   4582 &  30$\pm$2 &   0.19$\pm$0.02 &    1.0 &       2.17$\pm$0.05 &       10.94$\pm$0.62 &      -0.34 & 1 \\
D3 & 00:01:40.999 & +23:29:36.9 &   4584 &  25$\pm$2 &   0.07$\pm$0.01 &    1.0 &       1.75$\pm$0.05 &       12.17$\pm$0.65 &      -0.30 & 1 \\
D4 & 00:01:41.016 & +23:29:37.3 &   4569 &  48$\pm$4 &   0.18$\pm$0.02 &    1.0 &       2.15$\pm$0.05 &       14.73$\pm$0.71 &      -0.21 & 1 \\
D5 & 00:01:40.997 & +23:29:37.4 &   4553 &  25$\pm$2 &   0.13$\pm$0.01 &    1.0 &       2.01$\pm$0.05 &       16.37$\pm$0.74 &      -0.17 & 1 \\
D6 & 00:01:40.976 & +23:29:38.0 &   4542 &  73$\pm$6 &   0.36$\pm$0.04 &    1.0 &       2.25$\pm$0.05 &       11.76$\pm$0.64 &      -0.31 & 1 \\
D7 & 00:01:40.958 & +23:29:36.3 &   4585 &  38$\pm$3 &   0.26$\pm$0.03 &    1.0 &       2.31$\pm$0.05 &        7.72$\pm$0.53 &      -0.49 & 1 \\
D8 & 00:01:40.958 & +23:29:36.7 &   4588 &  39$\pm$3 &   0.26$\pm$0.03 &    1.0 &       2.31$\pm$0.05 &       15.96$\pm$0.73 &      -0.18 & 1 \\
D9 & 00:01:40.942 & +23:29:36.6 &   4590 &  43$\pm$3 &   0.43$\pm$0.05 &    1.0 &       2.53$\pm$0.05 &       21.99$\pm$0.86 &      -0.04 & 1 \\
D10 & 00:01:40.941 & +23:29:37.0 &   4586 &  51$\pm$4 &   0.37$\pm$0.04 &    1.0 &       2.47$\pm$0.05 &      42.96$\pm$1.18 &       0.25 & 1 \\
D11 & 00:01:40.959 & +23:29:37.3 &   4548 &  25$\pm$2 &   0.18$\pm$0.02 &    1.0 &       2.15$\pm$0.05 &      30.69$\pm$1.01 &       0.11 & 1 \\
D12 & 00:01:40.951 & +23:29:37.5 &   4546 &  27$\pm$2 &   0.22$\pm$0.02 &    1.0 &       2.24$\pm$0.05 &      26.08$\pm$0.93 &       0.03 & 1 \\
D13 & 00:01:40.935 & +23:29:37.5 &   4539 &  36$\pm$3 &   0.28$\pm$0.03 &    1.0 &       2.35$\pm$0.05 &      29.66$\pm$0.99 &       0.09 & 1 \\
D14 & 00:01:40.919 & +23:29:38.1 &   4546 &  38$\pm$3 &   0.22$\pm$0.02 &    1.0 &       2.24$\pm$0.05 &      36.31$\pm$1.09 &       0.18 & 1 \\
D15 & 00:01:40.902 & +23:29:38.0 &   4544 &  35$\pm$3 &   0.27$\pm$0.03 &    1.0 &       2.33$\pm$0.05 &      23.53$\pm$0.88 &      -0.01 & 1 \\
D16 & 00:01:40.888 & +23:29:38.1 &   4541 &  47$\pm$4 &   0.29$\pm$0.03 &    1.0 &       2.37$\pm$0.05 &      14.83$\pm$0.71 &      -0.21 & 1 \\
D17 & 00:01:40.856 & +23:29:37.6 &   4556 &  36$\pm$3 &   0.14$\pm$0.02 &    1.0 &       2.05$\pm$0.05 &       6.14$\pm$0.48 &      -0.59 & 1 \\
D18 & 00:01:40.818 & +23:29:37.7 &   4507 &  90$\pm$7 &   0.36$\pm$0.04 &    1.0 &       2.45$\pm$0.05 &       2.15$\pm$0.31 &      -1.05 & 1 \\
D19 & 00:01:40.797 & +23:29:37.5 &   4495 &  75$\pm$15 &   0.19$\pm$0.05 &    1.0 &      2.18$\pm$0.11 &       1.53$\pm$0.28 &      -1.20 & 1 \\
D20 & 00:01:40.850 & +23:29:38.0 &   4508 &  71$\pm$6 &   0.25$\pm$0.03 &    1.0 &       2.30$\pm$0.05 &       3.84$\pm$0.39 &      -0.80 & 1 \\
D21 & 00:01:40.830 & +23:29:38.0 &   4504 &  62$\pm$5 &   0.32$\pm$0.04 &    1.0 &       2.41$\pm$0.05 &       3.53$\pm$0.38 &      -0.83 & 1 \\
D22 & 00:01:40.812 & +23:29:38.0 &   4515 &  70$\pm$6 &   0.31$\pm$0.03 &    1.0 &       2.38$\pm$0.05 &       2.86$\pm$0.35 &      -0.92 & 1 \\
D23 & 00:01:40.845 & +23:29:38.3 &   4498 &  59$\pm$5 &   0.23$\pm$0.03 &    1.0 &       2.27$\pm$0.05 &       3.27$\pm$0.37 &      -0.87 & 1 \\
D24 & 00:01:40.824 & +23:29:38.3 &   4494 &  50$\pm$4 &   0.22$\pm$0.02 &    1.0 &       2.24$\pm$0.05 &       2.66$\pm$0.34 &      -0.96 & 1 \\
D25 & 00:01:40.836 & +23:29:38.8 &   4484 &  46$\pm$4 &   0.19$\pm$0.02 &    1.0 &       2.17$\pm$0.05 &       2.56$\pm$0.33 &      -0.97 & 1 \\
D26 & 00:01:40.818 & +23:29:38.6 &   4456 &  91$\pm$18 &   0.24$\pm$0.07 &    1.0 &       2.28$\pm$0.11 &      2.30$\pm$0.32 &      -1.02 & 1 \\
D27 & 00:01:40.799 & +23:29:38.5 &   4472 &  59$\pm$5 &   0.23$\pm$0.03 &    1.0 &       2.26$\pm$0.05 &       2.40$\pm$0.33 &      -1.00 & 1 \\
D28 & 00:01:40.793 & +23:29:37.0 &   4446 &  25$\pm$2 &   0.09$\pm$0.01 &    1.0 &       1.84$\pm$0.05 &       0.43  &      -1.75 & 0\\ 
D29 & 00:01:40.771 & +23:29:37.3 &   4489 &  46$\pm$4 &   0.21$\pm$0.02 &    1.0 &       2.22$\pm$0.05 &       0.43  &      -1.75 & 0\\ 
D30 & 00:01:40.759 & +23:29:37.2 &   4499 &  67$\pm$5 &   0.26$\pm$0.03 &    1.0 &       2.31$\pm$0.05 &       0.43  &      -1.75 & 0\\ 
D31 & 00:01:40.741 & +23:29:37.2 &   4480 &  69$\pm$6 &   0.26$\pm$0.03 &    1.0 &       2.32$\pm$0.05 &       0.43  &      -1.75 & 0\\ 
D32 & 00:01:40.723 & +23:29:37.3 &   4467 &  56$\pm$4 &   0.25$\pm$0.03 &    1.0 &       2.30$\pm$0.05 &       0.43  &      -1.75 & 0\\ 
D33 & 00:01:40.723 & +23:29:36.9 &   4467 &  63$\pm$5 &   0.31$\pm$0.03 &    1.0 &       2.39$\pm$0.05 &       0.43  &      -1.75 & 0\\ 
D34 & 00:01:40.706 & +23:29:36.6 &   4465 &  72$\pm$11 &   0.23$\pm$0.05 &    1.0 &       2.26$\pm$0.08 &      0.43  &      -1.75 & 0\\
D35 & 00:01:40.746 & +23:29:37.7 &   4476 &  78$\pm$6 &   0.30$\pm$0.03 &    1.0 &       2.37$\pm$0.05 &       0.43  &      -1.75 & 0\\ 
D36 & 00:01:40.728 & +23:29:37.7 &   4455 &  70$\pm$6 &   0.35$\pm$0.04 &    1.0 &       2.44$\pm$0.05 &       0.43  &      -1.75 & 0\\ 
D37 & 00:01:40.732 & +23:29:37.9 &   4490 & 104$\pm$8 &   0.42$\pm$0.05 &    1.0 &       2.52$\pm$0.05 &       0.43  &      -1.75 & 0\\ 
D38 & 00:01:40.715 & +23:29:37.8 &   4467 &  78$\pm$6 &   0.29$\pm$0.03 &    1.0 &       2.37$\pm$0.05 &       0.43  &      -1.75 & 0\\ 
D39 & 00:01:40.730 & +23:29:38.3 &   4512 &  69$\pm$6 &   0.28$\pm$0.03 &    1.0 &       2.34$\pm$0.05 &       7.31$\pm$0.51 &      -0.52 & 1 \\
D40 & 00:01:40.659 & +23:29:36.4 &   4429 &  53$\pm$8 &   0.17$\pm$0.03 &    1.0 &       2.13$\pm$0.08 &       0.43  &      -1.75 & 0\\ 
D41 & 00:01:40.674 & +23:29:35.6 &   4347 &  52$\pm$8 &   0.16$\pm$0.03 &    1.0 &       2.10$\pm$0.08 &       0.43  &      -1.75 & 0\\ 
E1 & 00:01:40.875 & +23:29:37.1 &   4509 &  27$\pm$4 &   0.09$\pm$0.02 &    1.0 &       1.83$\pm$0.08 &       14.86  &      -0.21 & 1 \\ 
E2 & 00:01:40.870 & +23:29:36.6 &   4463 &  57$\pm$5 &   0.21$\pm$0.02 &    1.0 &       2.21$\pm$0.05 &        0.43  &      -1.75 & 0\\ 
E3 & 00:01:40.851 & +23:29:36.5 &   4465 &  45$\pm$4 &   0.19$\pm$0.02 &    1.0 &       2.18$\pm$0.05 &        0.43  &      -1.75 & 0\\ 
E4 & 00:01:40.852 & +23:29:36.1 &   4445 &  57$\pm$5 &   0.21$\pm$0.02 &    1.0 &       2.23$\pm$0.05 &        0.43  &      -1.75 & 0\\ 
E5 & 00:01:40.859 & +23:29:35.8 &   4402 &  75$\pm$11 &   0.23$\pm$0.05 &    1.0 &       2.26$\pm$0.08 &       0.43  &      -1.75 & 0\\
E6 & 00:01:40.862 & +23:29:35.5 &   4383 &  57$\pm$5 &   0.27$\pm$0.03 &    1.0 &       2.33$\pm$0.05 &        0.43  &      -1.75 & 0\\ 
E7 & 00:01:40.820 & +23:29:36.6 &   4518 &  45$\pm$9 &   0.11$\pm$0.03 &    1.0 &       1.92$\pm$0.11 &        0.43  &      -1.75 & 0\\ 
E8 & 00:01:40.795 & +23:29:35.9 &   4426 &  48$\pm$7 &   0.15$\pm$0.03 &    1.0 &       2.08$\pm$0.08 &        0.43  &      -1.75 & 0\\ 
E9 & 00:01:40.780 & +23:29:35.7 &   4431 &  68$\pm$5 &   0.29$\pm$0.03 &    1.0 &       2.36$\pm$0.05 &        0.43  &      -1.75 & 0\\ 
E10 & 00:01:40.768 & +23:29:35.4 &   4411 &  83$\pm$17 &   0.22$\pm$0.06 &    0.5 &       2.24$\pm$0.11 &      0.43  &      -1.75 & 0\\
E11 & 00:01:40.904 & +23:29:34.9 &   4524 &  48$\pm$4 &   0.16$\pm$0.02 &    1.0 &       2.10$\pm$0.05 &       0.43  &      -1.75 & 0\\ 
E12 & 00:01:40.895 & +23:29:34.5 &   4521 &  66$\pm$5 &   0.28$\pm$0.03 &    1.0 &       2.34$\pm$0.05 &       0.43  &      -1.75 & 0\\ 
E13 & 00:01:40.883 & +23:29:34.3 &   4512 &  76$\pm$6 &   0.25$\pm$0.03 &    1.0 &       2.30$\pm$0.05 &       0.43  &      -1.75 & 0\\ 
E14 & 00:01:40.865 & +23:29:34.5 &   4534 &  50$\pm$4 &   0.17$\pm$0.02 &    1.0 &       2.12$\pm$0.05 &       0.43  &      -1.75 & 0\\ 
E15 & 00:01:40.902 & +23:29:34.2 &   4500 &  71$\pm$6 &   0.27$\pm$0.03 &    1.0 &       2.34$\pm$0.05 &       0.43  &      -1.75 & 0\\ 
E16 & 00:01:40.893 & +23:29:33.9 &   4501 &  57$\pm$9 &   0.17$\pm$0.03 &    1.0 &       2.13$\pm$0.08 &       0.43  &      -1.75 & 0\\ 
E17 & 00:01:40.800 & +23:29:34.4 &   4559 &  29$\pm$2 &   0.10$\pm$0.01 &    0.5 &       1.92$\pm$0.05 &       0.43  &      -1.75 & 0\\ 
E18 & 00:01:40.836 & +23:29:33.9 &   4507 &  29$\pm$6 &   0.07$\pm$0.02 &    0.5 &       1.77$\pm$0.11 &       0.43  &      -1.75 & 0\\ 
E19 & 00:01:40.923 & +23:29:33.6 &   4457 &  42$\pm$8 &   0.11$\pm$0.03 &    1.0 &       1.93$\pm$0.11 &       0.43  &      -1.75 & 0\\ 
E20 & 00:01:40.940 & +23:29:33.4 &   4437 &  71$\pm$14 &   0.15$\pm$0.04 &    0.5 &       2.08$\pm$0.11 &      0.43  &      -1.75 & 0\\
E21 & 00:01:40.918 & +23:29:33.3 &   4453 &  50$\pm$10 &   0.12$\pm$0.03 &    1.0 &       1.97$\pm$0.11 &      0.43  &      -1.75 & 0\\
F1 & 00:01:41.221 & +23:29:36.7 &   4535 &  53$\pm$4 &   0.24$\pm$0.03 &    1.0 &       2.29$\pm$0.05 &        0.43  &      -1.75 & 0\\ 
F2 & 00:01:41.210 & +23:29:36.5 &   4538 &  49$\pm$4 &   0.29$\pm$0.03 &    1.0 &       2.36$\pm$0.05 &        0.43  &      -1.75 & 0\\ 
F3 & 00:01:41.195 & +23:29:36.4 &   4526 &  48$\pm$4 &   0.28$\pm$0.03 &    1.0 &       2.34$\pm$0.05 &        0.43  &      -1.75 & 0\\ 
F4 & 00:01:41.177 & +23:29:36.4 &   4523 &  53$\pm$4 &   0.32$\pm$0.04 &    1.0 &       2.41$\pm$0.05 &        0.43  &      -1.75 & 0\\ 
F5 & 00:01:41.157 & +23:29:36.4 &   4522 &  38$\pm$3 &   0.23$\pm$0.03 &    1.0 &       2.25$\pm$0.05 &        0.43  &      -1.75 & 0\\ 
F6 & 00:01:41.173 & +23:29:36.1 &   4518 &  56$\pm$5 &   0.33$\pm$0.04 &    1.0 &       2.42$\pm$0.05 &        0.43  &      -1.75 & 0\\ 
F7 & 00:01:41.172 & +23:29:35.8 &   4506 &  47$\pm$4 &   0.17$\pm$0.02 &    1.0 &       2.14$\pm$0.05 &        0.43  &      -1.75 & 0\\ 
F8 & 00:01:41.244 & +23:29:36.0 &   4495 &  48$\pm$4 &   0.17$\pm$0.02 &    1.0 &       2.13$\pm$0.05 &        0.43  &      -1.75 & 0\\ 
F9 & 00:01:41.193 & +23:29:35.6 &   4559 &  79$\pm$16 &   0.21$\pm$0.06 &    1.0 &       2.21$\pm$0.11 &       0.43  &      -1.75 & 0\\
F10 & 00:01:41.164 & +23:29:35.4 &   4576 &  35$\pm$3 &   0.15$\pm$0.02 &    1.0 &       2.06$\pm$0.05 &       0.43  &      -1.75 & 0\\ 
F11 & 00:01:41.161 & +23:29:35.0 &   4576 &  37$\pm$3 &   0.14$\pm$0.02 &    1.0 &       2.04$\pm$0.05 &       0.43  &      -1.75 & 0\\ 
F12 & 00:01:41.127 & +23:29:34.8 &   4572 &  62$\pm$5 &   0.25$\pm$0.03 &    1.0 &       2.30$\pm$0.05 &       0.43  &      -1.75 & 0\\ 
F13 & 00:01:41.094 & +23:29:34.9 &   4535 & 103$\pm$15 &   0.32$\pm$0.06 &    1.0 &       2.40$\pm$0.08 &      0.43  &      -1.75 & 0\\
F14 & 00:01:41.103 & +23:29:34.6 &   4526 & 104$\pm$8 &   0.74$\pm$0.08 &    1.0 &       2.77$\pm$0.05 &       0.43  &      -1.75 & 0\\ 
F15 & 00:01:41.107 & +23:29:34.3 &   4522 &  47$\pm$4 &   0.35$\pm$0.04 &    1.0 &       2.44$\pm$0.05 &       0.43  &      -1.75 & 0\\ 
F16 & 00:01:41.089 & +23:29:34.3 &   4518 &  49$\pm$4 &   0.52$\pm$0.06 &    1.0 &       2.61$\pm$0.05 &       0.43  &      -1.75 & 0\\ 
F17 & 00:01:41.080 & +23:29:34.1 &   4510 &  54$\pm$4 &   0.54$\pm$0.06 &    1.0 &       2.63$\pm$0.05 &       0.43  &      -1.75 & 0\\ 
F18 & 00:01:41.076 & +23:29:33.8 &   4511 &  57$\pm$5 &   0.66$\pm$0.07 &    1.0 &       2.72$\pm$0.05 &       0.43  &      -1.75 & 0\\ 
F19 & 00:01:41.061 & +23:29:33.9 &   4505 &  56$\pm$4 &   0.44$\pm$0.05 &    1.0 &       2.54$\pm$0.05 &       0.43  &      -1.75 & 0\\ 
F20 & 00:01:41.078 & +23:29:33.5 &   4521 &  61$\pm$5 &   0.51$\pm$0.06 &    1.0 &       2.60$\pm$0.05 &       0.43  &      -1.75 & 0\\ 
F21 & 00:01:41.060 & +23:29:33.6 &   4506 &  59$\pm$5 &   0.38$\pm$0.04 &    1.0 &       2.48$\pm$0.05 &       0.43  &      -1.75 & 0\\ 
F22 & 00:01:41.076 & +23:29:33.3 &   4528 &  57$\pm$5 &   0.26$\pm$0.03 &    1.0 &       2.31$\pm$0.05 &       0.43  &      -1.75 & 0\\ 
F23 & 00:01:41.055 & +23:29:33.3 &   4511 &  58$\pm$5 &   0.28$\pm$0.03 &    1.0 &       2.35$\pm$0.05 &       0.43  &      -1.75 & 0\\ 
F24 & 00:01:41.058 & +23:29:33.1 &   4528 &  55$\pm$4 &   0.28$\pm$0.03 &    1.0 &       2.34$\pm$0.05 &       0.43  &      -1.75 & 0\\ 
F25 & 00:01:41.066 & +23:29:32.9 &   4545 &  63$\pm$5 &   0.29$\pm$0.03 &    1.0 &       2.36$\pm$0.05 &       0.43  &      -1.75 & 0\\ 
F26 & 00:01:41.046 & +23:29:32.8 &   4570 &  43$\pm$3 &   0.18$\pm$0.02 &    1.0 &       2.16$\pm$0.05 &       0.43  &      -1.75 & 0\\ 
F27 & 00:01:41.042 & +23:29:32.4 &   4584 &  28$\pm$2 &   0.16$\pm$0.02 &    1.0 &       2.10$\pm$0.05 &       0.43  &      -1.75 & 0\\ 
F28 & 00:01:40.988 & +23:29:35.2 &   4542 &  40$\pm$3 &   0.14$\pm$0.02 &    1.0 &       2.05$\pm$0.05 &       0.43  &      -1.75 & 0\\ 
F29 & 00:01:41.015 & +23:29:34.9 &   4557 &  66$\pm$5 &   0.25$\pm$0.03 &    1.0 &       2.30$\pm$0.05 &       0.43  &      -1.75 & 0\\ 
F30 & 00:01:41.002 & +23:29:34.7 &   4535 &  68$\pm$5 &   0.33$\pm$0.04 &    1.0 &       2.42$\pm$0.05 &       0.43  &      -1.75 & 0\\ 
F31 & 00:01:41.033 & +23:29:34.4 &   4481 &  36$\pm$3 &   0.13$\pm$0.01 &    1.0 &       2.00$\pm$0.05 &       0.43  &      -1.75 & 0\\ 
F32 & 00:01:41.058 & +23:29:31.6 &   4524 &  32$\pm$6 &   0.07$\pm$0.02 &    1.0 &       1.77$\pm$0.11 &       0.43  &      -1.75 & 0\\ 
F33 & 00:01:41.057 & +23:29:31.3 &   4520 &  30$\pm$2 &   0.13$\pm$0.02 &    1.0 &       2.03$\pm$0.05 &       0.43  &      -1.75 & 0\\ 
F34 & 00:01:41.040 & +23:29:31.0 &   4534 &  57$\pm$11 &   0.13$\pm$0.04 &    0.4 &       2.02$\pm$0.11 &      0.43  &      -1.75 & 0\\
F35 & 00:01:41.022 & +23:29:30.9 &   4554 &  38$\pm$8 &   0.09$\pm$0.03 &    1.0 &       1.85$\pm$0.11 &       0.43  &      -1.75 & 0\\ 
G1 & 00:01:40.938 & +23:29:31.6 &   4402 &  98$\pm$20 &   0.16$\pm$0.05 &    0.5 &       2.11$\pm$0.11 &       0.43  &      -1.75 & 0\\
G2 & 00:01:40.932 & +23:29:31.2 &   4417 &  88$\pm$18 &   0.20$\pm$0.06 &    1.0 &       2.19$\pm$0.11 &       0.43  &      -1.75 & 0\\
G3 & 00:01:40.812 & +23:29:30.1 &   4433 &  52$\pm$8 &   0.15$\pm$0.03 &    1.0 &       2.07$\pm$0.08 &        0.43  &      -1.75 & 0\\ 
G4 & 00:01:40.811 & +23:29:29.7 &   4428 &  57$\pm$5 &   0.22$\pm$0.02 &    1.0 &       2.23$\pm$0.05 &        0.43  &      -1.75 & 0\\ 
G5 & 00:01:40.812 & +23:29:29.4 &   4426 &  75$\pm$6 &   0.26$\pm$0.03 &    1.0 &       2.31$\pm$0.05 &        0.43  &      -1.75 & 0\\ 
G6 & 00:01:40.827 & +23:29:29.2 &   4437 &  65$\pm$5 &   0.26$\pm$0.03 &    1.0 &       2.32$\pm$0.05 &        0.43  &      -1.75 & 0\\ 
G7 & 00:01:40.807 & +23:29:29.1 &   4434 &  58$\pm$5 &   0.29$\pm$0.03 &    1.0 &       2.36$\pm$0.05 &        0.43  &      -1.75 & 0\\ 
G8 & 00:01:40.835 & +23:29:29.0 &   4426 &  65$\pm$5 &   0.26$\pm$0.03 &    1.0 &       2.32$\pm$0.05 &        0.43  &      -1.75 & 0\\ 
G9 & 00:01:40.762 & +23:29:30.4 &   4412 &  30$\pm$2 &   0.12$\pm$0.01 &    1.0 &       1.99$\pm$0.05 &        0.43  &      -1.75 & 0\\ 
G10 & 00:01:40.756 & +23:29:30.0 &   4448 &  44$\pm$4 &   0.18$\pm$0.02 &    1.0 &       2.16$\pm$0.05 &       0.43  &      -1.75 & 0\\ 
G11 & 00:01:40.751 & +23:29:29.7 &   4455 &  43$\pm$3 &   0.19$\pm$0.02 &    1.0 &       2.17$\pm$0.05 &       0.43  &      -1.75 & 0\\ 
G12 & 00:01:40.721 & +23:29:29.8 &   4463 &  38$\pm$3 &   0.21$\pm$0.02 &    1.0 &       2.22$\pm$0.05 &       0.43  &      -1.75 & 0\\ 
G13 & 00:01:40.733 & +23:29:29.5 &   4464 &  38$\pm$3 &   0.17$\pm$0.02 &    1.0 &       2.12$\pm$0.05 &       0.43  &      -1.75 & 0\\ 
G14 & 00:01:40.712 & +23:29:29.5 &   4473 &  48$\pm$4 &   0.18$\pm$0.02 &    1.0 &       2.15$\pm$0.05 &       0.43  &      -1.75 & 0\\ 
G15 & 00:01:40.721 & +23:29:29.3 &   4473 &  60$\pm$5 &   0.21$\pm$0.02 &    1.0 &       2.22$\pm$0.05 &       0.43  &      -1.75 & 0\\ 
G16 & 00:01:40.727 & +23:29:28.9 &   4459 &  44$\pm$4 &   0.15$\pm$0.02 &    1.0 &       2.08$\pm$0.05 &       0.43  &      -1.75 & 0\\ 
G17 & 00:01:40.736 & +23:29:28.7 &   4441 &  46$\pm$4 &   0.16$\pm$0.02 &    1.0 &       2.11$\pm$0.05 &       0.43  &      -1.75 & 0\\ 
H1 & 00:01:41.605 & +23:29:40.0 &   4437 &  47$\pm$9 &   0.11$\pm$0.03 &    0.4 &       1.96$\pm$0.11 &        0.43  &      -1.75 & 0\\
H2 & 00:01:41.650 & +23:29:40.9 &   4377 &  36$\pm$7 &   0.06$\pm$0.02 &    0.4 &       1.70$\pm$0.11 &        0.43  &      -1.75 & 0\\
H3 & 00:01:41.644 & +23:29:42.3 &   4341 &  39$\pm$3 &   0.14$\pm$0.02 &    1.0 &       2.04$\pm$0.05 &        0.43  &      -1.75 & 0\\ 
H4 & 00:01:41.648 & +23:29:42.7 &   4344 &  39$\pm$3 &   0.13$\pm$0.02 &    1.0 &       2.03$\pm$0.05 &        0.43  &      -1.75 & 0\\ 
H5 & 00:01:41.632 & +23:29:42.9 &   4341 &  25$\pm$2 &   0.11$\pm$0.01 &    1.0 &       1.92$\pm$0.05 &        0.43  &      -1.75 & 0\\ 
H6 & 00:01:41.620 & +23:29:43.1 &   4359 &  32$\pm$3 &   0.14$\pm$0.02 &    1.0 &       2.05$\pm$0.05 &        0.43  &      -1.75 & 0\\ 
H7 & 00:01:41.585 & +23:29:43.3 &   4380 &  30$\pm$2 &   0.15$\pm$0.02 &    1.0 &       2.07$\pm$0.05 &        1.03  &      -1.37 & 1 \\ 
H8 & 00:01:41.585 & +23:29:43.6 &   4386 &  30$\pm$2 &   0.14$\pm$0.02 &    1.0 &       2.06$\pm$0.05 &        2.22  &      -1.03 & 1 \\ 
H9 & 00:01:41.622 & +23:29:43.5 &   4375 &  57$\pm$5 &   0.37$\pm$0.04 &    1.0 &       2.47$\pm$0.05 &        3.02  &      -0.90 & 1 \\ 
H10 & 00:01:41.628 & +23:29:43.8 &   4384 &  72$\pm$6 &   0.42$\pm$0.05 &    1.0 &       2.52$\pm$0.05 &       3.06  &      -0.90 & 1 \\ 
H11 & 00:01:41.606 & +23:29:43.7 &   4389 &  68$\pm$5 &   0.43$\pm$0.05 &    1.0 &       2.54$\pm$0.05 &       5.37  &      -0.65 & 1 \\ 
H12 & 00:01:41.615 & +23:29:44.0 &   4390 &  82$\pm$7 &   0.41$\pm$0.05 &    1.0 &       2.51$\pm$0.05 &       3.07  &      -0.89 & 1 \\ 
H13 & 00:01:41.601 & +23:29:45.4 &   4403 &  35$\pm$3 &   0.23$\pm$0.03 &    1.0 &       2.26$\pm$0.05 &       ---          &        --- &  -1 \\ 
H14 & 00:01:41.599 & +23:29:45.7 &   4408 &  38$\pm$3 &   0.19$\pm$0.02 &    1.0 &       2.18$\pm$0.05 &       ---          &        --- &  -1 \\
H15 & 00:01:41.597 & +23:29:46.0 &   4419 &  56$\pm$4 &   0.28$\pm$0.03 &    1.0 &       2.35$\pm$0.05  &       ---          &        --- &  -1 \\
H16 & 00:01:41.588 & +23:29:46.4 &   4426 &  42$\pm$3 &   0.20$\pm$0.02 &    1.0 &       2.19$\pm$0.05  &       ---          &        --- &  -1 \\
H17 & 00:01:41.581 & +23:29:46.6 &   4429 &  60$\pm$5 &   0.21$\pm$0.02 &    1.0 &       2.22$\pm$0.05  &       ---          &        --- &  -1 \\
H18 & 00:01:41.577 & +23:29:47.0 &   4428 &  44$\pm$4 &   0.16$\pm$0.02 &    1.0 &       2.11$\pm$0.05  &       ---          &        --- &  -1 \\
H19 & 00:01:41.574 & +23:29:47.2 &   4461 & 104$\pm$8 &   0.34$\pm$0.04 &    0.7 &       2.43$\pm$0.05  &       ---          &        --- &  -1 \\
I1 & 00:01:41.779 & +23:29:43.2 &   4344 &  40$\pm$3 &   0.17$\pm$0.02 &    1.0 &       2.13$\pm$0.05  &       ---          &        --- &  -1 \\
I2 & 00:01:41.786 & +23:29:43.9 &   4505 &  25$\pm$2 &   0.11$\pm$0.01 &    1.0 &       1.93$\pm$0.05  &       ---          &        --- &  -1 \\
I3 & 00:01:41.772 & +23:29:44.5 &   4525 &  25$\pm$2 &   0.11$\pm$0.01 &    1.0 &       1.96$\pm$0.05  &       ---          &        --- &  -1 \\
I4 & 00:01:41.761 & +23:29:44.9 &   4534 &  25$\pm$2 &   0.11$\pm$0.01 &    1.0 &       1.95$\pm$0.05  &       ---          &        --- &  -1 \\
I5 & 00:01:41.754 & +23:29:45.2 &   4533 &  42$\pm$3 &   0.18$\pm$0.02 &    1.0 &       2.15$\pm$0.05  &       ---          &        --- &  -1 \\
J1 & 00:01:40.293 & +23:29:48.6 &   4411 &  82$\pm$7 &   0.27$\pm$0.03 &    1.0 &       2.33$\pm$0.05 &        0.43  &      -1.75 & 0\\ 
J2 & 00:01:40.265 & +23:29:48.6 &   4419 &  65$\pm$5 &   0.22$\pm$0.02 &    1.0 &       2.23$\pm$0.05 &        0.43  &      -1.75 & 0\\ 
J3 & 00:01:39.946 & +23:29:50.2 &   4534 &  56$\pm$11 &   0.10$\pm$0.03 &    0.4 &       1.89$\pm$0.11 &       0.43  &      -1.75 & 0\\
J4 & 00:01:39.856 & +23:29:51.8 &   4514 &  52$\pm$4 &   0.18$\pm$0.02 &    1.0 &       2.16$\pm$0.05  &       ---          &        --- &  -1 \\
J5 & 00:01:39.764 & +23:29:52.1 &   4468 &  64$\pm$5 &   0.24$\pm$0.03 &    1.0 &       2.28$\pm$0.05  &       ---          &        --- &  -1 \\
J6 & 00:01:39.740 & +23:29:52.3 &   4440 &  93$\pm$19 &   0.17$\pm$0.05 &    0.4 &       2.12$\pm$0.11 &       ---          &        --- &  -1 \\
J7 & 00:01:39.143 & +23:29:50.0 &   4154 &  53$\pm$4 &   0.21$\pm$0.02 &    1.0 &       2.22$\pm$0.05  &       ---          &        --- &  -1 \\
J8 & 00:01:39.157 & +23:29:48.0 &   4155 &  51$\pm$8 &   0.15$\pm$0.03 &    1.0 &       2.07$\pm$0.08  &       ---          &        --- &  -1 \\
AE1 & 00:01:40.743 & +23:29:22.3 &   4419 &  52$\pm$4 &   0.18$\pm$0.02 &    1.0 &       2.14$\pm$0.05  &       ---          &        --- &  -1 \\
AE2 & 00:01:40.075 & +23:29:21.1 &   4301 &  79$\pm$16 &   0.14$\pm$0.04 &    1.0 &       2.05$\pm$0.11 &       ---          &        --- &  -1 \\
AE3 & 00:01:40.049 & +23:29:21.1 &   4244 &  42$\pm$6 &   0.13$\pm$0.03 &    1.0 &       2.01$\pm$0.08  &       ---          &        --- &  -1 \\
AE4 & 00:01:40.212 & +23:29:15.0 &   4397 &  38$\pm$3 &   0.16$\pm$0.02 &    1.0 &       2.11$\pm$0.05  &       ---          &        --- &  -1 \\
AE5 & 00:01:40.108 & +23:29:12.0 &   4492 &  59$\pm$9 &   0.17$\pm$0.03 &    1.0 &       2.12$\pm$0.08  &       ---          &        --- &  -1 \\
AE6 & 00:01:39.841 & +23:29:09.0 &   4543 &  37$\pm$3 &   0.13$\pm$0.01 &    1.0 &       2.01$\pm$0.05  &       ---          &        --- &  -1 \\
AE7 & 00:01:39.805 & +23:29:08.5 &   4498 &  50$\pm$10 &   0.13$\pm$0.04 &    1.0 &       2.00$\pm$0.11 &       ---          &        --- &  -1 \\
AE8 & 00:01:38.706 & +23:29:20.2 &   4329 &  64$\pm$5 &   0.36$\pm$0.04 &    1.0 &       2.45$\pm$0.05  &       ---          &        --- &  -1 \\
AE9 & 00:01:38.585 & +23:29:20.0 &   4326 &  40$\pm$3 &   0.21$\pm$0.02 &    1.0 &       2.23$\pm$0.05  &       ---          &        --- &  -1 \\
\enddata
\tablenotetext{a}{Regions A1 to A4 have multiple CO components 1a, 1b, 2a, 2b.. etc. Their integrated Pa$\alpha$ flux and SFR properties are combined. }
\tablenotetext{b}{Heliocentric radial velocity using the optical definition.}
\tablenotetext{c}{FWHM corrected for instrumental resolution.}
\tablenotetext{d}{Quality factor of CO spectrum,  1 = good, 0.7 = weak detection, 0.5 = very weak, 0.4 = poor, 0 = undetected in CO line. Only quality 1 are shown in the figures.}
\tablenotetext{e}{Molecular surface density for nominal case of X$_{\rm CO}$ = 1/4 X$_{\rm CO,20}$, see text.}
\tablenotetext{f}{Surface flux and star formation rates estimated from the Pa$\alpha$ emission assuming minimal (Case 1) extinction (see \S 5.1). }
\tablenotetext{g}{Pa$\alpha$ detection flags; Flag = 0 indicates an upper limit, Flag = 1 indicates a detection, and Flag = -1 indicates that the region lies outside the  Pa$\alpha$ NICMOS coverage area.}
\end{deluxetable*}
\end{longrotatetable}

\begin{figure*}[ht]
\includegraphics[width=0.9\textwidth]{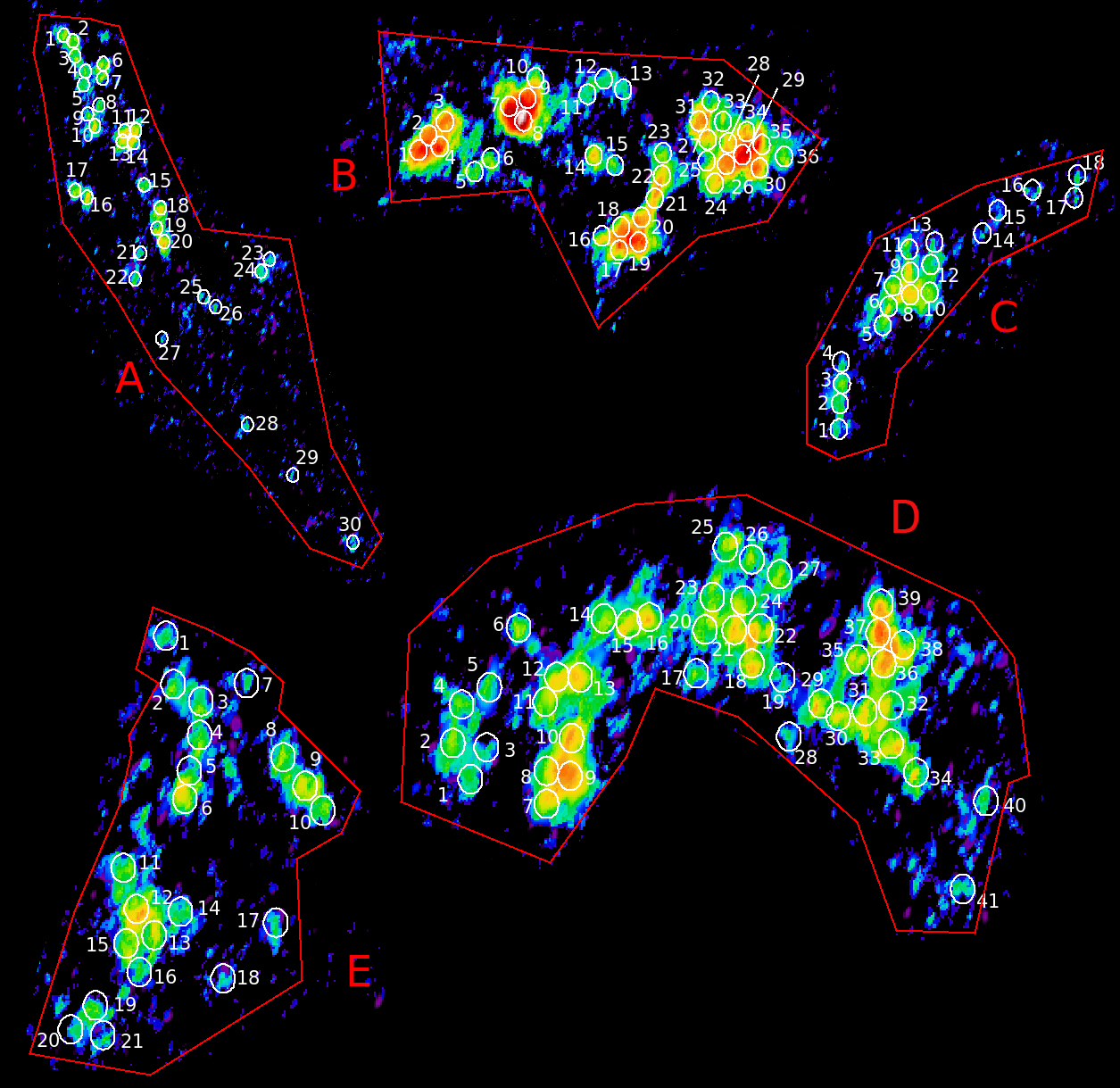}
\caption{Extraction apertures for regions A, B, C, D, and E superimposed on the integrated CO surface density image. }
\label{fig:A2}
\end{figure*} 

\begin{figure*}[ht]
\includegraphics[width=0.9\textwidth]{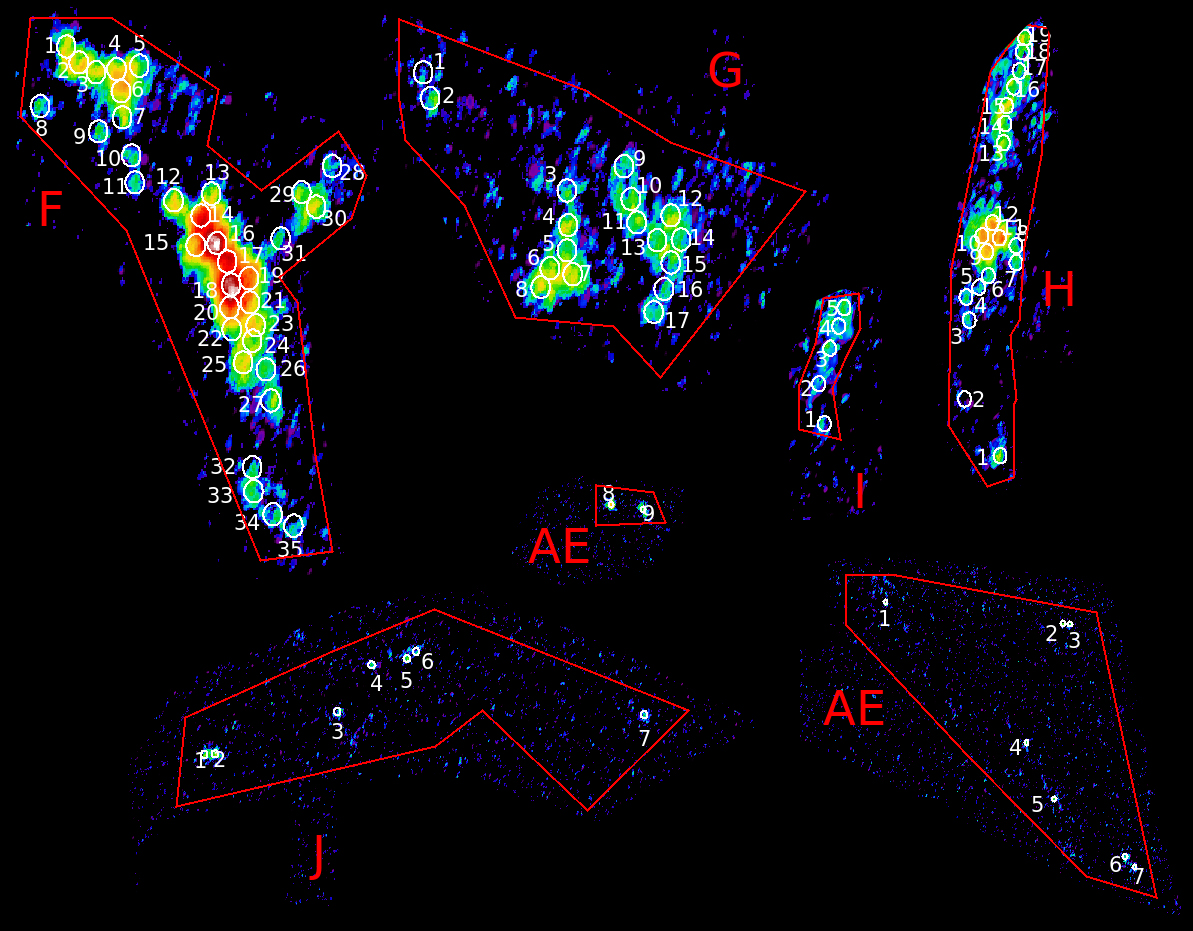}
\caption{Extraction apertures for regions F, G, H, I, J, AE1 and AE2 superimposed on the integrated CO surface density image. }
\label{fig:A3}
\end{figure*}





\end{document}